\documentclass[12pt]{article} 

\usepackage{amsmath,amssymb,amsfonts}
\usepackage{psfrag}
\usepackage{color}
\definecolor{darkblue}{rgb}{0.1,0.1,.7}
\definecolor{purple}{rgb}{0.6,0,0.6}
\definecolor{orange}{rgb}{0.9,0.6,0}
\usepackage[colorlinks, linkcolor=darkblue, citecolor=darkblue, urlcolor=darkblue, linktocpage]{hyperref} 
\usepackage[square, comma, compress,numbers]{natbib}
\usepackage[]{amsmath}
\usepackage[]{graphicx}
\usepackage[]{latexsym}
\usepackage[utf8]{inputenc}
\usepackage{geometry}
\usepackage{amscd}
\usepackage[all,cmtip]{xy}
\usepackage{mathrsfs}
\usepackage[margin=10pt,font=small,labelfont=bf]{caption}
\usepackage{dsdshorthand}
\usepackage{changepage}
\usepackage{setspace}
\usepackage{tikz}
\usepackage{subcaption}

\usepackage[titles]{tocloft}
\def\Dg{\Delta_{\rm gap}}
\def\O{{\cal O}}
\def\zb{\overline z}
\def\hb{\overline h}
\def\eqr{\eqref}

\def\ssec{\subsection}
\def\sssec{\subsubsection}
\def\rar{\rightarrow}
\def\la{\langle}
\def\ra{\rangle}
\newcommand{\es}[2] {\begin{equation} \label{#1} \begin{split} #2 \end{split} \end{equation}}
\newcommand{\eq}[2] {\begin{equation} \label{#1} #2 \end{equation}}
\def\o{\over}

\newcommand\Disc{\textrm{Disc}}

\newcommand\PFQ[5]{{}_#1 F_#2\left(\genfrac{}{}{0pt}{}{#3}{#4};#5\right)}

\numberwithin{equation}{section}

\begin{document}
\pagenumbering{gobble}
\vspace*{-.6in} \thispagestyle{empty}
\begin{flushright}
\hfill{\tt CALT-TH-2018-013 }\\
\hfill{\tt PUPT-2550}
\end{flushright}
\vspace{.2in} {\Large
\begin{center}
{\bf The Conformal Bootstrap at Finite Temperature \\\vspace{.1in}}
\end{center}
}
\vspace{.2in}
\begin{center}
{\bf 
Luca Iliesiu$^{1}$,
Murat Kolo\u{g}lu$^{2}$,
Raghu Mahajan$^{1,3}$,\\
Eric Perlmutter$^{2}$,
David Simmons-Duffin$^{2}$} 
\\
\vspace{.2in} 
$^1${\it Department of Physics, Princeton University, Princeton, NJ 08540, USA\/} \\
$^2${\it Walter Burke Institute for Theoretical Physics, Caltech, Pasadena, CA 91125, USA\/}\\
$^3${\it School of Natural Sciences, Institute for Advanced Study, Princeton, NJ 08540, USA\/}
\end{center}

\vspace{.2in}

\begin{abstract}
We initiate an approach to constraining conformal field theory (CFT) data at finite temperature using methods inspired by the conformal bootstrap for vacuum correlation functions. We focus on thermal one- and two-point functions of local operators on the plane. The KMS condition for thermal two-point functions is cast as a crossing equation. By studying the analyticity properties of thermal two-point functions, we derive a ``thermal inversion formula'' whose output is the set of thermal one-point functions for all operators appearing in a given OPE. This involves identifying a kinematic regime which is the analog of the Regge regime for four-point functions. We demonstrate the effectiveness of the inversion formula by recovering the spectrum and thermal one-point functions in mean field theory, and computing thermal one-point functions for all higher-spin currents in the critical $O(N)$ model at leading order in $1/N$. Furthermore, we develop a systematic perturbation theory for thermal data in the large spin, low-twist spectrum of any CFT. We explain how the inversion formula and KMS condition may be combined to algorithmically constrain CFTs at finite temperature. Throughout, we draw analogies to the bootstrap for vacuum four-point functions. Finally, we discuss future directions for the thermal conformal bootstrap program, emphasizing applications to various types of CFTs, including those with holographic duals.
\end{abstract}

\newpage

{\small\tableofcontents}

\pagenumbering{arabic}

\section{Introduction}
\label{sec:intro}

One of the basic operations in quantum field theory (QFT) is dimensional reduction on a circle. When we interpret the circle as Euclidean time (and impose appropriate boundary conditions) this corresponds to studying a QFT at nonzero temperature $T=1/\b$, where $\b$ is the length of the circle.\footnote{This notion of temperature is distinct from the temperature of a classical statistical theory  that one tunes to reach a critical point. The latter is simply a relevant coupling in the effective action. 
For example, the critical $O(2)$ model deformed by a relevant singlet describes an XY magnet (a $3$-dimensional classical theory) away from criticality. 
However, the critical $O(2)$ model compactified on a circle describes the nonzero temperature physics of the quantum critical point separating the superfluid and insulating phases of a thin film (a (2+1)-dimensional quantum theory) \cite{PhysRevB.44.6883,PhysRevLett.95.180603,Katz:2014rla,Kos:2015mba}.} When we interpret the circle as a spatial direction, this is Kaluza-Klein compactification.

In this work, we use bootstrap techniques to study conformal field theories (CFTs) on $S^1 \x \R^{d-1}$, focusing mostly on $d>2$. This setting is important for several reasons. Firstly, quantum critical points always have nonzero temperature in the laboratory, so it is crucial to compute  observables in this regime to make contact with experiment.\footnote{Note that we must analytically continue Euclidean correlators on $S^1 \x \R^{d-1}$ to describe real-time correlators of a Lorentzian theory at finite temperature.} More abstractly, $S^1\x\R^{d-1}$ is perhaps the simplest manifold not conformally-equivalent to $\R^d$ (when $d>2$). This poses an important challenge for bootstrap techniques. Ideally, any nonperturbative solution of a QFT should describe its observables on arbitrary manifolds.\footnote{At least when the theory on that manifold makes sense. See appendix~\ref{sec:subtletieswithdimred} for a discussion of subtleties that can arise in compactification on $S^1$ and other manifolds. See \cite{Nakayama:2016cim,Hasegawa:2016piv,Hasegawa:2018yqg} for previous work on the bootstrap in $d>2$ on nontrivial manifolds.} Finally, in the context of holography \cite{Maldacena:1997re,Gubser:1998bc,Witten:1998qj}, finite-temperature CFTs are dual to AdS black holes, and we obtain valuable information about both by translating between them.

CFT correlators on $S^1 \x \R^{d-1}$ are a limit of correlators on $S^1 \x S^{d-1}$, where we take the radius of the $S^{d-1}$ to be much larger than the length of the $S^1$. An advantage of this point of view is that states on $S^{d-1}$ are understood in principle via the state-operator correspondence. However, this limit is difficult to take in practice. Current bootstrap techniques work best at small twist $\tau=\De-J \sim \cO(1)$. However, the above limit requires knowledge of the spectrum and OPE coefficients at large dimension $\De$, which is usually out of reach. We would like an alternative approach that more directly constrains finite-temperature observables. We would also like an approach that could work for other compactifications, for instance, on the torus $T^d$.

In \cite{ElShowk:2011ag}, El-Showk and Papadodimas identified an interesting crossing equation for a two-point function on $S_\b^1 \x \R^{d-1}$ (here $\b$ denotes the length of the $S^1$). Because this geometry is conformally flat, one can compute two-point functions using the operator product expansion (OPE), assuming the points are sufficiently close together. The new data entering this computation are thermal one-point functions. For example, the one-point function of a scalar operator is
\be
\<\cO\>_\b \equiv \<\cO\>_{S_\b^1 \x \R^{d-1}} &= \frac{b_\cO}{\b^{\De_\cO}} = b_{\cO} T^{\Delta_\cO}.
\ee
The $\beta$ dependence of $\<\cO\>_\b$ is fixed by the scale symmetry, but the coefficient $b_\cO$ is not fixed by symmetry.
The OPE gives an expression for a thermal two-point function that can be schematically written as:
\be
\label{eq:opeexpansionfortwopt}
g(\tau) \equiv \<\f(\tau) \f(0)\>_{S_\b^1 \x \R^{d-1}} &\sim \frac{1}{|\tau|^{2\De_\phi}}\sum_{\cO\in \f\x\f} \frac{f_{\f\f\cO} b_\cO}{c_\cO} \left|\frac{\tau}{\b}\right|^{\De_\cO},
\ee
where $f_{\phi\phi\cO}$ is the OPE coefficient of $\cO$, $\Delta_{\cO}$ is the scaling dimension of $\cO$, and $c_\cO$ is the two-point coefficient of $\cO$ in the vacuum.\footnote{It is conventional to normalize $c_\cO$ to 1. However, some operators like the stress tensor have their own canonical normalization coming from Ward identities.} 
For simplicity, we have taken the operators to be separated only in the circle direction with distance $\tau$. (In section~\ref{sec:CFTs-finite-T-review} we study more general kinematics.) The KMS condition for the two-point function of identical bosonic operators separated only along Euclidean time reads
\be
g(\tau) &= g(\b-\tau).
\label{eq:kmstimeonly}
\ee
El-Showk and Papadodimas noted that \eqr{eq:opeexpansionfortwopt} does not manifestly satisfy \eqr{eq:kmstimeonly}. Imposing the KMS condition therefore gives a nontrivial ``thermal crossing equation". This constrains the $b_\cO$'s in terms of the other data of the CFT, namely the OPE coefficients $f_{\f\f\cO}$ and dimensions $\De_\cO$. Via the limit $S^1 \x S^{d-1}\to S^1\x\R^{d-1}$, this equation can be understood as a consequence of the usual crossing symmetry for four-point functions where we sum over some of the ``external" operators.

As we explain in section~\ref{sec:CFTs-finite-T-review}, the one-point coefficients $b_\cO$, together with the usual CFT data $f_{ijk}$, $\De_i$, determine all finite-temperature correlators. Thus, our focus will be on computing thermal one-point coefficients using nonperturbative methods. We should note however that many interesting finite-temperature observables, like e.g. the thermal mass (discussed in section~\ref{eq:kkquantization}), are difficult to extract from thermal one-point functions. Such observables are an even more challenging target for the future.

The thermal crossing equation is problematic for numerical bootstrap techniques because the expansion (\ref{eq:opeexpansionfortwopt}) has coefficients $f_{\f\f\cO} b_\cO/c_\cO$ that are not sign-definite. Sign-definiteness is crucial for the linear programming-based method of \cite{Rattazzi:2008pe} and its generalizations \cite{Poland:2011ey,El-Showk:2014dwa,Paulos:2014vya,Simmons-Duffin:2015qma}. In this sense, the thermal bootstrap is similar to the boundary bootstrap \cite{Liendo:2012hy,Gliozzi:2015qsa,Rastelli:2017ecj}, defect bootstrap \cite{Billo:2016cpy,Gadde:2016fbj,Liendo:2016ymz,Lauria:2017wav,Lemos:2017vnx}, and four-point bootstrap in non-unitary CFTs \cite{Gliozzi:2014jsa,Hikami:2017hwv}. Our strategy will be to develop analytical approaches to the thermal crossing equation, with the hope of eventually applying them (perhaps in conjunction with numerics) to CFTs whose spectrum and OPE coefficients are relatively well-understood, like e.g.\ the 3d Ising model \cite{ElShowk:2012ht,El-Showk:2014dwa,Kos:2014bka,Simmons-Duffin:2015qma,Kos:2016ysd,Simmons-Duffin:2016wlq}.\footnote{Our methods share many similarities with the recent defect bootstrap analysis in \cite{Lemos:2017vnx}.} We should note that most of our methods will apply with any choice of boundary conditions around the circle (perhaps with slight modifications). Although our focus will be on finite-temperature, one could also study supersymmetric compactifications, or compactifications with more general chemical potentials.

A general and powerful analytic bootstrap technique that can be applied to our problem is large-spin perturbation theory \cite{Simmons-Duffin:2016wlq,Fitzpatrick:2012yx,Komargodski:2012ek,Alday:2015eya,Alday:2015ota,Alday:2015ewa,Alday:2016njk}. 
Large-spin perturbation theory was recently reformulated by Caron-Huot in terms of a Lorentzian inversion formula \cite{Caron-Huot:2017vep} (inspired by a classic formula of Froissart and Gribov in the context of $S$-matrix theory \cite{Gribov:1961fr,Froissart:1961ux}). 
Caron-Huot's formula expresses OPE coefficients and dimensions in terms of an integral of a four-point function in a Lorentzian regime. 
Inserting the OPE expansion in the $t$-channel into the inversion formula, one obtains a systematic large-spin expansion for $s$-channel data. 
This process can be iterated to obtain further information about the solution to crossing symmetry \cite{Simmons-Duffin:2016wlq,lspt1, lspt2, lspt3, Alday:2017zzv,Alday:2017vkk, lspt4, lspt5, turi}.

In section~\ref{sec:lorentzian-inversion-formula}, we derive a Lorentzian inversion formula for thermal one-point functions as an integral of a thermal-two point function. The integral is over an interesting Regge-like Lorentzian regime that is more natural from the point of view of Kaluza-Klein compactification than finite-temperature physics. Our formula is very close to the Froissart-Gribov $S$-matrix formula, and in fact our derivation is almost identical. However, our result relies on some (well-motivated) assumptions about analyticity properties and asymptotics of thermal two-point functions that we discuss further in sections~\ref{sec:analyticity-w-plane} and~\ref{s35}. Our formula shows that thermal one-point functions in conformal field theory are also analytic in spin, in the same way as OPE coefficients and operator dimensions.

In sections~\ref{sec:MFT-example} and \ref{sec:largeN-example}, we apply our inversion formula in some examples, including Mean Field Theory and the critical $O(N)$ model in $d=3$ at large $N$. We also discuss some aspects of thermal correlators in general large-$N$ theories, especially holographic CFTs with a large gap to single-trace higher-spin operators. For the $O(N)$ model, by studying the two-point function $\la \phi_i\phi_i\ra_\beta$ we derive the thermal one-point functions $b_\cO$ for all single-trace operators $\cO$. This includes the singlet higher-spin currents, $J_\ell \sim \phi_i \partial^{\ell}\phi_i$, where $\ell$ is a positive even integer. The result, which to our knowledge is new for $\ell>2$, can be found in \eqr{jlsum}-\eqr{bell} and is reproduced here:
\eq{bellintro}{{b_{J_\ell}\o \sqrt{c_{J_\ell}}} = {\sqrt{N 2^{\ell+1}\ell}\o\ell!}\sum_{n=0}^\ell {2^{n}\o n!}{(\ell-n+1)_n\o (2\ell-n+1)_n} m_{\text{th}}^n \text{Li}_{\ell+1-n}(e^{-m_{\text{th}}}).}
where $m_{\text{th}}\,\beta = 2\log\left({1+\sqrt{5}\o2}\right)$ is the thermal mass of the critical $O(N)$ model to leading order in $1/N$. We have normalized $b_{J_\ell}$ by the square root of the norm of $J_\ell$. Interestingly, this result exhibits uniform transcendentality of weight $\ell+1$, a feature that would be worth understanding more deeply. 
For $\ell=2$, the case of the stress tensor, the result matches that of Sachdev \cite{Sachdev:1993pr}. 
We also derive sums of thermal coefficients for scalar composite operators with dimension $\De=2,4,6,\dots$. 

Together with the thermal mass, these higher-spin one-point functions have an interesting interpretation in the context of the holographic duality of the critical $O(N)$ model to Vasiliev higher-spin gravity in AdS$_4$ (see e.g. \cite{Vasiliev:1999ba,Klebanov:2002ja, Giombi:2012ms}). In particular, they determine the complete set of higher-spin charges of the putative black hole solution dual to the CFT state at finite temperature. Thus, we now have the full set of higher-spin gauge-invariant data necessary to check, or perhaps even construct, a candidate solution in the bulk.

In section~\ref{sec:large-spin-perturbation-theory}, we use our inversion formula to develop large-spin perturbation theory for thermal one-point functions. This allows us to study the thermal data of arbitrary, strongly-interacting CFTs. 
Crucially, thermal two-point functions have different OPE channels with overlapping regimes of validity. Inverting terms in one channel to the other relates thermal coefficients of operators in the theory in nontrivial ways: one-point functions determine terms in the large-spin expansion of other one-point functions. These relations can be posed to formulate an analytic bootstrap problem for the thermal data. The required calculations are similar to (but simpler than) those that arise in the context of vacuum four-point functions.
For example, the one-point functions of low-twist operators at large spin are dominated by an analog of the double-lightcone limit, and one is interested in the discontinuity of the correlator (as opposed to the ``double discontinuity" \cite{Caron-Huot:2017vep}) in this limit. In fact, we see that the large-spin perturbation theory of spectral and OPE data and of thermal data are intimately tied together.

As an example, we find a universal contribution to one-point functions of ``double-twist" operators $[\f\f]_{0,J}$\footnote{The operators $[\f\f]_{0,J}$ have twist $\tau = \De-J = 2\De_\f + \g(J)$, where $\g(J)$ is an anomalous dimension that vanishes as $J\to \oo$. They can be thought of schematically as $[\f\f]_{0,J}=\f \partial_{\mu_1}\cdots\partial_{\mu_J} \f$.} \cite{Alday:2007mf,Fitzpatrick:2012yx,Komargodski:2012ek}, proportional to the free-energy density, 
\be
\label{eq:universalcontribution}
{b_{[\f\f]_{0, J}}} &\sim \frac{{c_{[\f\f]_{0, J}}}}{f_{\f\f[\f\f]_{0, J}}}\frac{2^{J+1}\p{1+\tfrac 1 2 \g'(J)}}{\G(1+J+\tfrac 1 2 \g(J))} \nn\\
&\quad \x \left[\frac{\G(\De_\f + J + \tfrac 1 2 \g(J))}{\G(\De_\f)} - \p{\frac{f d\, \vol(S^{d-1}) \De_\f}{4} \frac{c_\mathrm{free}}{c_T}} \frac{\G(\De_\f - \frac{d-2}{2} +J+ \frac 1 2 \g(J))}{\G(\De_\f-\frac{d-2}{2})}+\dots\right].
\ee
Here, $f=F/T^d<0$, where $F$ is the free energy density, $c_T$ is the stress-tensor two-point coefficient, $c_\mathrm{free}$ is $c_T$ for a free boson, and $\g(J)$ is the anomalous dimension of $[\f\f]_{0,J}$. The OPE coefficients $f_{\f\f[\f\f]_0}(J)$ and anomalous dimensions $\g(J)$ can be computed using the lightcone bootstrap for vacuum four-point functions \cite{Simmons-Duffin:2016wlq,Fitzpatrick:2012yx,Komargodski:2012ek,Alday:2015eya,Alday:2015ota,Alday:2015ewa,Alday:2016njk,Caron-Huot:2017vep}.\footnote{Note that \cite{Fitzpatrick:2012yx} uses the convention $c_\cO=(-\tfrac 1 2)^{J_\cO}$.} Our precise result is that the two sides of (\ref{eq:universalcontribution}) match to all orders in an expansion in $1/J$. (Our inversion formula also produces nonperturbative corrections in $J$.) The leading term in brackets is due to the unit operator. The stress tensor contribution is the second term in brackets, and it falls off like $1/J^{\frac{d-2}{2}}$ relative to the leading term. 
The dots represent similar contributions from other operators $\cO$  that are suppressed by $1/J^{\frac{\tau_\cO}{2}}$ where $\tau_\cO=\De_\cO-J_\cO$ is the twist of $\cO$. In particular, the stress tensor gives the next large-$J$ correction after the unit operator if it is the lowest-twist operator in the $\phi\times\phi$ OPE (other than the unit operator).

Also in section~\ref{sec:large-spin-perturbation-theory}, we discuss subtleties associated with sums over infinite families of operators, and how crossing symmetry of four-point functions is embedded in the thermal crossing equations. As an example, we apply our large-spin technology to the 3d Ising model. We conclude with discussion and comments on future directions in section~\ref{sec:conclusion}.  In appendix~\ref{app:trytogetbT}, we pursue the independent direction of studying the partition function on $S^1_\b\x S^{d-1}$, give a rough estimate of $b_T$ in the 3d Ising model from the $\b\to 0$ limit, and discuss further aspects of this limit in appendix~\ref{sec:detailsonthermalsdminus1}. The next appendices further elaborate on technical details in the main text.

\section{CFTs at nonzero temperature}
\label{sec:CFTs-finite-T-review}

\subsection{Low-point functions and the OPE}
\label{sec:OPE-and-1-pt}

Any CFT correlation function on $\R^d$ can be computed using the operator product expansion (OPE). Beginning with an $n$-point function $\<\cO_1 \cdots \cO_n\>$, we  recursively use the OPE to reduce it to a sum of $1$-point functions, for example
\be
\<\cO_1 \cdots \cO_n\> &= \sum_{k_1} C_{12k_1} \<\cO_{k_1} \cO_3 \cdots \cO_n\> \nn\\
&= \sum_{k_1} \cdots \sum_{k_{n-1}} C_{12k_1}C_{k_1 3 k_2}\cdots C_{k_{n-2}n k_{n-1}} \<\cO_{k_{n-1}}\>.
\ee
Here the $C_{ijk}$ are differential operators, and we have suppressed the position dependence of the $\cO_i$ for brevity. Each time we apply the OPE we must find a pair of operators $\cO_i,\cO_j$ and a sphere surrounding them such that all other operators lie outside this sphere. This is always possible for generic configurations of points in $\R^d$.
Finally, by translation invariance and dimensional analysis,\footnote{Here, we assume unitarity, which implies that only the unit operator has scaling dimension $0$.} one-point functions on $\R^d$ are given by
\begin{align}
\<\cO\>_{\R^d} &=
\begin{cases}
1 & \textrm{if $\cO=\mathbf{1}$}, \\
0 & \textrm{otherwise.}
\end{cases}
\end{align}

The same procedure works on any conformally-flat manifold $\cM^d$, but with two additional complications. Firstly, non-unit operators can have nonzero one-point functions. Secondly, depending on the configuration of operator insertions, it may not always be possible to perform the OPE. More specifically, to compute $\cO_i \x \cO_j$, we must find a sphere containing only $\cO_i$ and $\cO_j$ whose interior is flat (possibly after performing a Weyl transformation). However, the geometry of $\cM^d$ may make this impossible. 

In this work, we study CFTs on the manifold $\cM_\b = S^1_\b\x \R^{d-1}$. We use coordinates $x=(\tau,\bx)$ on $S^1_\b \x \R^{d-1}$, where $\tau$ is periodic $\tau\sim\tau+\b$. One-point functions on $\cM_\b$ are constrained by symmetries as follows. To begin, translation-invariance implies that descendant operators have vanishing one-point functions:
\be
\<P^\mu \cO(x)\>_\b &= \ptl^\mu \<\cO(x)\>_\b = \ptl^\mu \<\cO(0)\>_\b = 0.
\ee
(The notation $\<\cdots\>_\b$ denotes a correlator on $\cM_\b$.)
Thus, let us consider a primary operator $\cO$ with dimension $\De$ and $\SO(d)$ representation $\rho$.

The geometry $S^1 \x \R^{d-1}$ is clearly invariant under $\SO(d-1)$. It also has a discrete symmetry under which $\tau\leftrightarrow -\tau$. In general, our CFT may not be parity-invariant, so to get a symmetry of the theory, we should accompany $\tau\leftrightarrow -\tau$ with a reflection in one of the $\R^{d-1}$ directions. This combines with $\SO(d-1)$ to give the symmetry group $\mathrm{O}(d-1)\subset \SO(d)$, where  a reflection in $\mathrm{O}(d-1)$ also flips the sign of $\tau$.\footnote{In a parity-invariant theory, the rotational symmetry group would be $\mathrm{O}(d-1)\x \Z_2$, and we would have the restriction that only parity-even operators could have nonzero one-point functions.}  For $\<\cO\>_\b$ to be nonzero, the restriction of $\rho$ under $\mathrm{O}(d-1)\subset \SO(d)$ must contain the trivial representation
\be
\mathrm{Res}^{\SO(d)}_{\mathrm{O}(d-1)} \rho &\supset \mathbf{1}.
\ee
This implies that $\rho$ must be a symmetric traceless tensor (STT), with even spin $J$. Finally, the one-point function of a spin-$J$ operator $\cO$ is fixed by symmetry and dimensional analysis, up to a single dimensionless coefficient $b_\cO$:
\be
\label{eq:thermalonepoint}
\<\cO^{\mu_1\cdots \mu_J}(x)\>_{\b} &= \frac{b_\cO}{\b^\De}(e^{\mu_1}\cdots e^{\mu_J} - \textrm{traces}).
\ee
Here, $e^\mu$ is the unit vector in the $\tau$-direction. Here and in what follows, we are implicitly normalizing our correlators by the partition function, $Z(\b)$. 

\begin{figure}
    \centering
    \includegraphics[width=0.7\textwidth]{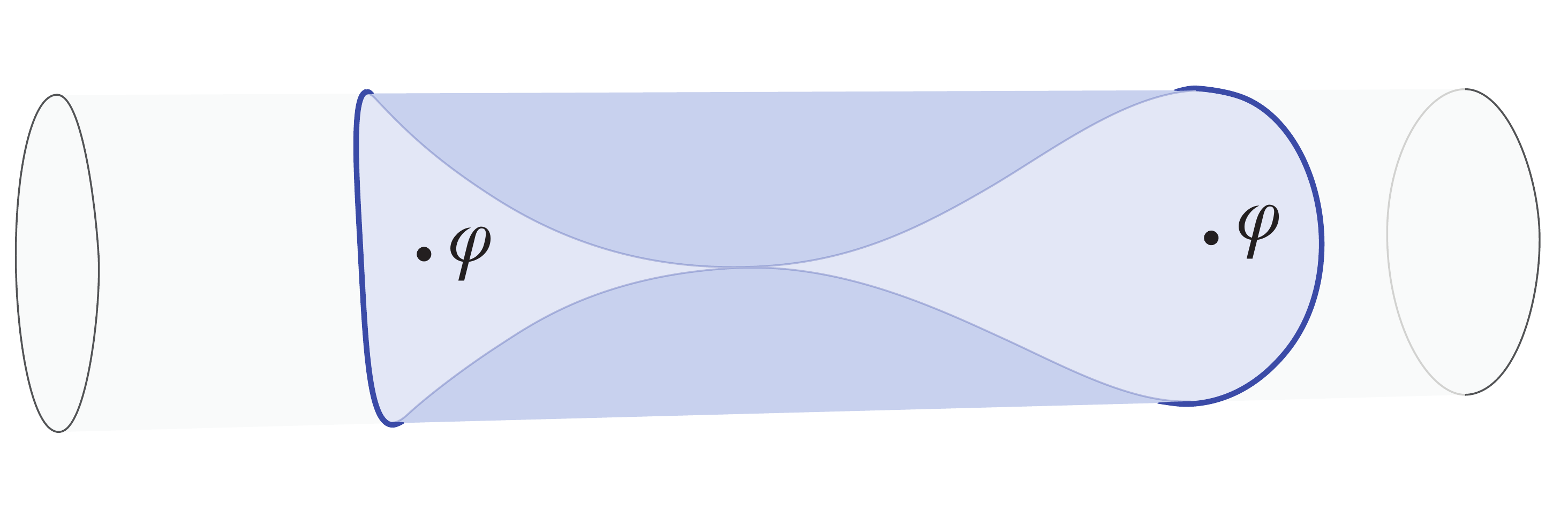}
    \caption{The OPE on $S^1_\b \x \R^{d-1}$ is valid if the two operators lie inside a sphere. The largest possible sphere has diameter $\beta$, wrapping entirely around the $S^1$ such that it is tangent to itself. Here, we illustrate such a sphere (blue) in $d=2$.}
    \label{fig:spheretangenttoself}
\end{figure}

We will find it convenient to study two-point functions, which encode the $b_\cO$'s via the OPE.\footnote{For previous discussions of the OPE for CFT two-point functions at finite temperature, see \cite{Katz:2014rla,Witczak-Krempa:2015pia}.} Note that, unlike in $\R^d$, two-point functions of non-identical operators may be nonvanishing on $\cM_\b$. However, for simplicity, we focus on two-point functions of identical operators,
\eq{}{g(\tau,\bx)=\<\f(x) \f(0)\>_{\b}.}
 The OPE is valid whenever both operators lie within a sphere whose interior is flat.
 In our case, the largest such sphere has diameter $\b$: it wraps entirely around the $S^1$ and is tangent to itself (figure~\ref{fig:spheretangenttoself}).  The condition for both $x$ and $0$ to lie within such a sphere is
\be
\label{eq:opeconvergencecondition}
|x|=\sqrt{\tau^2 + \bx^2} < \b, \qquad(\textrm{OPE convergence}).
\ee

Assuming $|x|<\b$, we can use the OPE to obtain
\be
\label{eq:usetheopetoobtain}
g(\tau,\bx)= \sum_{\cO\in \f\x\f} \frac{f_{\f\f\cO}}{c_\cO} |x|^{\De-2\De_\f-J}x_{\mu_1}\cdots x_{\mu_J} \<\cO^{\mu_1\cdots \mu_J}(0)\>_{\b}.
\ee
Here, $c_\cO$ is the coefficient in the two-point function of $\cO$,
\be
\<\cO^{\mu_1\cdots \mu_J}(x) \cO_{\nu_1\cdots \nu_J}(0)\>
&=
c_\cO\frac{I^{(\mu_1}_{(\nu_1} \cdots I^{\mu_J)}_{\nu_J)}-\textrm{traces}}{x^{2\De_\cO}},\quad I_\nu^\mu(x) = \de_\nu^\mu - \frac{2x_\nu x^\mu}{x^2},
\ee
and $f_{\phi\phi\cO}$ is the three-point coefficient
\be
\<\f(x_1)\f(x_2) \cO^{\mu_1\cdots \mu_J}(x_3)\> &= f_{\f\f\cO}\frac{Z^{\mu_1}\cdots Z^{\mu_J} - \textrm{traces}}{x_{12}^{2\De_\f-\De_\cO} x_{23}^{\De_\cO} x_{13}^{\De_\cO}},\quad
Z^\mu = \frac{x_{13}^\mu}{x_{13}^2} - \frac{x_{23}^\mu}{x_{23}^2}.
\ee
We often normalize $\cO$ so that $c_\cO=1$.
Note that because descendants have vanishing one-point functions, we need only the leading (non-derivative) term in the OPE for each multiplet.
Plugging (\ref{eq:thermalonepoint}) into (\ref{eq:usetheopetoobtain}), the index contraction is given by a Gegenbauer polynomial,\footnote{When the operators in the two-point function have spin, the appropriate generalization of the Gegenbauer polynomial is described in \cite{Kravchuk:2017dzd}.}
\be
|x|^{-J}(x_{\mu_1}\cdots x_{\mu_J})(e^{\mu_1}\cdots e^{\mu_J}-\textrm{traces}) &= \frac{J!}{2^J(\nu)_J} C^{(\nu)}_J(\eta),
\ee
where $\nu=\frac{d-2}{2}$, $(a)_n=\frac{\G(a+n)}{\G(a)}$ is the Pochhammer symbol, and $\eta = \frac{\tau}{|x|}$.   Thus, we obtain
\be
g(\tau,\bx) &= \sum_{\cO\in \f\x\f} \frac{a_\cO}{\b^\De} C_J^{(\nu)}(\eta) |x|^{\De-2\De_\f},\qquad\textrm{where}\nn\\
a_\cO &\equiv \frac{f_{\f\f\cO} b_\cO}{c_\cO} \frac{J!}{2^J (\nu)_J}.
\label{eq:blockdecomposition}
\ee
We can think of $|x|^{\De-2\De_\f} C_J^{(\nu)}(\eta)$ as a two-point conformal block on $S^1\times \R^{d-1}$.\footnote{Note that $a_\cO$ is independent of the normalization of $\cO$. We sometimes quote values for the combination $b_\cO/\sqrt{c_\cO}$, which changes sign under a redefinition $\cO\to -\cO$. We usually fix this ambiguity by choosing a sign for some OPE coefficient $f_{\f\f\cO}$ involving $\cO$.}

\subsubsection{Free energy density}
\label{sec:freenergydensity}

One of the most important thermal one-point coefficients is $b_T$, associated to the stress tensor $T^{\mu\nu}$. This is related to the free energy density of the thermal CFT as follows.
From (\ref{eq:thermalonepoint}), the energy density is given by\footnote{\label{foot:wick}The minus sign is because we are using conventions appropriate for Euclidean field theory. When Wick rotating from Euclidean to Lorentzian signature, it is conventional to include factors of $i$ in the $0$ components of tensor operators. This ensures that they go from tensors of $\SO(d)$ in Euclidean signature to tensors of $\SO(d-1,1)$ in Lorentzian signature. For the stress tensor, this means $T^{00}_\mathrm{Lorentzian} = i^2 T^{00}_\mathrm{Euclidean}$, so the expectation value of $T^{00}_\mathrm{Lorentzian}$ is positive as it should be, see e.g.\ \cite{Simmons-Duffin:2016gjk}.}
\begin{align}
    E(\beta) = -\langle T^{00} \rangle_\beta = -\p{1-\frac 1 d} b_T T^d >0\, .
\end{align}
 In particular, note that $b_T$ must be negative, by positivity of energy. By dimensional analysis, the free energy density $F$ must take the form $F=f T^d$, where $f$ is a dimensionless quantity. Using the thermodynamic relations $F = E - TS = E + T dF/dT$, we find
\begin{align}
    f = \frac{b_T}{d} < 0.
\end{align}
The Ward identity fixes
\begin{align}
f_{\phi\phi T} = - \frac{d}{d-1} \frac{\Delta_\phi}{S_d},\qquad S_d =\vol(S^{d-1}) = \frac{2\pi^{d/2}}{\G(d/2)}.    
\end{align}
Consequently, the coefficient of $T$ in the thermal block expansion of $\<\f\f\>_\b$ (\ref{eq:blockdecomposition}) is
\be
a_T &= -f S_d\frac{2\Delta_\f}{d-2} \frac{c_\mathrm{free}}{c_T},
\label{eq:afrelation}
\ee
where $c_\mathrm{free}=\frac{d}{d-1}\frac{1}{S_d^2}$ is stress tensor two-point coefficient for the free boson in $d$-dimensions \cite{Osborn:1993cr}.
For a single free (real) scalar, $b_T = -2 d \zeta(d) / S_d$, as can be checked by computing its free energy
\begin{align}
   F = f T^d = T \int \frac{d^{d-1}\mathbf{k}}{(2\pi)^{d-1}} \log[ 1 - \exp(-\beta \vert \mathbf{k} \vert)] = - \frac{2}{S_d}\z(d) T^d\, .
\end{align}
For the convenience of the reader, we now collect some known results for $b_T$ in various theories.
\begin{enumerate}
    \item For the free scalar in three dimensions, we have $b_T^\text{free} = -6\zeta(3)/(4\pi) \approx -0.57394$.
    \item For the $O(N)$ model in three dimensions at leading order in $1/N$, $b_T = 4N/5 \times  b_T^\text{free} \approx -0.45915 N$ \cite{csy, Sachdev:1993pr}. We will derive this from our inversion formula in section~\ref{sec:O(N)-model-example}.
    \item In the Monte Carlo literature, the quantity $f$ is known as the ``Casimir Amplitude". For the Ising model, Monte Carlo results give $f \approx -0.153$ \cite{PhysRevE.79.041142, casimir2, casimir3}, with numerical errors in the third digit. 
    This translates to $b_T^\text{Ising} \approx -0.459$.
    Note that $b_T^\text{Ising}$ is remarkably close to the value of $b_T/N$ for the $O(N)$ model at large $N$.\footnote{We estimate $b_T^\mathrm{Ising}$ using the known part of the spectrum of the 3d Ising model in appendix~\ref{app:trytogetbT}.} 
\end{enumerate}
 
 \subsubsection{Two dimensions}\label{s22d}
In $d=2$, $S^1_\b\times \R$ is conformal to the plane, so thermal correlators on the cylinder are determined by symmetry. All one-point functions vanish except for those of operators living in the Virasoro vacuum module:
\eq{1pt2d}{\la \cO\ra_{S^1_\b\times \R}=0~~\forall~~ \cO\notin \lbrace\mathbf{1},T^{\mu\nu},:\!T^{\mu\nu}T^{\rho\sigma}\!:\,,\ldots\rbrace\, .}
Likewise, two-point functions on $S^1_\b\times \R$ are determined via a conformal transformation as
\eq{2pt2d}{\la \cO_i(z,\zb)\cO_j(0,0)\ra_{S^1_\b\times \R} = \left({\b\o\pi}{\sinh\left({\pi z\o\b}\right)}\right)^{-2h_\cO} \left({\b\o\pi}{\sinh\left({\pi \zb\o\b}\right)}\right)^{-2\hb_\cO}\delta_{ij}.}
Unlike in $d>2$,  two-point functions of distinct operators vanish. It follows from Virasoro symmetry and \eqr{1pt2d} that the right-hand side of \eqr{2pt2d} is the two-point Virasoro $\times$ Virasoro vacuum block on the cylinder. By expanding \eqr{2pt2d} --- or using formulae of section~\ref{sec:inversion-d=2} --- one may extract the (weighted) sum of one-point coefficients $a_\O$ of all Virasoro descendants at a given level above the vacuum. These are, of course, determined by the action of the Schwarzian derivative \cite{difr, Gaberdiel:1994fs, Chen:2013dxa}. 

\subsubsection{From the sphere to the plane}
\label{sec:spheretoplane}

Thermal correlation functions are also naturally computed on $S^1_\b\times S_L^{d-1}$, owing to the role of spherical slicing in the state-operator correspondence. Due to the presence of the $S^{d-1}$ curvature radius $L$, these thermal correlators are less constrained by conformal invariance than their counterparts on $S^1_\b\times \R^{d-1}$. However, they must obey the flat space limit
\eq{}{\lim_{L\to \oo}\<\cO_1 \cdots \cO_n\>_{S^1_\b \times S^{d-1}_L}  = \<\cO_1 \cdots \cO_n\>_{S^1_\b \times \R^{d-1}}.}

One-point functions are fixed by dimensional analysis and spherical symmetry to take the form
\be
\label{eq:thermalonepointsphere}
\<\cO^{\mu_1\cdots \mu_J}(x)\>_{S^1_\b \times S^{d-1}_L} &= \frac{b_\cO f_{\cO}({\b\o L})}{\b^\De}(e^{\mu_1}\cdots e^{\mu_J} - \textrm{traces}).
\ee
where $f_\cO({\b\o L})$ is a function that is not determined by conformal symmetry; it obeys the boundary condition $f_\cO(0) =1$. On the other hand, employing radial quantization,
\eq{1ptope}{\<\cO^{\mu_1\cdots \mu_J}(x)\>_{S^1_\b \times S^{d-1}_L}  = \frac{1}{Z(\b)}\sum_{\cO'}e^{-\b \De_{\cO'}}\la \cO'|\cO^{\mu_1\cdots \mu_J}(x)|\cO'\ra,}
where the sum runs over all local operators $\cO'$ (not just primaries) and $Z(\beta) = \sum_{\cO'} e^{-\beta \Delta_{\cO'}}$ is the partition function. 
It is useful to introduce one-point thermal conformal blocks on the sphere via
\eq{1ptsphere}{\<\cO^{\mu_1\cdots \mu_J}(x)\>_{S^1_\b \times S^{d-1}_L}=  \frac{1}{Z(\b)}\sum_{\text{Primary }\cO'}f_{\cO\cO'\cO'}F(h_{\cO},\hb_{\cO};h_{\cO'},\hb_{\cO'}|\b)(e^{\mu_1}\cdots e^{\mu_J} - \textrm{traces}),}
where $F(h_{\cO},\hb_{\cO};h_{\cO'},\hb_{\cO'}|\b)$ captures all contributions of the conformal family of $\cO'$ to $\<\cO^{\mu_1\cdots \mu_J}(x)\>_{S^1_\b \times S^{d-1}_L}$. 
We have set $L=1$, and introduced the left- and right-moving conformal weights
\eq{}{h_\cO={\De_\cO-J\o2}~, \quad \hb_\cO={\De_\cO+J\o2},}
and likewise for $\cO'$. These blocks were recently computed in any $d$, for scalar $\cO$ and scalar $\cO'$, in \cite{maloney}. Two-point functions may also be written using the OPE and a sum over states, although we refrain from showing the details here. 

Consistency of \eqr{1ptsphere} with the flat space limit \eqr{eq:thermalonepoint} can in principle be established by taking $\b\rar 0$, which involves contributions from all high-energy states. We discuss further details of this limit and thermal blocks on $S^1\x S^{d-1}$ in appendix~\ref{sec:detailsonthermalsdminus1}. The general lesson is that exact computation of $\la \cO\ra_{S^1_\b\times \R^{d-1}}$ by passage from $S^1_\b\times S^{d-1}_L$ is challenging. Perhaps the simplest observable to compute using these methods is $b_T$, and we explore this possibility in appendix~\ref{app:trytogetbT} with the free boson and 3d Ising model as examples. 
The rest of this paper is devoted to developing new methods directly on $S^1_\b\times \R^{d-1}$. 

\subsection{The KMS condition and crossing}

\label{sec:KMS&crossing}

Let us now review the derivation of the KMS condition. Consider a thermal two-point function in Euclidean time $\<A(\tau)B(0)\>_\b$, and let us assume $\tau>0$. This is given by
\be
\label{eq:tracemanip}
\<A(\tau)B(0)\>_\b &= \Tr(e^{-\b H}e^{\tau H}A(0) e^{-\tau H} B(0)) = \Tr(e^{-(\b-\tau) H}A(0)e^{-\tau H} B(0)),
\ee
where $H$ is the Hamiltonian. Note that convergence of the exponential factor $e^{-\tau H}$ requires $\tau>0$ and convergence of the exponential factor $e^{-(\beta-\tau)H}$ requires $\tau < \beta$. Thus, the above expression defines the thermal two-point function for $\tau \in (0,\beta)$. From cyclicity of the trace, one immediately finds that
\be
 \<A(\tau)B(0)\>_\b &= \<B(\b-\tau)A(0)\>_\b.
\ee
This is the KMS condition. 

Taking $A(\tau) = \phi(\tau, \mathbf{x})$, $B(0)=\phi(0, 0)$ and $\tau = \beta/2 + \widetilde \tau$, with $\widetilde \tau \in \left[-\frac{\b}2, \frac{\b}2\right] $ we get
\be
\label{eq:periodicity-KMS}
g(\beta/2 + \widetilde \tau, \mathbf{x}) = g(\beta/2 - \widetilde \tau, -\mathbf{x}).
\ee
By $\SO(d-1)$-invariance, the correlator depends only on $|\bx|$, so is invariant under $\bx\to-\bx$.  Thus, we can further conclude that
\be
\label{eq:periodicity-and-parity}
g(\b/2+\widetilde \tau, \mathbf{x})= g(\b/2-\widetilde \tau, \mathbf{x})  .
\ee
The fact that the scalar thermal two-point function is even in $\mathbf{x}$ is built into the conformal block decomposition (\ref{eq:blockdecomposition}).
Another approach to understand (\ref{eq:periodicity-and-parity}) is to note that Euclidean thermal correlators are computed by a path integral on the geometry $S_\b^1\x\R^{d-1}$, and then (\ref{eq:periodicity-and-parity}) is evident from the $\mathrm{O}(d-1)$ symmetries of the geometry discussed in section~\ref{sec:OPE-and-1-pt}.

Note that the thermal conformal block decomposition (\ref{eq:blockdecomposition}) can be constrained by the KMS condition (\ref{eq:periodicity-and-parity}) due to the lack of manifest  periodicity for the thermal conformal block, in  a similar way in which the four-point functions conformal blocks are not manifestly crossing-symmetric. This constraint is well-defined within the OPE radius of convergence, whenever both $\b/2+ \widetilde \tau,\text{ and } \b/2-\widetilde \tau \in [0, \beta]$ (\ref{eq:opeconvergencecondition}). Thus, in analogy to the crossing equation for vacuum four-point functions, we will interpret (\ref{eq:periodicity-and-parity}) as a constraint equation for all the thermal coefficients $a_\cO$ appearing in  (\ref{eq:blockdecomposition}). This observation was made in \cite{ElShowk:2011ag}. The analog of expanding four-point functions around the crossing-symmetric point $z=\zb=1/2$ is, using the reflection property \eqr{eq:periodicity-and-parity}, to enforce that
\eq{}{{\partial^{n+m}\o \partial\tau^n\partial^m |\bx|}g(\tau, \mathbf{x})\Big|_{\tau={\b\o2}, \bx = 0}=0~~\text{for odd }n\text{ and even } m~.}

This philosophy extends naturally to thermal $n$-point functions, which are expectation values of Euclidean time-ordered products\footnote{Note that time ordering is the only sensible ordering when operators are at different Euclidean times (i.e.\ ``imaginary" times, although here it corresponds to real $\tau$). This is because if the operators weren't ordered appropriately, the exponential factors $e^{-(\tau_i-\tau_j)H}$ would be divergent. By contrast, if some operators are at the same Euclidean time but different Lorentzian (``real") times, we can consider different orderings among those operators, and these orderings correspond to different analytic continuations of the Euclidean correlator. For example, in real time thermal physics (where $\tau_i=i t_i$ with $t_i\in \R$), one can study arbitrary orderings of the operators.}
\be
\<A_1(\tau_1)\cdots A_n(\tau_n)\>_\b &= \Tr(e^{-\b H}T\{A_1(\tau_1)\cdots A_n(\tau_n)\}) \nn\\
&= \Tr(e^{-\b H}A_1(\tau_1)\cdots A_n(\tau_n)) \theta(\Re(\tau_1-\tau_2)) \cdots \theta(\Re(\tau_{n-1}-\tau_n))\nn\\
& \quad + \textrm{permutations}.
\ee
The above representation of the correlator is valid if $\Re(\tau_1-\tau_n)\leq \b$.
If $\tau_n$ is the earliest time, then a similar manipulation to (\ref{eq:tracemanip}) using cyclicity of the trace implies
\be
\label{kmsn}
\la A_1(\tau_1)\cdots A_n(\tau_n) \ra_\b 
&= \la A_n(\tau_n+\b)A_1(\tau_1)\cdots A_{n-1}(\tau_{n-1}) \ra_\b \nn\\
&= \la A_1(\tau_1)\cdots A_{n-1}(\tau_{n-1}) A_n(\tau_n+\b)\ra_\b.
\ee
(In the second line we used that operators trivially commute inside the time-ordering symbol.)
It follows that the thermal expectation value of a Euclidean time-ordered product
is periodic in each of the $\tau_i$ (since we can decrease $\tau_i$ until it becomes the earliest time and then apply (\ref{kmsn})). This is again obvious from the geometry. We may regard these periodicity conditions as crossing equations. In this work, we focus on the case $n=2$. (See \cite{ranga} for recent study of KMS conditions for $n$-point functions.)

While the KMS condition imposes constraints on the $a_\cO$, there is an immediate obstacle to an efficient bootstrap: the OPE expansion \eqr{eq:blockdecomposition} is linear in the OPE coefficients, nor must the $a_\cO$ be sign-definite. Thus, the resulting expression lacks manifest positivity.  This is more analogous to the bootstrap in the presence of a conformal boundary or defect, rather than the vacuum four-point function bootstrap \cite{Lemos:2017vnx}. To proceed, we need to develop some complementary tools; this will be the content of section~\ref{sec:lorentzian-inversion-formula}.

\subsection{Away from the OPE regime}
\label{eq:kkquantization}

The OPE representation (\ref{eq:blockdecomposition}) comes from interpreting the two-point function $g(\tau,\mathbf{x})$ in radial quantization around a point in $S^1_\b \x \R^{d-1}$. As discussed in section~\ref{sec:OPE-and-1-pt}, this representation is only valid when the points satisfy $|x|<\b$. Other ways of quantizing the theory give other representations with their own regimes of validity (possibly overlapping).

Perhaps the most familiar way to study thermal correlators is to quantize the theory on $\R^{d-1}$-slices, where $S^1$ is interpreted as a Euclidean time direction. This quantization leads to expressions for thermal correlation functions like (\ref{eq:tracemanip}). It is also the most natural choice from the point of view of the limit $S^1 \x S^{d-1} \to S^1 \x \R^{d-1}$ discussed in section~\ref{sec:spheretoplane}. 

Another way of quantizing the theory (that will prove useful in the next section) is to choose a direction in $\R^{d-1}$, say $x^1$, as the time direction. States then live on slices with geometry $S^1 \x \R^{d-2}$. This quantization is natural if we imagine a Kaluza-Klein compactification of a $d$-dimensional QFT on a spatial $S^1$. In the compactified theory, the momentum generator around the $S^1$, which we call $P_{\mathrm{KK}}$, becomes a global $U(1)$ symmetry with a discrete spectrum. The Hamiltonian $H_{\mathrm{KK}}$ generates translations in $x^1$. Explicitly, the generators are given by
\be
P_\mathrm{KK} &= \int_0^\b d\tau \int_{-\oo}^\oo dx^2 \cdots dx^{d-1} \p{-iT^{10}(\tau,\bx)},\nn\\
H_\mathrm{KK} &= \int_0^\b d\tau \int_{-\oo}^\oo dx^2 \cdots dx^{d-1} \p{-T^{11}(\tau,\bx) - \frac{b_T}{d}\frac{1}{\b^d}},
\ee
where the factor of $-i$ in $P_\mathrm{KK}$ and the minus sign in $H_\mathrm{KK}$ come from Wick rotation as discussed in footnote~\ref{foot:wick}. Because the charges are conserved, we can evaluate them at any value of $x^1$.
In $H_\mathrm{KK}$, we have chosen to subtract off the vacuum energy by hand so that it annihilates the vacuum on $S^1 \x \R^{d-2}$.\footnote{Note that the $d-1$-dimensional vacuum energy density, equivalently the Casimir energy density of the CFT on a circle, is simply $\frac{b_T}{d}\frac{1}{\b^{d-1}} = \b F$. In particular, it is negative since $b_T$ is negative by the discussion in section~\ref{sec:freenergydensity}.}

In our two-point function $g(\tau,\mathbf{x})$, we can use $\SO(d-1)$-invariance to set $\bx=(x^1,0,\dots,0)$ with $x^1>0$. Interpreting the correlator in KK quantization, we obtain
\be
\label{eq:KKrepresentation}
g(\tau,x^1) &= \<0|\f(0) e^{-H_{\mathrm{KK}} x^1 + i \tau P_{\mathrm{KK}}} \f(0)|0\>,
\ee
where $|0\>$ is the ground-state on $S^1\x \R^{d-2}$.
Note that $H_{\mathrm{KK}}$ is Hermitian and bounded from below, so the factor $e^{-H_{\mathrm{KK}} x^1}$ leads to exponential suppression. We discuss the regime of validity of (\ref{eq:KKrepresentation}) in section~\ref{sec:analyticity-w-plane}. 

The behavior of the correlator at large $x^1$ (with fixed $\tau$) is determined by the mass gap of the compactified theory, i.e.\ the smallest nonzero eigenvalue of $H_{\mathrm{KK}}$ which we call $m_{\text{th}}$ (the ``thermal mass"). By dimensional analysis, $m_{\text{th}}$ is a constant times $1/\b$. It is a folk-theorem that dimensional reduction on a circle with thermal boundary conditions produces a massive theory, i.e.\ $m_{\text{th}}>0$.\footnote{We thank Zohar Komargodski for discussions on this point.} Assuming this folk-theorem is true, the correlator approaches a factorized form exponentially quickly at large $|\mathbf{x}|$
\be
\label{eq:clusterdecomposition}
g(\tau,\mathbf{x}) &\sim \<\f\>_\b^2 + O(e^{-m_{\text{th}} |\mathbf{x}|}).
\ee
Like the KMS condition, the decay (\ref{eq:clusterdecomposition}) is not at all obvious from the OPE.
In free theories, supersymmetric compactifications, or in the presence of nontrivial chemical potentials, we could have $m_\mathrm{gap}=0$ and the behavior of the long-distance correlator would be different. 

Finally, let us note that the representation (\ref{eq:KKrepresentation}) does not use the full $(d-1)$-dimensional Poincare invariance of the compactified theory. To do so, we insert a complete set of states and classify them according to their $(d-1)$-dimensional invariant mass and KK momentum. This leads to a version of the K\"all\'en-Lehmann spectral representation
\be
\label{eq:kallenlehman}
g(\tau,\mathbf{x}) &= \sum_{n=-\oo}^\oo e^{in\tau}  \int_0^\oo dm^2 \rho_n(m^2) G_F(\mathbf{x},m^2),
\ee
where $n$ is the KK momentum and $G_F(\mathbf{x},m^2)$ is the Feynman propagator in $(d-1)$-dimensions. The decomposition (\ref{eq:kallenlehman}) comes from going to momentum space in the compact direction, and then applying the usual K\"all\'en-Lehmann representation in each momentum sector, yielding a density of states $\rho_n(m^2)$ for each $n$. For real $\tau,\mathbf{x}$, the expression (\ref{eq:kallenlehman}) is valid whenever $|\mathbf{x}|>0$, so it has an overlapping regime of validity with the OPE. It would be interesting to study the equality of these two representations.

\section{A Lorentzian inversion formula}
\label{sec:lorentzian-inversion-formula}

Inversion formulas provide an efficient way to study the operator content of vacuum four-point functions in flat space. The starting point is an expansion of the four-point function in a complete set of single-valued conformal partial waves, which are solutions to the conformal Casimir equations on $\R^d$. This basis is natural because physical four-point functions are single-valued in Euclidean space. The expansion also follows on general grounds from the Plancherel theorem for the conformal group $\SO(d+1,1)$ \cite{Dobrev:1977qv}. One may then invert the expansion using orthogonality and completeness, to extract the exchanged operator data from integrals of the four-point function.

In \cite{Caron-Huot:2017vep}, Caron-Huot derived a remarkable inversion formula for four-point functions that involves an integral in Lorentzian signature.\footnote{See \cite{Simmons-Duffin:2017nub} for another derivation.} In this section, we will derive a Lorentzian inversion formula for thermal two-point functions. However, our formula will not have a similarly clean group-theoretic interpretation as in the case of four-point functions. The reason is that $C_J^{(\nu)}(\eta) |x|^{\De-2\De_\f}$, the two-point thermal blocks on $S^1_\b\times \R^{d-1}$, are not a complete set of solutions to a differential equation (for any choices of $(\De,J)$), simply because they are not single-valued functions on $\R^{d-1}\x S^1$.\footnote{One can restrict to a disc $|x|<r$ and introduce boundary conditions at $|x|=r$ to obtain a completeness relation. However, such boundary conditions are not satisfied by physical two-point functions. Alternatively, one can lift two-point functions to the universal cover of $\R^{d-1}\x S^1$ and study completeness relations on this space.} Still, we will make progress without a completeness relation by focusing on the OPE limit and writing a formula that picks out the data in this limit.

\subsection{Euclidean inversion}
\label{sec:lorentzian-inversion-general-grounds}

In analogy with the conformal partial wave expansion for four-point functions, we complexify $\De$, and write the thermal block expansion (\ref{eq:blockdecomposition}) as a spectral integral:
\be
g(\tau,\bx) = 
\sum_{J=0}^\oo \oint_{-\epsilon - i \oo}^{-\epsilon + i \oo} \frac{d\De}{2\pi i} a(\De,J) C_J^{(\nu)}(\eta) |x|^{\De-2\De_\f}.
\label{eq:integralfortwopointfunction}
\ee
For simplicity, we set $\b=1$ for the remainder of the paper. 
The full dependence on $\b$ can be restored by dimensional analysis.
The function $a(\De,J)$ should have simple poles at the physical operator dimensions, with residues proportional to the coefficients $a_\cO$, 
\be
\label{eq:polesandresidues}
a(\De,J) &\sim -\frac{a_\cO}{\De-\De_\cO}.
\ee
We also require that $a(\De,J)$ not grow exponentially in the right $\De$-half-plane.
When $|x|<1$, we can close the $\De$ contour to the right to encircle the poles clockwise (hence the minus sign in (\ref{eq:polesandresidues})) and recover the usual thermal conformal block decomposition. The position of the $\De$ contour is arbitrary as long as the integral converges. We have chosen it to lie to the left of all physical poles, including the one from the unit operator.

It is simple to write an inversion formula that produces $a(\De,J)$ from $g(\tau,\mathbf{x})$, by integrating against a Gegenbauer polynomial to pick out the contribution from spin $J$, and then Laplace transform in $|x|$ to pick out poles in $\De$,
\be
a(\De,J) &= \frac{1}{N_J}\int_{|x|<1} d^d x\, C_J^{(\nu)}(\eta) |x|^{2\De_\f-\De-d} g(\tau,\bx).
\label{eq:bootleginversion}
\ee
This choice of $a(\De,J)$ is not unique, since we only demand that it have poles and residues consistent with (\ref{eq:polesandresidues}). To obtain the correct poles in $\De$, it suffices to integrate $x$ over any neighborhood of the origin. We call (\ref{eq:bootleginversion}) a ``Euclidean inversion formula" because it involves an integral over Euclidean space. For simplicity, we have chosen to integrate over a circle with radius $1$.\footnote{We would like to ensure that poles in $a(\De,J)$ correspond only to operators in the OPE. To do this, we can restrict the range of integration to be $|x|<r$ for some $r<1$, so that the integral remains strictly inside the regime of convergence of the OPE. The only singularity in this region is the OPE singularity, and this is the only place poles can come from. We have written $r=1$ for simplicity, but a careful reader can imagine $r=1-\e$ for positive $\e$.} The factor $N_J$ is defined by
\be
\label{eq:orthogonality-condition}
\int_{S^{d-1}} d\Omega\, C_J^{(\nu)}(\eta)C_{J'}^{(\nu)}(\eta) &= N_J \de_{JJ'},
\ee
where
\be
N_J &= \frac{4^{1-\nu } \pi ^{\nu +\frac{3}{2}} \Gamma (J+2 \nu )}{J! (J+\nu ) \Gamma (\nu )^2 \Gamma (\nu +\frac{1}{2})}
\label{eq:mathnl},
\qquad\qquad \nu=\frac{d-2}{2}.
\ee
This standard normalization of the Gegenbauer polynomial is unfortunately singular when $d=2$, so we will treat that as a special case. 

\subsection{Continuing to Lorentzian signature}

The angular dependence of a two-point block on $S^1_\b\times \R^{d-1}$ is precisely the same as the angular dependence of a partial wave in a $2\to 2$ scattering amplitude --- both are given by Gegenbauer polynomials. In the case of amplitudes, the Froissart-Gribov formula \cite{Gribov:1961fr,Froissart:1961ux} expresses partial wave coefficients as an integral of the amplitude over a Lorentzian regime of momenta. A standard derivation of the Froissart-Gribov formula (see e.g.\ \cite{Caron-Huot:2017vep,Paulos:2017fhb,Lemos:2017vnx}) carries over essentially unchanged to our case, where it gives $a(\De,J)$ as an integral over a Lorentzian region in position space. Note that the Lorentzian region we find does not correspond to the usual real time dynamics at finite temperature (where $\tau$ gets complexified). Instead, one of the components of $\mathbf{x}$ gets complexified and plays the role of Lorentzian time.

\sssec{Kinematics}
Before giving the derivation, let us discuss the Lorentzian region that will appear in our formula. Using $\SO(d-1)$ invariance, we can restrict $\mathbf{x}$ to a line and denote the coordinate along this line as $x_E$. Let us introduce coordinates
\be\label{zzbar}
z = \tau + i x_E, \quad \bar z = \tau - i x_E.
\ee
It will also be useful to introduce polar coordinates $r$ and $w=e^{i\th}$ such that
\be
z = r w,\qquad \bar z = r w^{-1}.
\ee
In Euclidean signature, $w$ lies on the unit circle and represents the angle of the two operators relative to the $\tau$-direction, and $z,\bar z$ are complex conjugates of each other.

\begin{figure}[t]
\begin{center}
\includegraphics[width=0.9\textwidth]{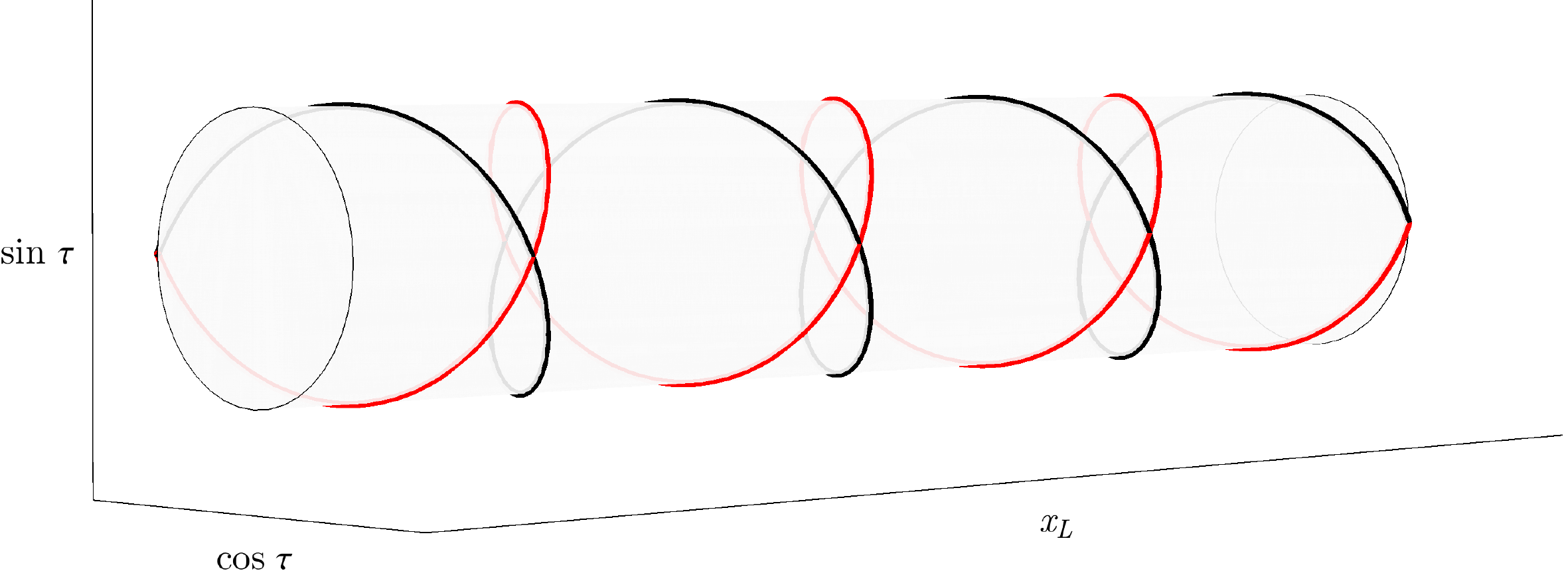}
\end{center}
\caption{\label{fig:lorentzian-cylinder} Lightlike trajectories moving in the $x_L$ direction and around the thermal circle. One trajectory is $z=0$ and the other is $\bar z=0$. Poles in the Lorentzian inversion formula come from the neighborhood of these trajectories.}
\end{figure}

We will continue to Lorentzian signature by Wick-rotating $x_E = -i x_L$, so that $z,\bar z$ become independent real variables. In particular, the direction $\tau$ along the thermal circle remains Euclidean and retains the periodicity $\tau\sim \tau+\b$, and $w$ is real. This configuration is best interpreted in terms of a Lorentzian theory, one of whose spatial directions has been compactified on $S^1$. It is {\it not\/} the Lorentzian kinematics usually considered in thermal field theory, where one considers complex $\tau$. Instead, $x_L$ plays the role of time. Poles in $a(\De,J)$ corresponding to physical one-point functions will come from the regime $z\sim 0$ or $\bar z \sim 0$, where one of the operators is following a lightlike trajectory around the thermal circle. These lightlike trajectories are depicted in figure~\ref{fig:lorentzian-cylinder}.

The regime of small or large $w$ will play an important role in what follows. In the limit of $r$ fixed and $w\rar \infty$, say, we have $\tau \rar x_L$ and $x_L\rar\infty$. Given the periodicity $\tau\sim \tau+\b$, the separation between the operators approaches a lightlike-trajectory along the cylinder at asymptotically large $x_L$. In terms of $(z,\zb)$, this limit corresponds to $z\rar\infty, \zb\rar 0$ with $z\zb$ fixed. This limit of large boost ($w\to 0$ or $w\to \oo$) is analogous to the Regge limit in flat space. 
\subsubsection{The inversion formula in $d=2$}
\label{sec:inversion-d=2}

Let us first present the derivation in $d=2$, where it is particularly simple. As noted in section~\ref{s22d}, thermal two-point functions in $d=2$ are related by a Weyl transformation to flat-space two-point functions, so this analysis is not necessary. However, the discussion in this subsection will generalize to higher dimensions. 

\begin{figure}[t]
\begin{center}
 \begin{subfigure}[b]{0.5\textwidth}
 \centering
\includegraphics[width=1.00\textwidth]{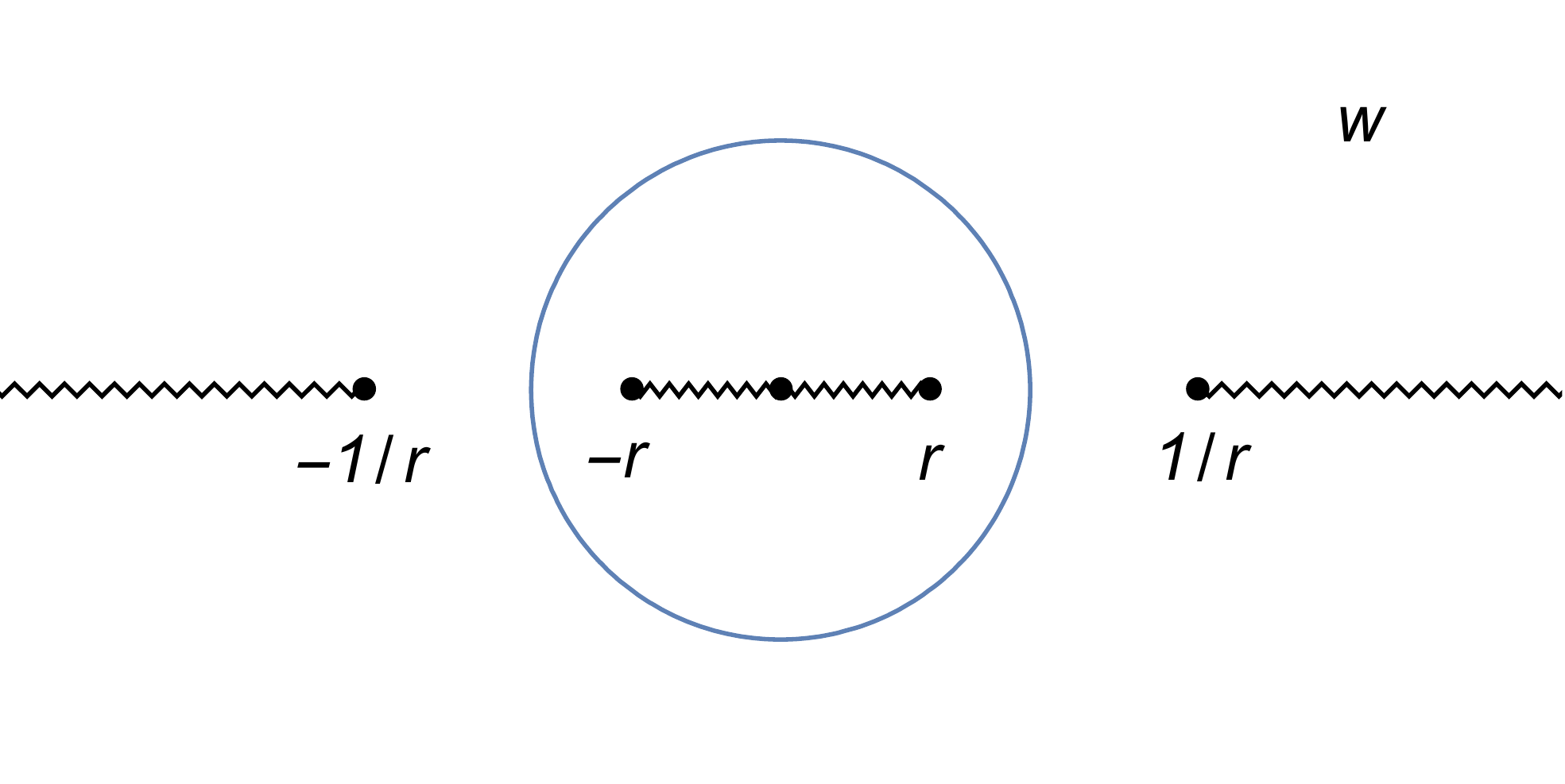}
\vspace{-0.13cm}
\caption{\label{fig:contoura}Initial integration contour $\vert w \vert = 1$.}
\end{subfigure}
\\\vspace{0.1cm}
 \begin{subfigure}[b]{0.40\textwidth}
 \centering
\includegraphics[width=1.00\textwidth]{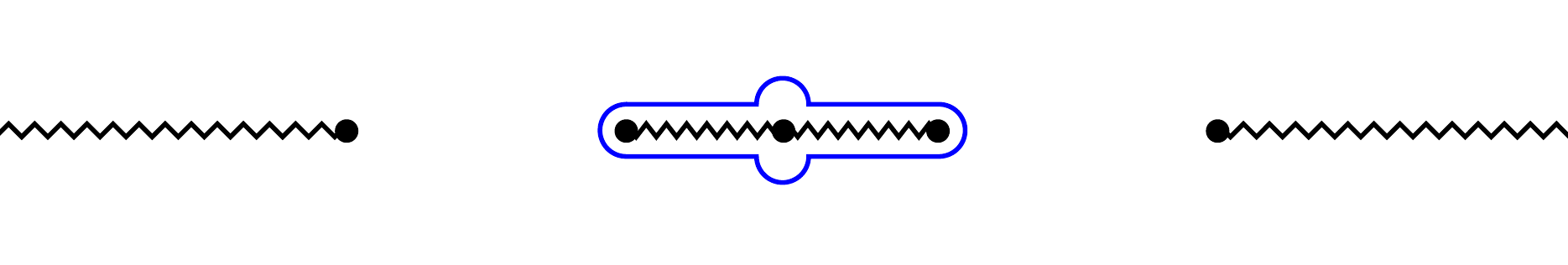}
\vspace{0.8cm}
\caption{\label{fig:contourb} The deformed contour for terms that behave as $w^J$.}
\end{subfigure}~\quad
 \begin{subfigure}[b]{0.40\textwidth}
\includegraphics[width=1.00\textwidth]{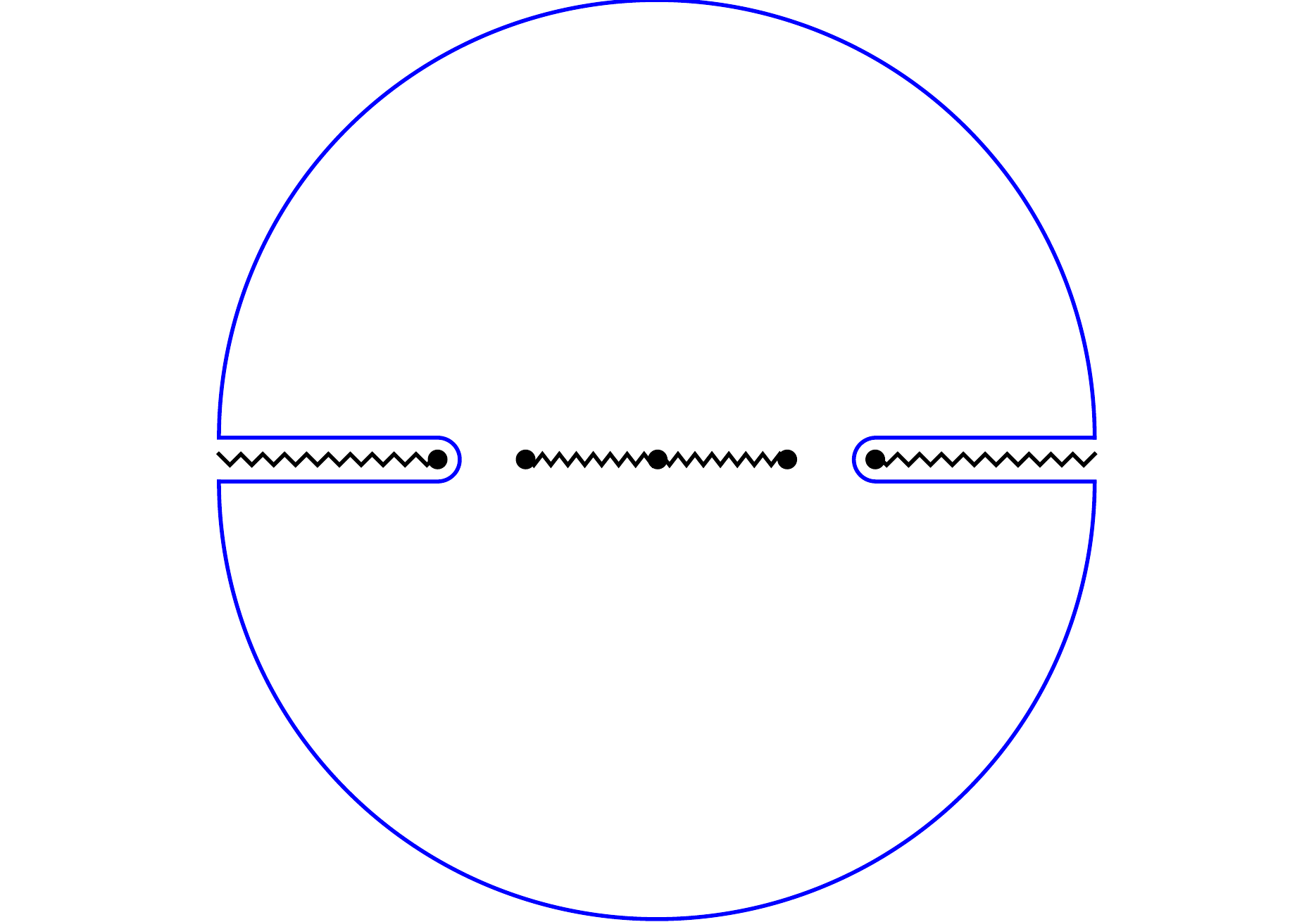}
\caption{\label{fig:contourc} The deformed contour for terms that behave as $w^{-J}$.}
\end{subfigure}
\end{center}
\caption{\label{fig:contour-manipulation} Contour manipulations for the inversion formula in the $w$ plane. In (a) we show the original contour which lies along the circle $\vert w\vert=1$. For the $w^J$ terms in \eqref{eq:euclideaninversion}, we deform the contour as in (b), and for the $w^{-J}$ terms in \eqref{eq:euclideaninversion}, we deform the contour as in (c). }
\end{figure}

In two dimensions, Gegenbauer polynomials are given by\footnote{We are considering only external scalars, which have $h = \overline{h}$. Equation (\ref{2pt2d}) explicitly shows that the correlator is symmetric under the interchange of $z$ and $\overline{z}$, which gives the symmetry of the block under the exchange of $J$ and $-J$. For spinning thermal correlators, a chirally-asymmetric block should be used.}
\be
\cos(J \theta) = \frac 1 2 (w^J + w^{-J}).
\ee
With this normalization, we have $N_J=\pi$. 
Viewed as a cylinder two-point block, the first term comes from the exchange of a vacuum Virasoro descendant having weights $(h,\overline{h})$ and spin $J = h - \overline{h}$.
The second term comes from the exchange of the conjugate state having weights $(\overline{h},h)$ and spin $-J$. 
The Euclidean inversion formula becomes
\be
a(\De,J) &= \frac{1}{\pi } \int_0^1 \frac{dr}{r} r^{2\De_\f - \De} \oint \frac{dw}{i w} \frac 1 2(w^{J} + w^{-J}) g(z=rw,\bar z= r w^{-1}),
\label{eq:euclideaninversion}
\ee
where the $w$-contour is along the unit circle as pictured in figure~\ref{fig:contoura}. Note that $J$ must be an integer in (\ref{eq:euclideaninversion}) in order for the integrand to be single-valued along the contour.

Now, the crucial claim is that $g(z=rw,\bar z=r w^{-1})$ satisfies the following properties as a function of $w$:
\begin{itemize}
\item It is analytic in the $w$ plane away from the cuts $(-\oo,-1/r)$, $(-r,0)$ $(0,r)$, and $(1/r,\oo)$. 

\item Its growth at large $w$ is bounded by a polynomial $w^{J_0}$ for some constant $J_0$. Similarly, by symmetry under $w\to w^{-1}$, the growth at small $w$ is bounded by $w^{-J_0}$.
\end{itemize}
We discuss these properties in the next section. For now, let us assume them and proceed with the derivation.

By analogy with the Froissart-Gribov formula, we now deform the $w$ contour away from the unit circle. We must do this separately for the two terms $w^J$ and $w^{-J}$. The term $w^J$ dies as $w\to 0$. Assuming $J>J_0$, we can deform the contour towards zero for that term to obtain the contour~\ref{fig:contourb}. Similarly, the term $w^{-J}$ dies as $w\to\oo$, so we can deform the contour towards infinity for that term to obtain the contour~\ref{fig:contourc} (again assuming $J>J_0$).

Let us focus on the $w^{-J}$ term, where we deform the contour towards infinity. By our analyticity assumption, we first encounter a branch cut at $w=r^{-1}$, or equivalently $z=1$ (we comment on the contribution of the $z=-1$ branch-cut shortly). We thus obtain an integral of the discontinuity of the two-point function $g(z, \bar z)$ across this cut,
\be
\label{eq:contour-neglect}
 \oint \frac{dw}{iw} w^{-J} g(z=rw,\bar z = r w^{-1}) &\supset \int_{w=r^{-1}}^\oo \frac{dw}{w}\, w^{-J} \textrm{Disc}[g(z=rw,\bar z = r w^{-1})],
 \ee
 where
 \be
\textrm{Disc}[g(z,\bar z)] &\equiv \frac{1}{i} \p{g(z+i\e,\bar z) - g(z-i\e,\bar z)}.
\ee
Here, we have assumed that $J>J_0$, so we can drop the arcs at infinity in figure~\ref{fig:contourc}. If instead $J \leq J_0$, we must keep the contribution from these arcs. The arc contributions are the analogs of finite subtractions in the case of dispersion relations for amplitudes. 

Because $g(z,\bar z)=g(-z,-\bar z)$, the branch cut from $(-\oo,-1/r)$ contributes the same as the cut from $(1/r,\oo)$, up to a factor of $(-1)^J$. 
Finally, because of symmetry under $w\to w^{-1}$, the contribution from deforming the contour for $w^J$ towards the origin is the same as the contribution from deforming the contour for $w^{-J}$ towards infinity, giving an overall factor of $2$.

Putting everything back in \eqref{eq:euclideaninversion}, we obtain
\be
\label{eq:thermfg}
a(\De,J) 
&= (1+(-1)^J)\frac 1 \pi \int_0^1 \frac{dr}{r} r^{2\De_\f-\De} \int_{w=r^{-1}}^\oo \frac{dw}{w}\,w^{-J} \Disc[g(z,\bar z)]+\theta(J_0-J) a_{\rm arcs}(\De,J)\nn\\
&= (1+(-1)^J)\frac 1 {2\pi} \int_0^{1} \frac{d\bar z}{\bar z}\int_1^{1/\bar z} \frac{dz}{z} \, z^{\De_\f-\bar h} \bar z^{\De_\f - h} \Disc[g(z,\bar z)]+\theta(J_0-J) a_{\rm arcs}(\De,J)\,,
\ee
where
\be h=\frac{\De-J}{2} \qquad \text{ and } \qquad \bar h = \frac{\De+J}{2}.
\ee
We have explicitly indicated the presence of non-trivial contributions from the arcs when $J\leq J_0$. 
These are given by the large $w$ region of \eqr{eq:euclideaninversion}. 
Their detailed form depends on the correlator in question.
We will see some explicit examples in the next section.

\subsubsection{The inversion formula in $d>2$}

\label{sec:inversion-d>2}

To perform the same derivation in $d>2$ dimensions, we must find the higher-dimensional analog of the decomposition $\cos(J \th)=\frac 1 2 (w^J+w^{-J})$. The role of $w^J$ will be played by the solution to the Gegenbauer differential equation that vanishes as $w\to 0$ (for positive $J$). This is given by\footnote{The function $F_J$ is proportional to $B_J$ defined in \cite{Simmons-Duffin:2017nub} and $Q_J$ defined in \cite{Paulos:2017fhb}. In $d=3$, it is proportional to a Legendre $Q$-function.} 
\be
F_J(w) &= w^{J+d-2}{}_2F_1\p{J+d-2,\frac d 2 - 1, J+\frac d 2, w^2}.
\ee
The Gegenbauer differential equation is symmetric under $w\to w^{-1}$ (because the equation depends only on $\cos(\th)=\frac 1 2 (w+w^{-1})$), so the solution that vanishes as $w\to \oo$ is $F_J(w^{-1})$.

Because the Gegenbauer differential equation is second-order, the two functions $F_J(w^{\pm 1})$ span the space of solutions. In particular, a Gegenbauer polynomial can be expressed as a linear combination
\be
C^{(\nu)}_J\p{\frac 1 2(w+w^{-1})} &=
\frac{\G(J+d-2)}{\G(\frac d 2 - 1)\G(J+\frac d 2)}\p{F_J(w^{-1})e^{i\pi\frac{d-2}{2}} + F_J(w) e^{-i \pi \frac{d-2}{2}}}, \quad \Im(w)>0.
\ee
The above representation is correct for $w$ in the upper half-plane. Because $F_J(w)$ has cuts along $(-\oo,-1]$ and $[1,\oo)$ and $F_J(w^{-1})$ has cuts along $[-1,1]$, the representation is different when $w$ is in the lower half-plane (the phases in front of the two terms swap). Note that when $w=e^{i\th}$ is on the unit circle, the two terms are complex-conjugates of each other, so their sum is real.

Plugging this representation of the Gegenbauer polynomial into the Euclidean inversion formula (\ref{eq:bootleginversion}), we can run the same contour argument as in $d=2$. The measure contributes an extra factor of $(z-\bar z)^{2\nu}$, but otherwise the derivation is essentially unchanged. We find
\be
a(\De,J) 
&=
(1+(-1)^J) K_J \int_0^1 \frac{d\bar z}{\bar z} \int_1^{1/\bar z} \frac{dz}{z} (z \bar z)^{\De_\f-\frac \De 2-\nu}(z-\bar z)^{2\nu} F_J\p{\sqrt{\frac {\bar z} {z}}} \Disc[g(z,\bar z)]\nn\\ &\quad+\theta(J_0-J) a_{\rm arcs}(\De,J),
\label{eq:inversionformulahigherd}
\ee
where
\be
K_J &\equiv 
 \frac{\G(J+1)\G(\nu)}{4\pi \G(J+\nu)}.
\ee
It is easy to check that this agrees with (\ref{eq:thermfg}) in $d=2$ after accounting for the proper normalization of the $d=2$ Gegenbauer polynomials.

\subsection{Comments on the Lorentzian formula}
\label{sec:commentsoninversion}

Like the Froissart-Gribov formula, our Lorentzian inversion formula (\ref{eq:inversionformulahigherd}) has the interesting property that it can be analytically continued in spin $J$. As explained e.g.\ in \cite{Caron-Huot:2017vep}, analyticity in spin is a consequence of polynomial boundedness in the $w\rar\infty$ limit --- specifically our assumption that the correlator does not grow faster than $w^{\pm J_0}$. Because each partial wave with nonzero spin grows in this limit, boundedness is only possible if there is a delicate balance, due to analyticity, between each partial wave with $J > J_0$. This state of affairs is precisely analogous to the Regge limit of vacuum four-point functions.

The integral (\ref{eq:inversionformulahigherd}) is over a Lorentzian regime of the two-point function. We will see shortly that poles in $\De$ come from the region $\bar z \sim 0$ where the factor $\bar z^{\De_\f - \frac{\De}{2}-\nu}$ is singular. The residues are then determined by a one-dimensional integral over $z$. Because poles come from infinitesimal $\bar z$, we can replace the upper limit of the $z$ integral with $\oo$. In other words, the locus that contributes to CFT one-point functions is $\tau \sim x_L$ (cf. \eqr{zzbar}), which is a lightlike trajectory moving around the thermal circle while moving forward in ``time" $x_L$. This trajectory is pictured in figure~\ref{fig:lorentzian-cylinder}.

The fact that physical data comes from an integral over $z$ with $\bar z\sim 0$ is also true in Caron-Huot's Lorentzian inversion formula for four-point functions. However, in that case, the integral remains entirely within the regime of convergence of both the $s$ and $t$-channel OPEs. An important difference in our case is that the $z$-integral extends  outside the regime of convergence of any OPE. In our conventions, the $s$-channel OPE is an expansion around $z=\bar z= 0$ and the $t$-channel OPE is an expansion around $z=\bar z = 1$. Their regimes of convergence are:
\be 
\text{s-channel OPE: } &|z|,|\bar z| < 1,\nn\\
\text{t-channel OPE: } &|1-z|,|1-\bar z|<1.
\ee
Our integral is within the regime of convergence of the $t$-channel OPE for $1\leq z < 2$. But for $z>2$ it exits this regime. Thus, we can only obtain partial information about $a(\De,J)$ from the $t$-channel OPE expansion alone. However, as we will see in more detail in section~\ref{sec:large-spin-perturbation-theory}, corrections coming from the region $z>2$ are exponentially suppressed in $J$.

Another interesting similarity between our inversion formula and Caron-Huot's is the significance of a double lightcone limit. We will see in section~\ref{sec:large-spin-perturbation-theory} that a systematic expansion for thermal one-point functions in $1/J$ requires understanding the thermal two-point function in the regime $\bar z\sim 0$, $z\sim 1$. 
This corresponds to a physical configuration where the second operator is approaching the first intersection of light rays from the first operator that wrap halfway around the thermal circle. 
In the context of four-point functions, the same regime $\bar z\sim 0$, $z\sim 1$ corresponds to all four operators approaching the corners of the square $(z,\zb) \in [0,1]$, and is dubbed the ``double lightcone" limit.\footnote{This is to be distinguished from the specialized terminology of \cite{Alday:2015ota}, where ``double lightcone" limit means fixed ${\bar z\over 1-z}$.} Because our limit plays a similar role in large-spin perturbation theory, we will adopt the same terminology.

\subsection{Analyticity in the $w$-plane}
\label{sec:analyticity-w-plane}

\begin{figure}
\begin{center}
\includegraphics[width=0.8\textwidth]{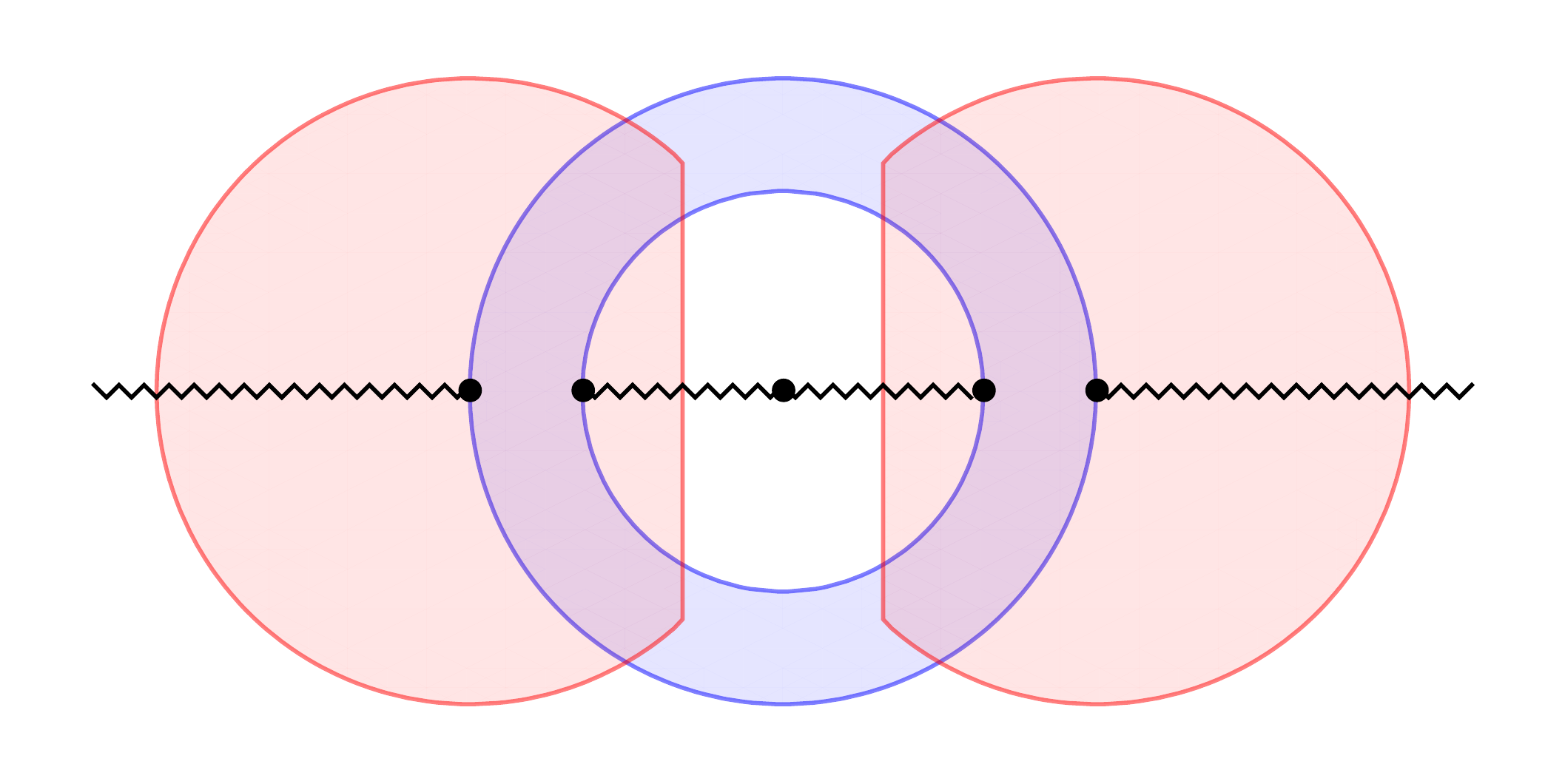}
\end{center}
\caption{\label{fig:opeconvergeneregionswplane} For fixed $r\in(0,1)$, the $s$-channel OPE (expansion around $z=\bar z = 0$) implies that the thermal two-point function $g(z,\bar z)$ is analytic in an annulus in the $w$ plane between radii $r$ and $1/r$ (shaded blue). The $t$-channel OPE (expansion around $z=\bar z = 1$), together with symmetry under $w\leftrightarrow -w$, implies analyticity in the red-shaded regions, except for cuts running along $(-\oo,-1/r), (-r,0), (0,r), (1/r,\oo)$ (indicated with zig-zags). In this section, we argue for analyticity everywhere in the upper and lower half planes.}
\end{figure}

To complete our derivation, we must justify the assumptions stated in section~\ref{sec:inversion-d>2}, namely analyticity of the two-point function in $w$ outside the cuts pictured in figure~\ref{fig:contour-manipulation}, and polynomial boundedness in $w$. First note that convergence of the $s$-channel OPE guarantees analyticity in the annulus
\be
r < |w| < r^{-1}.
\ee
Convergence of the OPE around $z,\bar z = 1$ and $z,\bar z = -1$ additionally guarantees analyticity in more complicated regions around $w=\pm 1$ (figure~\ref{fig:opeconvergeneregionswplane}).

To argue for analyticity in an even larger region, we will use the KK representation discussed in section~\ref{eq:kkquantization},
\be
\label{eq:kallenlehmen}
g(\tau, x_E) &= \<\Psi|e^{-x_E H_{\mathrm{KK}}+i\tau P_{\mathrm{KK}}}|\Psi\> = \<\Psi|e^{\frac i 2 (H_{\mathrm{KK}}+P_{\mathrm{KK}}) z - \frac i 2 (H_{\mathrm{KK}}-P_{\mathrm{KK}}) \bar z}|\Psi\>,
\ee
where
\be
|\Psi\> &= \f(0)|0\>_{S^1 \x \R^{d-2}}.
\ee
Here, we have quantized the Euclidean theory on spatial slices with geometry $S^1 \x \R^{d-2}$. The Hamiltonian $H_{\mathrm{KK}}$ generates translations in the noncompact direction parameterized by $x_E$, while $P_{\mathrm{KK}}$ generates translations in $\tau$ (the periodic direction). In this way of quantizing the theory, both $H_{\mathrm{KK}}$ and $P_{\mathrm{KK}}$ are Hermitian.

We first claim that $g(\tau,x_E)$ is bounded whenever $\Im(z)>0$ and $\Im(\bar z)<0$. Our goal will be to relate a general configuration with $\Im(z)>0$ and $\Im(\bar z)<0$ to a standard configuration where we know that the correlator is bounded. We begin by splitting the exponential into a positive Hermitian operator $V$ and a unitary operator $U$,
\be
e^{\frac i 2 (H_{\mathrm{KK}}+P_{\mathrm{KK}}) z - \frac i 2 (H_{\mathrm{KK}}-P_{\mathrm{KK}}) \bar z} &= V^{1/2} U V^{1/2} \nn\\
V &= e^{-\frac 1 2 (H_{\mathrm{KK}}+P_{\mathrm{KK}}) \Im(z) + \frac 1 2 (H_{\mathrm{KK}}-P_{\mathrm{KK}}) \Im(\bar z)},\nn\\
U &= e^{\frac i 2 (H_{\mathrm{KK}}+P_{\mathrm{KK}}) \Re(z) - \frac i 2 (H_{\mathrm{KK}}-P_{\mathrm{KK}}) \Re(\bar z)}.
\ee
The Cauchy-Schwartz inequality implies
\be
|g(\tau,x_E)| = |\<\Psi|V^{1/2} U V^{1/2}|\Psi\>| &\leq \<\Psi|V^{1/2}V^{1/2}|\Psi\>^{1/2}\<\Psi|V^{1/2}U^\dag UV^{1/2}|\Psi\>^{1/2}\nn\\
&= \<\Psi|V|\Psi\>.
\ee
This essentially allows us to assume that $z,\bar z$ are pure imaginary.

Next, we claim that $H_\mathrm{KK} \pm P_{KK}$ are bounded from below by a constant. To argue this, first let us forget about periodicity and consider a theory in $\R^d$. Then we have $H\pm P>0$ as a simple consequence of positivity of energy. (If either $H\pm P$ had a negative eigenvalue, Lorentz invariance would imply the existence of a state with negative energy.)

Now consider the case where $\tau$ is periodic. The spectrum of $P_{\mathrm{KK}}$ is quantized, with eigenvalues given by Kaluza-Klein (KK) momenta $n\in \Z$. In general, energies of excitations with KK-momentum $n$ are different from energies of excitations in $\R^d$ with $|\mathbf{p}|=n$. (For example, in the $n=0$ sector, the lowest nonzero eigenvalue of $H_{\mathrm{KK}}$ is the thermal mass $m_{\text{th}}$, while the Hamiltonian in $\R^d$ is gapless at zero momentum.) In particular, it is not obvious whether $H_{\mathrm{KK}}\pm P_{\mathrm{KK}}$ are positive operators. Positivity of $H_{\mathrm{KK}}\pm P_{\mathrm{KK}}$ does not follow immediately from positivity of energy because there is no Lorentz boost relating $H_\mathrm{KK}$ and $P_\mathrm{KK}$.  However, for sufficiently large $|n|$, periodicity of the $\tau$ direction becomes unimportant, and the spectrum of $H_{\mathrm{KK}}$ approaches the flat-space spectrum.  Thus, we expect $H_{\mathrm{KK}}\pm P_{\mathrm{KK}}$ are bounded from below for all $n$ by some $n$-independent constant $\l$.\footnote{Note that $\l$ is not the same thing as the thermal mass. The thermal mass is a lower bound on the nonzero eigenvalues of $H_\mathrm{KK}$ in the $n=0$ sector, whereas $\l$ is a lower bound on all eigenvalues of $H_\mathrm{KK}\pm P_\mathrm{KK}$ across all sectors. Note also that it's not necessary for $\l$ to be positive for the remainder of the argument in this section to work.} This is the key claim in this section, and we have not established it rigorously. However, we believe it is a physically reasonable assumption.

Thus, let us pick $\l$ such that $H_{\mathrm{KK}}\pm P_{\mathrm{KK}} > \l$, and let $\z = \min(\Im(z),-\Im(\bar z))$. We have
\be
\<\Psi|V|\Psi\> &= \<\Psi|e^{-\frac 1 2 (H_{\mathrm{KK}}+P_{\mathrm{KK}}) \Im(z) + \frac 1 2 (H_{\mathrm{KK}}-P_{\mathrm{KK}}) \Im(\bar z)}|\Psi\> \nn\\
&\leq \<\Psi|e^{-\frac 1 2 (H_{\mathrm{KK}}+P_{\mathrm{KK}}) \z - \frac 1 2 (H_{\mathrm{KK}}-P_{\mathrm{KK}}) \z}|\Psi\> \x e^{-\frac \l 2(\Im(z)-\z)-\frac \l 2(-\Im(\bar z) - \z)} \nn\\
&= g(0,\z)\, e^{-\frac \l 2(\Im(z)-\z)-\frac \l 2(-\Im(\bar z) - \z)}.
\ee
To summarize,
\be
|g(\tau,x_E)| &\leq g(0,\z)\, e^{-\frac \l 2(\Im(z)-\z)-\frac \l 2(-\Im(\bar z) - \z)}.
\ee
The correlator $g(0,\z)$ is simply a Euclidean correlator at nonzero $|\mathbf{x}|$ and time $\tau=0$. This is a nonsingular configuration, so the right-hand side is bounded.

Finally, note that this derivation did not actually depend on $\f(0)$ being primary. Thus, it applies to all correlators of descendants of $\f$, so all derivatives of $g(\tau,\mathbf{x})$ are bounded as well. It follows that $g(\tau,\mathbf{x})$ is analytic if $\z>0$, i.e.\ $\Im(z)>0$ and $\Im(\bar z)<0$. These conditions hold for $w$ in the upper half-plane. Symmetry under $w\to -w$ then implies that $g(\tau,x_E)$ is analytic in the lower $w$-half-plane as well.

\subsection{Behavior at large $w$}
\label{s35}

Our derivation of the Lorentzian inversion formula relies on the assumption that $g(z,\bar z)$ grows no faster than $w^{J_0}$ at large $w$ (anywhere in the upper half plane) and fixed $r$, for some fixed $J_0$. We have not been able to prove this claim or establish a rigorous upper bound on $J_0$ (analogous to the bound on chaos for thermal four-point functions \cite{Maldacena:2015waa}). In this section, we discuss the claim in more detail.

First, one can check explicitly that thermal two-point functions in $d=2$, given in \eqr{1pt2d}, are exponentially damped at large $w$. Thus, our inversion formula implies analyticity in spin for all $J\geq 0$ in 2d. This is perhaps unsurprising given that only members of the Virasoro multiplet of the identity get thermal expectation values, and such operators lie on simple trajectories as a function of $J$.

For $d>2$, we might hope to determine $J_0$ by studying perturbative examples. However, we should be aware that na\"ive perturbative expansions may not commute with the large-$w$ limit. For example, in the critical $O(N)$ model at leading order in $N$ (see section~\ref{sec:O(N)-model-example}), we find $J_0=0$. However, at each order in $1/N$ the correlator may grow more quickly.\footnote{Precisely this phenomenon happens for the Regge limit of four-point functions in large-$N$ theories. One can show on general grounds that the four-point function is bounded in the Regge limit \cite{Maldacena:2015waa}. However, each order in $1/N$ contributes faster and faster growth in the Regge limit. In holographic theories, this fact is related to the necessity of regularization and renormalization in the bulk effective theory \cite{Aharony:2016dwx}.}  It would be very interesting to set up a perturbative analysis specially adapted to the large-$w$ limit, perhaps analogous to \cite{Stanford:2015owe}.

Let us guess the behavior of the two-point function at large $w$ by studying two interesting physical regimes. Firstly, consider $w=i W$ with $W$ large and real. This corresponds to a Lorentzian two-point function at finite temperature, in the limit where both operators are highly boosted.\footnote{We thank Juan Maldacena for suggesting we consider this regime.} In the absence of the thermal bath, such a correlator would be independent of $W$ because a boost is a symmetry. The correlator can, roughly speaking, be interpreted as the amplitude for excitations created by the first operator to be absorbed by the second. There is no clear reason why this amplitude should be enhanced by the thermal bath, and thus we expect the correlator to grow no faster than $W^0$ in this regime. In fact, we might expect that the thermal bath destroys correlations between the operators, so the correlator actually decays at large $W$.\footnote{We have checked that a boundary thermal two-point function computed in AdS$_4$ using the geodesic approximation decays at large $W$.}

Another interesting physical regime is $w=(1+i\epsilon) W$ with $W$ large and real (i.e.\ $w$ on top of one of the cuts in figure~\ref{fig:opeconvergeneregionswplane}). This is the configuration that appears in the Lorentzian inversion formula. It corresponds to one of the operators moving on a nearly lightlike trajectory around the thermal circle, with one of the noncompact directions as increasing Lorentzian time $x_L$ (figure~\ref{fig:lorentzian-cylinder}). A physical picture is that the first operator creates excitations that move around the thermal circle. They repeatedly collide at $x_L=\b/2,\b,3\b/2,\dots$. Finally, some of them are absorbed by the second operator. We expect each collision to reduce the amplitude for excitations to reach the second operator. Thus, we conjecture that the correlator grows no faster than $W^0$ in this regime as well. If the collisions have a large inelastic component, the correlator should decay at large $W$.

These arguments are far from rigorous.
It would be nice to understand --- either from examples or a general argument --- what the nonperturbative behavior of the two-point function can be in the entire upper half-plane at large $w$.

\section{Applications I: Mean Field Theory}
\label{sec:MFT-example}

In this and the next section, we will perform some checks of the inversion formula, derive some new results and demonstrate its mechanics. We begin here with application to mean field theory.

In MFT, the operators appearing in the $\phi \times \phi$ OPE for some scalar primary $\phi$ are the unit operator and double-twist operators of schematic form
\be
\label{eq:MFT-operators}
[\f \f]_{n, \ell}  = \f \partial^{\mu_1} \dots \partial^{\mu_\ell} \partial^{2n}  \f~-~(\text{traces})\,,
\ee
where $\ell$ is even, with dimensions $\De_{n,\ell} = 2\De_\f + 2n+\ell$. 
Note that the free theory is the MFT with $\De_\f = \nu$, where the $[\phi\phi]_{0,\ell}$ are identified with spin-$\ell$ currents $J_\ell$. The thermal two-point function can be computed by using the method of images,
\be
\label{eq:two-pt-func-MFT}
g(z,\bar z) &= \sum_{m=-\oo}^\oo \frac{1}{((m -z)(m-\bar z))^{\De_\f}}\,.
\ee
Using this, we will perform a brute-force expansion of the two-point functions into thermal conformal blocks and compare that with the thermal one-point coefficients generated by the inversion formula.

\subsubsection*{Expanding the thermal two-point function}

We start by explicitly expanding the thermal two-point function without using the inversion formula in order to provide a non-trivial check for the entire methodology.  Going back to the $\bx$ and $\tau$ coordinates, we can write each term in (\ref{eq:two-pt-func-MFT}) as\footnote{Here, we use the identity $
\frac{1}{(1-2x y + y^2)^\a} = \sum_{j=0}^\oo C_j^{(\a)}(x)y^n.
$
}
\be
\frac{1}{((\tau+m)^2 + \bx^2)^{\De_\f}} &= \sum_{j=0}^\oo (-1)^j C_j^{(\De_\f)}(\eta)\mathrm{sgn}(m)^j \frac{|x|^{j}}{|m|^{2\De_\f+j}}\,,
\ee
where $\mathrm{sgn}(m)=\frac{m}{|m|}$ and $\eta=\frac{\tau}{|x|}$.
Thus, the two-point function is
\be
\label{eq:MFT-brute-force-expansion}
g(\tau,\bx) &= \frac 1 {|x|^{2\De_\f}} + \sum_{j=0}^\oo (-1)^j\p{\sum_{m\neq 0}  \frac{\mathrm{sgn}(m)^j}{|m|^{2\De_\f+j}}} C_j^{(\De_\f)}(\eta) |x|^{j}\nn\\
&= \frac 1 {|x|^{2\De_\f}} +  \sum_{j=0,2,\dots} 2\z(2\De_\f+j) C_j^{(\De_\f)}(\eta)|x|^{j}\,,
\ee
where $\z(s)$ is the Riemann $\z$-function. The Gegenbauer polynomials $C_j^{(\De_\f)}(\eta)$ have an expansion in terms of the correct Gegenbauer polynomials $C_j^{(\nu)}(\eta)$ appearing in section~\ref{sec:CFTs-finite-T-review} for the thermal OPE on $S^1_\beta \times \R^{d-1}$,
\be
C_j^{(\De_\f)}(\eta) &= \sum_{\ell=j,j-2,\dots,j\,\textrm{mod}\,2}  \frac{(\ell+\nu ) (\De_\f )_{\frac{j+\ell}{2}} (\De_\f -\nu )_{\frac{j-\ell}{2}}}{\p{\frac{j-\ell}{2}}! (\nu )_{\frac{j+\ell+2}{2}}} C^{(\nu)}_\ell(\eta), 
\ee
where $(a)_n=\frac{\G(a+n)}{\G(a)}$ is the Pochhammer symbol. Plugging this into (\ref{eq:MFT-brute-force-expansion}), and replacing $j=2n+\ell$, we get
\be
g(\tau,\bx) &= 
\frac{1}{|x|^{2\De_\f}} + \sum_{n=0}^\oo \sum_{\ell=0,2,\dots}  \frac{2\z(2\De_\f+2n+\ell)(\ell+\nu ) (\De_\f )_{\ell+n} (\De_\f -\nu )_{n}}{n! (\nu )_{\ell+n+1}} C^{(\nu)}_\ell(\eta)|x|^{2n+\ell}\,.
\ee
This has precisely the form of the thermal conformal block decomposition given by (\ref{eq:blockdecomposition}), with support only on the unit operator and double-twist operators \eqr{eq:MFT-operators}, whose one-point functions are given by
\be
\label{eq:MFT-thermal-one-pt}
a_{\mathbf 1} &= 1~, \\
a_{[\f\f]_{n,\ell}} &=
2\z(2\De_\f+2n+\ell)\frac{(\ell+\nu ) (\De_\f )_{\ell+n} (\De_\f -\nu )_n}{n! (\nu )_{\ell+n+1}}\nn\,.
\ee
In the free theory where $\De_\f = \nu$, the spin-$\ell$ currents $J_\ell\equiv [\f\f]_{0,\ell}$ have
\be
a_{J_\ell} &= 2\z(d-2+\ell)\,,
\label{eq:oneptfreetheory}
\ee
Note that when $d=3$, the coefficient $a_{J_0}$ is divergent. This is because the zero mode is badly behaved under dimensional reduction to $d=2$, which is related to the fact that the free boson in $d=2$ with noncompact target space is pathological.

We can now compare the above results to those predicted by the inversion formula, starting with the case $d=2$ where the Gegenbauer polynomials take a simpler form. 

\subsubsection*{Inversion in $d=2$ MFT}

As required in the inversion formula (\ref{eq:inversionformulahigherd}), we should be looking at discontinuities across the real $z$ axis for each term in (\ref{eq:two-pt-func-MFT}),
\be
\label{eq:MFT-disc-formula}
\Disc\left[\frac{1}{((m-z)(m-\bar z))^{\De_\f}}\right] &= 2\sin(\pi \De_\f)\frac{1}{(m-\bar z)^{\De_\f}( z- m)^{\De_\f}}\th( z - m)\,.
\ee
Plugging (\ref{eq:MFT-disc-formula}) into the $d=2$ inversion formula (\ref{eq:thermfg}), we find
\be
\label{eq:2d-MFT-inversion-integral}
a(\De,\ell) &= \frac{(1+(-1)^\ell)}{2\pi} 2\sin(\pi \De_\f) \sum_{m=1}^{\oo} \int_0^1 d\bar z \frac{\bar z^{\De_\f-h-1}}{(m-\bar z)^{\De_\f}}  \int_m^{\max(m,1/\bar z)} dz \frac{z^{\De_\f-\bar h - 1}}{(z-m)^{\De_\f}}.
\ee
Because we will be interested in poles coming from the region of infinitesimal $\bar z$, we can replace the upper limit of the $z$ integral with $\oo$. 
The $\zb$ and $z$ integrals in (\ref{eq:2d-MFT-inversion-integral}) then factorize and become
\be\label{zbint}
\int_0^1 d\bar z \frac{\bar z^{\De_\f-h-1}}{(m-\bar z)^{\De_\f}} &=  \sum_{n=0}^\oo \frac{\G(\De_\f+n)}{\G(n+1)\G(\De_\f)}\frac{1}{m^{\De_\f+n}}\frac{1}{\De_\f+n-h}\,,
\ee
and 
\be
\int_m^\oo dz \frac{z^{\De_\f-\bar h - 1}}{(z-m)^{\De_\f}} &= 
\frac{\Gamma (1-\Delta_\phi ) \Gamma (\bar h)}{\Gamma (\bar h-\Delta_\phi +1)}\frac{1}{m^{\bar h}},
\ee
respectively. As expected, \eqr{zbint} has poles at $h=\De_\f+n$, corresponding to MFT operators (\ref{eq:MFT-operators}). Computing the residues of each pole, we find
\be
a(\Delta, \ell) &= \sum_{n=0}^\oo \frac{-1}{\De-(2\De_\f+2n+\ell)}\p{\frac{2(1+(-1)^\ell) \G(\De_\f+n)\G(\De_\f+n+\ell)}{\G(n+1)\G(\De_\f)^2\G(n+\ell+1)} \z(2\De_\f+2n+\ell)}.
\label{eq:fgmftapplication}
\ee
Note that we get an extra factor of two when we write the pole in $h$ as a pole in $\De$. The thermal one-point function is minus the residue of the pole in $\De$. Thus, the thermal coefficient for double-twist operators of even spin can be read off as, 
\be 
a_{[\f\f]_{n, \ell}} = 4\zeta (2 \Delta _{\phi }+\ell+2 n){(\De_\phi)_{n+\ell}(\De_\phi)_n\o n! \Gamma(n+\ell+1)}, 
\ee
which is in agreement with (\ref{eq:MFT-thermal-one-pt}) when $\nu = 0$.

\subsubsection*{Inversion in $d>2$ MFT}
While in $d=2$ one can obtain the contribution of all double-twist families through simple integral manipulations, the $d>2$ case will require a more careful series of approximations to get the residues corresponding to each family's pole. 

Plugging in the discontinuity (\ref{eq:MFT-disc-formula}) into the inversion formula (\ref{eq:inversionformulahigherd}), we are left to compute
\be
\int_0^1 \frac{d\bar z}{\bar z} \int_m^{\max (m,1/\bar z)} \frac{dz}{z} (z\bar z)^{\De_\f-\frac \De 2 - \nu}(z-\bar z)^{2\nu} F_J\p{\sqrt{\frac {\bar z} {z}}} \frac{1}{(m-\bar z)^{\De_\f}(z-m)^{\De_\f}}\,. \label{eq:MFT-integral}
\ee
Again, poles for double-twist operators (\ref{eq:MFT-operators}) come from the region of integration near $\bar z\sim 0$. Thus, we can replace the upper limit of the $z$ integral with $\oo$.  We are also free to rescale $\bar z \rightarrow z \bar z$ and set the integration range for $\bar z$ back to $[0,1]$, since we will obtain the same pole location with the same residue.  By also rescaling $z\to m z$ we find
\be\label{zzresc}
m^{-\De}\int_0^1 \frac{d\bar z}{\bar z} \int_1^\oo \frac{dz}{z} (z^2\bar z)^{\De_\f-\frac \De 2 - \nu}z^{2\nu}(1-\bar z)^{2\nu} F_J\p{\sqrt{\bar z}} \frac{1}{(1-z\bar z)^{\De_\f}(z-1)^{\De_\f}}.
\ee
The $z$ integral can be done explicitly, leaving the $\bar z$ integral
\be \label{mftint}
&m^{-\De} \frac{\G(1-\De_\f)\G(\De-\De_\f)}{\G(1+\De-2\De_\f)}
\nn\\
&\x\int_0^1 \frac{d\bar z}{\bar z} \bar z^{\De_\f-\frac \De 2 - \nu}(1-\bar z)^{2\nu}{}_2 F_1(\De_\f,-\De+2\De_\f,1-\De+\De_\f,\bar z) F_J\p{\sqrt{\bar z}}.
\ee
We can now expand in $\bar z$ and get a series of poles. 

Let us focus on the first sets of poles in the integrand of \eqr{mftint} corresponding to the $[\f\f]_{0, \ell}$ and $[\f\f]_{1, \ell}$ operators, 
\be
\label{eq:MFT-barZ-expansion}
&\bar z^{\De_\f-\frac \De 2 - \nu}(1-\bar z)^{2\nu}{}_2 F_1(\De_\f,-\De+2\De_\f,1-\De+\De_\f,\bar z) F_J\p{\sqrt{\bar z}} \nn\\
&\sim \bar z^{\frac {J-\De} 2+\De_\f}\p{1+\bar{z} \left(\frac{\Delta_\phi  (\Delta -2 \Delta_\phi )}{\Delta -\Delta_\phi -1}-\frac{(J+2) \nu }{J+\nu +1}\right)+\dots}.
\ee
Multiplying this by the factor $(1+(-1)^J)2\sin(\pi \De_\f) K_J$, left out in \eqref{eq:MFT-integral} for clarity, gives the full contribution of these poles to $a(\De,J)$. The first term gives a pole at $h=\De_\f$ of the form
\be
a(\De,J) &\supset (1+(-1)^J)2\sin(\pi \De_\f) K_J \sum_{m=1}^\oo \frac{1}{m^{\De}} \frac{\G(1-\De_\f)\G(\De-\De_\f)}{\G(\De-2\De_\f+1)} \frac{1}{\De_\f-h}\nn\\
 &= -(1+(-1)^J)\z(2\De_\f+J)\frac{\G(J+\De_\f)\G(\nu)}{\G(\De_\f)\G(J+\nu)} \frac{1}{\De-(2\De_\f+J)},
\ee
where we've set $\De=2\De_\f+J$ in the last step to obtain the correct value of the residue. This agrees with \eqr{eq:MFT-thermal-one-pt} at $n=0$. The order $\bar z$ term in (\ref{eq:MFT-barZ-expansion}) gives
\be
a(\Delta, J) \supset -(1+(-1)^J)\z(2\De_\f+2+J) \frac{(J+\nu)(\De_\f)_{J+1}(\De_\f-\nu)}{(\nu)_{J+2}}\frac{1}{\De-J-(2\De_\f+2)}.
\ee
This agrees with \eqr{eq:MFT-thermal-one-pt} at $n=1$.

Note that the unit operator pole was absent in the above manipulations. This is resolved by the presence of the ``arc terms'' in the Lorentzian inversion formula, $a^{\rm arcs}(\De,J)$, which we have neglected here. In the limit $|w|\rightarrow \infty$ the MFT correlator is simply given by, $\lim_{|w|\rightarrow \infty} g(r, w) = 1/r^{2\De_\f}$. The contribution of the contours given by (\ref{eq:contour-neglect}) precisely yields a pole corresponding to the unit operator at $J=0\, ,\Delta=0$ with residue equal to $1$. We will witness a more intricate balance between the arc and non-arc contributions when studying the $O(N)$ vector model in the subsection below.

\section{Applications II: Large $N$ CFTs}
\label{sec:largeN-example}

Consider a CFT with large central charge $c_T\sim N^2\rar\infty$. In the OPE regime, we may organize two-point functions on $S^1_\b\times \R^{d-1}$ by powers of $1/N$. Let us use the canonical normalization
\eq{stnorm}{\la \O\O\ra_{\R^d} \sim N^0~, ~~ \la \O_1\O_2\O_3\ra_{\R^d} \sim {1\o N}~, ~~\ldots,}
where $\O_i$ are single-trace operators. Then the thermal scalar two-point function $g(\tau,\mathbf{x})$ receives the following types of contributions, organized by powers of $1/N$ appearing in the OPE coefficients:
\es{nexp}{g(\tau,\mathbf{x}) &\approx \left(\la \mathbf{1}\ra_\b + \sum_{n,\ell}\la [\phi\phi]_{n,\ell} \ra_\b\right)+ {1\o N}\left(\sum_{\O\in\phi\times\phi} \la \O\ra_\b\right)\\
&+ {1\o N^2}\left(\sum_{n,\ell}\la [\phi\phi]_{n,\ell} \ra_\b + \sum_{n,\ell}\sum_{[\O_i\O_j]_{n,\ell}\in\phi\times\phi}\la [\O_i\O_j]_{n,\ell}\ra_\b\right)+O\left({1\o N^3}\right),}
where we have again defined the double-trace composite operators $[AB]_{n,\ell}$, of schematic form
\eq{}{[AB]_{n,\ell} = A\partial^{2n}\partial_{\mu_1}\ldots \partial_{\mu_{\ell}} B~ -~(\text{traces}).}
The first group of operators in \eqr{nexp} represents the two-point function of MFT, in which the $[\phi\phi]_{n,\ell}$ appear with the MFT OPE coefficients, which can be found in \cite{Fitzpatrick:2011dm}; the second group represents single-trace operators; the third group represents double-trace operators, including the $1/N^2$ corrections to the MFT exchanges; and so on. However, this way of organizing the contributions is not terribly useful because the one-point functions themselves scale with positive powers of $N$. In particular, in the normalization \eqr{stnorm}, one-point functions of $n$-trace operators exhibit the leading-order scaling
\eq{}{\la [A_1\ldots A_n]\ra_\b \sim N^n+\ldots.}
This implies an infinite set of contributions to $g(\tau,\mathbf{x})$ at order $N^0$, which poses an obvious challenge to computing $g(\tau,\mathbf{x})$, in contrast to the familiar $1/N$ counting used in vacuum four-point functions.

We now study the inversion formula in the critical $O(N)$ vector model, and discuss some features of its application to CFTs with weakly coupled holographic duals.

\subsection{$O(N)$ vector model at large $N$}
\label{sec:O(N)-model-example}

The critical $O(N)$ model at large $N$ has been studied in detail before \cite{Sachdev:1993pr, chubukov1994theory,ZinnJustin:2002ru}. 
This theory has $c_T= N c_{\rm free}$ to leading order in $1/N$. 
The main feature we will need is the value of the thermal mass.
In the $O(N)$ model at large $N$, the thermal mass is equal to the expectation value of the IR operator $\sigma$, which appears in the action after applying a Hubbard-Stratanovich transformation to the $\phi^4$ coupling:
\be
\label{eq:O(N)-lagrange}
\mathcal L = \frac{1}2 (\partial_\mu \phi_i)^2 + \frac{1}2 \sigma \phi_i \phi_i  - \frac{\sigma^2}{4\lambda}\,.
\ee
The critical point is obtained by taking $\lambda \rightarrow \infty$ as $\sigma^2$ becomes irrelevant in the IR. 
In appendix~\ref{app:O(N)-thermal-mass}, we review the derivation of the following result \cite{Sachdev:1992py, ZinnJustin:2002ru},
\be
\label{eq:O(N)-thermal-mass}
 \langle \sigma \rangle_\beta = m_{\text{th}}^2 = \beta^{-2} \left[ 2 \log \left(\frac{1+\sqrt{5}}2\right)\right]^2+\cO\left({1/ N}\right)\,.
\ee 
As we shall see later in this section, the above formula for the thermal mass is intimately related to correctly reproducing the $O(N)$ singlet spectrum from the inversion formula. Thermal properties of the $O(N)$ model were also studied in \cite{Petkou:1998fb,Petkou:1998fc}.

Let us enumerate the $O(N)$ singlets of the critical $O(N)$ model whose thermal expectation values we will compute (any non-singlet has vanishing thermal one-point function). 
The ``single-trace" singlets are the scalar $\sigma$, with $\De=2+\cO(1/N)$, and the higher-spin currents $J_\ell$, with $\ell\in 2\mathbb{Z}^+$ and $\De=\ell+1+\cO(1/N)$. In the $\phi_i\x\phi_i$ OPE, one generates the larger family of operators\footnote{Note that, due to the equation motion for $\sigma$, schematically of the form $\partial^2 \phi_i  \sim \sigma \phi_i$, the $n>0$ families in (\ref{eq:O(N)-operators}) may be related to families involving both $\phi_i$ and $\sigma$. For instance, $[\phi_i \sigma \phi_i]_{0, \ell}\equiv \phi_i  \partial^{\mu_1} \dots \partial^{\mu_\ell} \sigma\f_i, \sim [\phi_i \phi_i]_{1,\ell}$. There are still other families of primary singlet operators which are not of this form. For instance,
$[\f_i \sigma \f_i]_{n,k, \ell} = \phi_i  \partial^{\mu_1} \dots \partial^{\mu_\ell} (\partial^{2n} \sigma) \partial^{2k}\f_i$, with $\De_{n,k,\ell} = 1+2n+2k+2+\ell$. Note that such operators are degenerate in $h$ and $\bar h$ for different values of $n$ and $k$ and may have degenerate dimensions with some $n>0$ operators in (\ref{eq:O(N)-operators}); in the presence of degeneracies, the inversion method as presented here yields linear combinations of thermal one-point functions. \label{foot27} 
}
\eq{eq:O(N)-operators}{\ell>0:\quad [\f_i \f_i]_{n, \ell} =\f_i \partial^{\mu_1} \dots \partial^{\mu_\ell}\partial^{2n}  \f_i\quad\text{where}\quad \De_{n,\ell} = 1+2n+\ell+\g_{n,\ell}~.}
where the anomalous dimensions are suppressed as $\g_{n,\ell} \sim \cO(1/N)$. For $n=0$, these operators are the slightly-broken higher-spin currents, 
\eq{eqonmodelfamilies}{J_\ell \equiv [\phi_i\phi_i]_{0,\ell}~, \quad \text{where}\quad  \De_\ell = \ell+1+\cO(1/N). }
The families (\ref{eqonmodelfamilies}) do not analytically continue down to $\ell =0$; instead, $\sigma$ plays the role of $\phi_i\phi_i$ in the IR. Accordingly, the most basic scalar operators are powers of $\sigma$, 
\be
\label{eq:O(N)-scalar-operators}
\ell=0: \quad \sigma^m~, \quad \text{where} \quad \De_m = 2m +\cO(1/N)\,.
\ee
In what follows, we will compute thermal one-point functions of $J_\ell$ and $ \sigma^m$, and exhibit the algorithm for computing the one-point functions of $[\phi_i\phi_i]_{n,\ell}$ for all $(n,\ell)$. 

As discussed below \eqr{nexp}, the thermal coefficients $a_\cO$ in large $N$ CFTs receive contributions from an infinite set of operators in the $\phi\phi$ OPE, due to the opposite large $N$ scaling of OPE coefficients $f_{\phi\phi\cO}$ and thermal one-point functions $b_\cO$. That discussion was for single-trace operators $\phi$, but the same scaling holds for the $\phi_i$ fields in the $O(N)$ model, i.e.\footnote{\label{f30}We assume canonical normalization for the operators $\f_i$ and $\sigma$:  $\langle \phi_i(x) \phi_j(0) \rangle = \delta_{ij}/|x|$ and $\langle \s(x) \s(0) \rangle = \delta_{ij}/|x|^4$. } \be
\label{eq:O(N)-large-N-scalings}
&\frac{f_{\f_i \f_i \sigma^m}}{\sqrt{c_{\sigma^m}}} \sim \cO\left(\frac{1}{N^{m/2}}\right)\,, \quad \frac{b_{\sigma^m}}{\sqrt{c_{\sigma^m}}} \sim \cO(N^{m/2}) \,, \quad \Rightarrow \quad a_{\sigma^m} \sim \cO(1),\nn\\
&\frac{f_{\f_i \f_i [\phi_i \phi_i]_{n, \ell}}}{\sqrt{c_{[\phi_i \phi_i]_{n, \ell}}}} \sim \cO\left(\frac{1}{N^{\frac{n+1}2}}\right)\,, \frac{\quad b_{[\phi_i \phi_i]_{n, \ell}}}{\sqrt{c_{[\phi_i \phi_i]_{n, \ell}}}} \sim \cO(N^{\frac{n+1}2}) \quad \Rightarrow \quad a_{[\phi_i \phi_i]_{n, \ell}} \sim \cO(1),
\ee
where in order to derive the second set of scalings we have used the schematic operator relation $\partial^2\phi_i \sim \sigma\phi_i$. We emphasize that the computation of $a_{\sigma^m}$ and $a_{[\phi_i \phi_i]_{n, \ell}}$ (which we will show below) gives a window into arbitrarily high orders in $1/N$ perturbation theory: for instance, to derive $ f_{\f_i \f_i \sigma^m}$ would require going to $(m-1)^{\rm th}$ order in large-$N$, which is intractable using standard perturbative methods. 

\subsubsection*{Thermal two-point function review}
\label{sec:O(N)-model-thermal-review}

The propagator for the field $\phi_i$ in fourier space is given by, 
  \be
     \label{eq:O(N)-propagator-in-momentum-space}
  G_{ij}(\omega_n, \mathbf k)  = \langle \phi_i \phi_j \rangle (\omega_n, \mathbf k)=  \frac{\delta_{ij}}{\omega_n^2 + \bk^2 + \sigma} =  \frac{\delta_{ij}}{\omega_n^2 + \bk^2 + m_{\text{th}}^2}.
  \ee
   At the saddle-point, the non-zero expectation of $\sigma$ thus acts like a mass term which is absent when considering the MFT propagator considered in section~\ref{sec:MFT-example}.  We can now use the $ G_{ij}(\omega_n, \mathbf k) $ to express the propagator in position-space as\footnote{To derive this, we use the Poisson resummation formula to turn a sum over Matsubara  frequencies $\w_n=2\pi n$ into a sum over shifts in $\tau$:
   \be
   \sum_{n\in \Z} \tl f(\w_n) e^{-i\w_n \tau}
   &= \int d\w \sum_{n\in \Z} \de(\w-\w_n) \tl f(\w) e^{-i\w \tau}
   = \int \frac{d\w}{2\pi} \sum_{m\in \Z} e^{-i\w(\tau-m)} \tl f(\w)
   = \sum_{m\in \Z} f(\tau-m).
   \ee
   Here, $\tl f$ is the Fourier transform of $f$. Thus, we can Fourier transform $G_{ij}(\w,\mathbf k)$ (treating $\w,k$ as continuous), which gives the Yukawa potential $\frac{e^{-m_\mathrm{th}|x|}}{|x|}$. Then we sum over integer shifts in $\tau$ to obtain (\ref{eq:O(N)-propagator-in-position-space}).}
   \be 
   G_{ij}(\tau, \mathbf x) &= \delta_{ij} \sum_{n=-\infty}^{\infty} \int \frac{d^2k}{(2\pi)^2} \frac{e^{-i \mathbf k \cdot \mathbf x- i \omega_n \tau }}{\omega_n^2 + \bk^2 + m_{\text{th}}^2} 
   \nn\\
   &= \delta_{ij} \sum_{m = -{\infty}}^{\infty} \frac{1}{\left[(m-z)(m-\bar z)\right]^{1/2}} e^{-m_{\text{th}}\left[(m-z)(m-\bar z)\right]^{1/2}}.
      \label{eq:O(N)-propagator-in-position-space}
   \ee
This is similar to the MFT propagator (\ref{eq:two-pt-func-MFT}), but with an exponentially decaying factor multiplying each term. While in the MFT study in section~\ref{sec:MFT-example}, each term in  (\ref{eq:two-pt-func-MFT}) could be expanded in Gegenbauer polynomials, to our knowledge an expansion for each term in (\ref{eq:O(N)-propagator-in-position-space}) cannot be found in the literature. Thus, we will seek to find it using the Lorentzian inversion formula.

\subsubsection*{Inversion I: Higher-spin currents}
\label{sec:O(N)-model-apply-inversion-ff_0}

We now use the inversion formula (\ref{eq:inversionformulahigherd}) to recover the thermal one-point functions of the currents $J_\ell$, and give implicit results for the higher families $[\phi_i\phi_i]_{n,\ell}$ with $n>0$. 

First one has to understand the discontinuities along the axis $\text{Im}\, z = 0$ with $\text{Re } z>1$ for each term in (\ref{eq:O(N)-propagator-in-position-space}):
  \be
  \label{eq:O(N)-propagator-disc}
  \text{Disc}\,\frac{e^{-m_{\text{th}}\left[(m-z)(m-\bar z)\right]^{1/2}}}{\left[(m-z)(m-\bar z)\right]^{1/2}} = \frac{2\cos \left(m_{\text{th}}\left[(z-m)(m-\bar z)\right]^{1/2}\right)}{\left[(z-m)(m-\bar z)\right]^{1/2}} \theta(z-m)\,.
  \ee
  We now apply the inversion formula (\ref{eq:inversionformulahigherd}) to get the contribution of each term in (\ref{eq:O(N)-propagator-in-position-space}). We focus on the integral, multiplying overall factors at the end. We also denote the spin as $\ell$, rather than $J$. For terms with $m\geq 1$ we find, following the same approximation scheme as in section~\ref{sec:MFT-example} for $d>2$ MFT (see around \eqr{zzresc}),
  \be
  \label{eq:O(N)-inversion-indiv-term}
  & 2 \int_0^1 \frac{d \bar z}{\bar z}\int_m^{\max (m,1/\bar z)} \frac{dz}{z} (z \bar z)^{-\frac \De 2}(z-\bar z) F_\ell\p{\sqrt{\frac {\bar z} {z}}} \frac{\cos \left(m_{\text{th}}\left[(z-m)(m-\bar z)\right]^{1/2}\right)}{\left[(z-m)(m-\bar z)\right]^{1/2}} \nn \\ 
  &\xrightarrow[\substack{\bar z\rightarrow mz\bar z\\z\rightarrow mz}]{} \frac{2}{m^\Delta}  \int_1^\oo dz \, z^{-\De} \int_0^{z} \frac{d \bar z}{\bar z} \bar z^{-\frac{\Delta} 2}(1-\bar z) F_\ell\p{\sqrt{\bar z}}\frac{\cos \left(m_{\text{th}} m \left[(z-1)(1-z \bar z)\right]^{1/2}\right)}{\left[(z -1)(1- z\bar z)\right]^{1/2}}\,.
  \ee
Expanding the integrand in (\ref{eq:O(N)-inversion-indiv-term}) at small $\bar z$, 
 \be 
\label{eq:O(N)-smallzBar-expansion}
(1-\bar z) F_\ell(\sqrt{\bar z}) \frac{\cos \left(m_{\text{th}} m \sqrt{(z-1)(1-z \bar z)}\right)}{\left[(z -1)(1- z\bar z)\right]^{1/2}} &= \frac{\bar z^{\frac{\ell+1}2}}{\sqrt{z-1}}\big(\cos\left(m_{\text{th}} m \sqrt{z-1}\right)+ \cO(\bar z)\big)\,.
 \ee
The $\bar z$ integral at leading order gives rise to the contribution of the first double twist family with $h = 1/2$, via $\int_0^{z} d \bar z~ \bar z^{-(\Delta-\ell-1)/2} = {z^{\frac{1}2-h}}/({\frac{1}2-h})$ with a pole at $h=1/2$. Plugging this into \eqr{eq:O(N)-inversion-indiv-term}, we now perform the $z$-integral to extract the residue at $h=1/2$, which is found to be
\be
\Res_{h={1\o 2}} \eqr{eq:O(N)-inversion-indiv-term} = - {2^{{5\o 2} - \Delta } \sqrt{\pi}\o\Gamma(\De)} m_{\text{th}}^{\Delta-{1\o2}}m^{-{1\o2}} K_{ \Delta-{1\o2}}(m_{\text{th}} m),
\ee
where $\De=1+\ell$ and $K_{\De-{1\o2}}$ is the modified Bessel function. The full result requires a sum over $m$ as in \eqr{eq:O(N)-propagator-in-position-space}; performing this sum, and appending overall factors from the inversion formula, we find that the thermal coefficient $a_{J_\ell}$ for the higher-spin currents $J_\ell$ is
\be\label{ajl}
a_{J_\ell} = (1+(-1)^\ell)\frac{ 2^{-{1\o2}-\ell} (m_{\text{th}})^{\frac 1 2+\ell}}{\Gamma\left({1\o2}+\ell\right)} \sum_{m=1}^{\infty} m^{-\frac 1 2} K_{\frac 1 2+\ell}(m_{\text{th}} m).
\ee 
This sum can be performed to yield the following result:
\eq{jlsum}{a_{J_\ell} = \sum_{n=0}^\ell {2^{n+1}\o n!}{(\ell-n+1)_n\o (2\ell-n+1)_n} m_{\text{th}}^n \text{Li}_{\ell+1-n}(e^{-m_{\text{th}}}).}
We can translate this to a result for the thermal one-point function, $b_{J_\ell}$, itself, using known results in the literature for the OPE coefficients $f_{\f_i\f_i J_\ell}$, together with our $\phi_i$ normalization in footnote \ref{f30}. From e.g.\ \cite{skv}, in $d=3$ we have
\eq{}{{f_{\f_i\f_i J_\ell}\o \sqrt{c_{J_\ell}}} ={1\o \sqrt{N}} \Gamma\left(\ell+{1\o2}\right)\sqrt{2^{\ell+1}\o\pi \ell}.}
Using the relation \eqr{eq:blockdecomposition} between $a_\cO$ and $b_\cO$, we find 
\eq{bell}{{b_{J_\ell}\o \sqrt{c_{J_\ell}}} = {\sqrt{N 2^{\ell+1}\ell}\o\ell!}\sum_{n=0}^\ell {2^{n}\o n!}{(\ell-n+1)_n\o (2\ell-n+1)_n} m_{\text{th}}^n \text{Li}_{\ell+1-n}(e^{-m_{\text{th}}}).}
which is the ratio which is independent of the norm of $J_\ell$. 

This is an elegant result. The case $\ell=2$ corresponds to the stress tensor. In this case, the sum may be further simplified to yield
\eq{}{a_{T} = {8\over 5}\zeta(3)~.}
Using (\ref{eq:afrelation}), we see that this agrees with a previous result of \cite{csy, Sachdev:1993pr}. For the higher-spin currents $\ell>2$, we are not aware of previous results in the literature for the thermal one-point functions, so \eqr{bell} are new. Intriguingly, $a_{J_\ell}$ is a transcendental function of uniform transcendental weight $\ell+1$, where we note that $m_{\text{th}}$ is itself of transcendental weight one. It would be fascinating to understand this transcendentality better. 

As mentioned in the introduction, this result has implications for higher-spin black hole solutions of Vasiliev higher-spin gravity in AdS$_4$. The translation invariance of thermal one-point functions means that \eqr{bell} are proportional to the higher-spin charges of the CFT at finite temperature. Together with the thermal mass $m_{\rm th} \sim \sqrt{\la \sigma\ra_\beta}$, these charges fully determine the ``higher-spin hair'' of the putative black hole solution dual to the CFT thermal state with vanishing higher-spin chemical potentials. This black hole has not yet been constructed, due to difficulties in interpreting and solving Vasiliev's equations. Our result provides a benchmark, both for any explicit candidate black hole, and for a physical interpretation of proposed constructions of higher-spin gauge-invariant charges (see e.g. \cite{hs1, hs2, hs3, hs4, hs5}).

It is not much more difficult to derive the one-point functions of the $n>0$ families appearing in (\ref{eq:O(N)-operators}). One simply has to keep higher orders in $\bar z$ in (\ref{eq:O(N)-smallzBar-expansion}): a term of $\cO(\zb^n)$ gives a pole at $h={1\o2}+n$, i.e.\ for spin-$\ell$ operators with $\De=1+\ell+2n$.

\subsubsection*{Inversion II: Scalars}
\label{sec:O(N)-model-correction-to-inversion} 

The above results are incomplete in the scalar sector, and present a small puzzle. Note that our final expression \eqr{ajl} was actually valid all the way down to $\ell=0$, which would correspond to a scalar with dimension $\Delta  =1$, even though such an operator is absent. The same would happen for the poles with higher $n>0$, which would seem to indicate the presence of spurious scalars with odd integer dimension $\Delta \in 2 \,\mathbb Z^+ - 1$. Moreover, we did not recover the $\De=2m$ scalar poles corresponding to $\sigma^m$ exchanges, nor the unit operator. As we now show, these issues are remedied by considering the arc contributions to the inversion formula. 

Following the notation in section~\ref{sec:inversion-d=2} where $z= r w$ and $\bar z= rw^{-1}$, we are interested in computing the $w\rightarrow e^{i \phi} \infty$ behavior for each term in the propagator (\ref{eq:O(N)-propagator-in-position-space}). In this limit, the only surviving term is given by the $m=0$ term, 
\be
 G_{ij}(r, |w| \rightarrow \infty) = \frac{\delta_{ij}}{r} e^{-m_{\text{th}} r}\,.
\ee
The contribution of the integral correction to the inversion formula is given by (\ref{eq:thermfg}). When plugging in the asymptotic value of the propagator this becomes 
\be
a^{\text{arcs}}(\Delta, \ell) = 2K_\ell(1+(-1)^\ell) &\int_0^1 {dr} r^{-\Delta}\oint \frac{dw}{i w} \lim_{|w|\rightarrow \infty} \bigg[\p{\frac 1 {i}(w-w^{-1})}^{2\nu} \nn\\ &\p{F_\ell(w) e^{-i \pi \nu} + F_\ell(w^{-1})e^{i\pi \nu}}\bigg] \frac{1}r e^{-m_{\text{th}} r}\,.
\ee  
The integral over $w$ in the limit in which $|w|\rightarrow \infty$ is trivial and simply gives a factor of $2\pi$ when $\ell=0$ and $0$ when $\ell>0$. This indeed confirms that the thermal coefficients quoted above for the currents with $\ell>0$ are correct. For $\ell=0$, we are left with an integral over $r$: 
\be 
a^{\text{arcs}}(\Delta, 0) =\int_0^1 \frac{dr}{r^{1+\Delta}} e^{-m_{\text{th}} r}\,,
\ee
where we note that the factor $ 4\pi K_\ell (1+(-1)^\ell)  = 2$ in the case $\ell=0$. 
The poles in $\Delta$ of $a(\Delta, \ell)$ are independent of the upper bound of the integral. Changing the upper bound of the integral to $\infty$, we find the extremely simple formula, 
\be
a^{\text{arcs}}(\Delta, 0) = m_{\text{th}}^{\Delta} \,\Gamma(-\Delta)\,.
\ee
Since $\Gamma(x)$ has poles at each negative integer value of $x$, we can express the function $a(\Delta, 0) $ around each $m \in \mathbb Z_{\geq 0}$	as
\be
\label{eq:O(N)-arc-contribution}
a^{\text{arcs}}(\Delta, 0) \sim \frac{1}{\Delta - m} \frac{(-1)^{m+1}m_{\text{th}}^m}{\Gamma(m+1)}\,,
\ee
These poles do two things. First, they cancel all spurious scalar poles of $a(\De,0)$ at $\De\in 2\Z^+-1$. Second, they give the correct poles for the actual scalar operators of the theory, which have $\De\in 2\Z^+$, as well as the unit operator pole.

Let us first analyze the case $m=0$. This simply returns a thermal one-point function of $1$, corresponding to a correctly normalized unit operator.  

Next we take $m=1$. This pole at $\Delta =1$ and $\ell=0$ should cancel the spurious scalar pole of the previous analysis. From \eqr{jlsum}, we get
\be
a(\Delta, 0) \sim -\frac{2  \log \left(1-e^{-m_{\text{th}}}\right)}{\Delta-1}.
\ee
This only cancels $a^{\text{arcs}}(\Delta, 0)$ when 
\be 
-2 \log(1-e^{-m_{\text{th}}}) =m_{\text{th}}.
\ee
The solution of this equation is uniquely given by the saddle point value of the thermal mass in (\ref{eq:O(N)-thermal-mass})! Thus, a correct value for the thermal mass in (\ref{eq:O(N)-propagator-in-position-space}) is what yields the precise cancellation of the $\De=1$ scalar contribution from the non-arc terms to the thermal one-point function obtained via inversion. Alternatively, this may be viewed as a novel derivation of $m_{\rm th}$. Likewise, the spurious scalar poles  with $\De=3,5,\ldots$ that would arise from $n>0$ terms in \eqr{eq:O(N)-inversion-indiv-term} should cancel against the $m=3,5,\ldots$ terms in \eqr{eq:O(N)-arc-contribution}. 

Finally, by taking $m=2\Z_+$ in (\ref{eq:O(N)-arc-contribution}), we find the residue
\be\label{sigmam}
\Res_{\De=2m}a^{\rm arcs}(\De,0) = -\frac{m_{\text{th}}^{2m}}{\Gamma(2m+1)} \,.
\ee
This gives a linear combination of the $a_\cO$ coefficients for all scalar operators $\cO$ of dimension $\De=2m$. For $m=2$, there is only a single operator, $\sigma^2$. For higher values of $m$, there are possible degeneracies, as briefly discussed in footnote \ref{foot27}.

\ssec{Holographic CFTs}
\label{sec:holographic-CFTs}
We now make some comments on large $N$ CFTs in the context of AdS/CFT. 

A universal set of contributions to the OPE expansion \eqr{nexp} comes from the stress tensor, $T_{\mu\nu}$, and its multi-traces, $[T\ldots T]$, which necessarily appear in the $\phi\times\phi$ OPE for any $\phi$. In a CFT with a weakly coupled gravity dual, these terms represent the purely gravitational interactions between the bulk field $\Phi$, dual to $\phi$, and the thermal geometry. The form of these contributions is sensitive to the gap scale to single-trace higher-spin operators ($J>2$), $\Dg$. We would like to understand how. 

First, consider the case $\Dg\gg1$, where the bulk dual is general relativity plus small corrections, coupled to low-spin matter \cite{Camanho:2014apa, Afkhami-Jeddi:2016ntf, Costa:2017twz,Meltzer:2017rtf}. In this case, the thermal state on $S^1_\b \times \R^{d-1}$ is dual to an AdS$_{d+1}$-Schwarzchild black brane geometry with inverse Hawking temperature $\b$,\footnote{At infinite spatial volume, thermal AdS is thermodynamically disfavored.} and the stress tensor contributions in the OPE decomposition \eqr{nexp} are dual to the exchange of arbitrary numbers of gravitons between $\Phi$ and the black brane. In a heavy probe limit $1\ll\De_\f\ll M_{\rm pl}L_{\rm AdS}$, the connected two-point function may be computed as the exponential of a geodesic length, $\la \phi(x)\phi(0)\ra_\b\sim e^{-\De_\phi x}$. The disconnected component of the correlator, $\sim \la\phi\ra_\b^2$, is computed as an infall of each particle into the black brane horizon.\footnote{This interaction requires a nonzero cubic coupling between two gravitons and $\Phi$; this is forbidden at the two-derivative level, but may appear at the four-derivative level in the form of a $\phi C_{\mu\nu\rho\sigma}^2$ coupling, where $C_{\mu\nu\rho\sigma}$ is the Weyl tensor (see e.g.\ \cite{Myers:2016wsu} for an application). Such couplings are, however, suppressed by the mass scale of higher-spin particles in the bulk \cite{Meltzer:2017rtf} and, in more general theories of gravity, by universal bounds \cite{Cordova:2017zej, Meltzer:2017rtf}.} This disconnected contribution goes to a constant plus $e^{-m_{\text{th}} x}$ corrections, and thus becomes more important at sufficiently long distances.

It is instructive to examine the classic case of strongly coupled $\cN=4$ super-Yang-Mills (SYM), with $SU(4)_R$ symmetry. The single-trace scalar spectrum consists of the Lagrangian operator, as well as the 1/2-BPS operators $\O_p$ with $p=2,3,4,\ldots$, which live in the $[0,p,0]$ representation of $SU(4)_R$ and have conformal dimensions $\De=p$. The $R$-symmetry constrains the thermal two-point functions $\la \O_p\O_p\ra_\b$ to take the form
\eq{}{\la \O_p(x)\O_p(0)\ra_\b = \la \O_p(x)\O_p(0)\ra_\b^{\rm MFT} + (\text{stress tensor terms}).}
That is, in the OPE decomposition of $\la \O_p\O_p\ra_\b$ for any $p$, the stress tensor terms are the {\it only} terms besides the MFT contributions at leading order in $1/N$. This follows from the absence of  $R$-singlets in the single-trace spectrum besides the identity operator, the Lagrangian operator and the stress tensor, and the fact that the Lagrangian carries charge under a emergent $U(1)_Y$ bonus symmetry \cite{Intriligator:1998ig}. The stress tensor contribution, $a_T$, exhibits a famous dependence on the 't Hooft coupling $\l$ \cite{gkp2}, 
\eq{}{{a_T|_{\l\rar\infty}\o a_T|_{\l\rar0}} = {b_T|_{\l\rar\infty}\o b_T|_{\l\rar0}}= {3\o4}\, .}
In relating $a_T$ to $b_T$, we have used that $f_{\O_p\O_p T_{\mu\nu}}$ and $C_T$ are $\l$-independent.

In more general theories with large $\Dg$ and a sparse spectrum of light operators, there can be single-trace global symmetry singlets, so contributions from operators other than the stress tensor to the scalar two-point function $g(\tau,\mathbf{x})$ are possible. As $\Dg$ decreases, there are different possible sources of $\Dg$ corrections to thermal correlation functions. First, the low-spin OPE data receive power-law corrections in $\Dg$. This includes the OPE coefficients of double-trace operators (see \cite{Goncalves:2014ffa} for an $\cN=4$ example). In addition, there are $e^{-\Dg}$ corrections due to new contributions of massive string states with $\De\sim \Dg$. At finite $\Dg$, there are many possible behaviors. 

Finally, note that if instead we examine thermal two-point functions of the stress tensor, $\la T_{\mu\nu} T_{\rho\sigma}\ra_\b$, the effects of large $\Dg$ are more visible. For instance, $\la T_{\mu\nu}T_{\rho\sigma}T_{\lambda\eta}\ra$ and $\la T_{\mu\nu}T_{\rho\sigma}\cO\ra$ couplings scale with inverse powers of $\Dg$ \cite{Camanho:2014apa, Afkhami-Jeddi:2016ntf, Costa:2017twz,Meltzer:2017rtf}, thus suppressing various possible contributions to $\la T_{\mu\nu} T_{\rho\sigma}\ra_\b$ in the OPE limit. For many reasons, it would be interesting to extend the methods discussed herein to the case of spinning external operators, and to $T_{\mu\nu}$ in particular; this would allow us to study the purely gravitational physics of the thermal geometry in AdS, without the need for a probe scalar field.

 \section{Large-spin perturbation theory}
\label{sec:large-spin-perturbation-theory}

So far, our discussion of thermal two- and one-point functions has been in theories where we have enough analytic control to explicitly compute the thermal two-point functions, which we can invert to obtain one-point functions. How can we analyze theories for which we don't have any direct method of computing two- or one-point functions, such as the 3d Ising CFT? Inspired by studies of CFT four-point functions in Minkowski space, we use the inversion formula to set up a bootstrap algorithm for the thermal data in any CFT. The inversion formula takes in the two-point function, and returns its decomposition in the $s$-channel OPE. Crucially, any presentation of the two-point function could be inserted into the inversion formula, and its inversion would yield how the given presentation is related to the $s$-channel data. Here, we will invert the $t$-channel OPE to the $s$-channel data. This relates the one-point functions of operators in the theory in a highly non-trivial fashion. By iterating these relations, we solve the thermal bootstrap in an all-orders asymptotic expansion in large spin, $J$. 

We can make explicit use of our tools in the 3d Ising model. As in MFT and the $O(N)$ model, low-twist operators can be arranged into double-twist families with relatively small anomalous dimensions. By combining our analytic tools with previous results from the four-point function bootstrap, we will estimate the thermal one-point functions of the operators in the lowest-twist family, $[\s\s]_0$, as a function of $b_\e$ and $b_T$.

\subsection{Leading double-twist thermal coefficients}
\label{sec:leading-double-twist-pert-theory}

Let's study the two-point function of identical scalars 
\begin{align}
g(z,\bar z) = \langle \phi(z,\bar z) \phi(0,0) \rangle_\beta\,, \label{eq:2pt function}
\end{align} in some CFT at finite temperature. We want to understand the ``contributions'' to the thermal coefficients of operators in the $[\phi\phi]_{0}$ family, $a_{[\phi\phi]_{0,J}}$, from other thermal data of the theory. Our starting point is the $t$-channel OPE ($z\sim \bar z \sim 1$),
\begin{align}
g(z,\bar z) &= \sum_{\mathcal{O} \in \phi \times \phi} a_{\mathcal{O}} ((1-z)(1-\bar z))^{\frac{\Delta_{\mathcal{O}}}{2} -\Delta_\phi} C_{\ell_{\mathcal{O}}}^{(\nu)} \left(\frac{1}{2} \left(\sqrt{\frac{1-z}{1-\bar z}}+ \sqrt{\frac{1-\bar z}{1-z}} \right) \right) \label{eq:2pt function t-channel} \,,
\end{align} which we will systematically invert to the $s$-channel data $a(\Delta,J)$. Expanding the Gegenbauer polynomials yields a power series in $1-z$ and $1-\overline{z}$:
\begin{align}
g(z,\bar z) &= \sum_{\mathcal{O} \in \phi \times \phi} a_{\mathcal{O}} \sum_{k=0}^{\ell_{\mathcal{O}}} \frac{\Gamma(\ell_{\mathcal{O}} -k +\nu) \Gamma(k+\nu)}{\Gamma(\ell_{\mathcal{O}}-k+1)\Gamma(k+1)}\frac{1}{ \Gamma(\nu)^2} (1-z)^{h_{\mathcal{O}}-\Delta_\phi +k} (1-\bar z)^{\bar h_{\mathcal{O}} -\Delta_\phi -k}. \label{eq:2pt function t-channel series exp}
\end{align} 
For future convenience, let's define the coefficients
\begin{align}
p_k(\ell_\cO) \equiv \frac{\Gamma(\ell_{\mathcal{O}} -k +\nu) \Gamma(k+\nu)}{\Gamma(\ell_{\mathcal{O}}-k+1)\Gamma(k+1)}\frac{1}{ \Gamma(\nu)^2}\,. \label{eq:def t-channel series coefficients}
\end{align}
Massaging the inversion formula \eqref{eq:inversionformulahigherd}, we rewrite it as a series in $z$ and $\bar z$,
\begin{align}
a(\Delta,J) = (1+(-1)^J) K_J \int_0^1 \frac{d\bar z}{\bar z} \int_1^{1/\bar z} \frac{dz}{z} \sum_{m=0}^\infty q_m(J) z^{\Delta_\phi-\bar h-m}\bar z^{\Delta_\phi-h+m} {\rm Disc}[g(z,\bar z)]\,, \label{eq:inversion series expansion}
\end{align} with coefficients 
\begin{align}
q_m(J) &\equiv  
(-1)^m \frac{(J+2 m)}{J} \frac{ (J)_{m} (-m+\nu +1)_m}{m! (J+\nu +1)_m}\,. \label{eq:def inversion series coefficients}
\end{align}  Let's suppose we are considering $J$ large enough so that the contributions of the arcs in (\ref{eq:inversionformulahigherd}) vanish.

Before inverting the $t$-channel OPE into $a(\Delta,J)$, let's analyze a few key features of the inversion formula:
\begin{itemize}
\item First, as discussed previously, recall that poles of $a(\Delta,J)$, associated with physical operators, come from the region near $\bar z = 0$. A term like $\bar z^a$ in the expansion around $\bar z=0$ inverts to terms of the form
\begin{align}
\int_0^1 \frac{d\bar z}{\bar z} \bar z^{\Delta_\phi-h+m} \bar z^a = \frac{1}{\Delta_\phi+a+m-h}\,, \label{eq:pole-from-single-zbar-term}
\end{align} and gives poles at $h= \Delta_\phi+a+m$. Such poles represent infinite families of operators with unbounded spin $J$ and scaling dimensions $\Delta = 2\Delta_\phi + 2a + 2m +J$. Of course, in interacting CFTs, operators should have anomalous dimensions. We discuss the effects that shift the locations of these na\"ive poles to their correct values in section~\ref{sec:pole-locations}.

\item  Next, let's imagine a term of the form $\bar z^a (1-z)^c$ in $g(z,\bar z)$ expanded around the double-lightcone limit ($\bar z=0$ and $z=1$), and invert it. The $z$ integral determines the residue of the poles in \eqref{eq:pole-from-single-zbar-term} as a function of $\bar h$ (recall that $\bar h = h+J$). Typical $z$ integrals are of the form
\begin{align}
\int_1^\infty \frac{dz}{z} z^{\Delta_\phi -\bar h -m} \Disc[(1-z)^c],
\end{align}
where we've set the upper limit on the $z$ integral equal to $\oo$ because $\bar z$ is infinitesimal.
The discontinuity is $\Disc[(1-z)^c] = 2\sin(-\pi c) (z-1)^c$, so the integral gives
\begin{align}
2\sin(-\pi c) \frac{\Gamma(1+c) \Gamma(\bar h +m -\Delta_\phi -c)}{\Gamma(\bar h + m- \Delta_\phi+1)}\,.
\end{align} Note that this is naturally a term in a large-$\bar h$ expansion, since
\begin{align}
\frac{\Gamma(\bar h +m -\Delta_\phi -c)}{\Gamma(\bar h + m- \Delta_\phi+1)} = \frac{1}{\bar h^{c+1}} +\cO\left(\frac{1}{\bar h^{c+2}}\right)\,.
\end{align} Thus, we see that terms in the double-lightcone expansion of $g(z,\bar z)$ correspond to power law corrections in $1/\bar h$, or equivalently in $1/J$, to the thermal coefficients of families of operators (in the $s$-channel). 

\item As highlighted in section~\ref{sec:commentsoninversion}, the $t$-channel OPE in \eqref{eq:2pt function t-channel series exp} is valid for the range $0\le z, \bar z \le 2$. Thus, we are only justified in integrating the $t$-channel OPE between $1\le z \le 2$ for the $z$ integral. In general, we don't have an expression for $g(z,\bar z)$ that is valid in the region $0\le\bar z \le 1$ and $z>2$. Luckily, for the integrands of interest, the $z$ integral in the range $z>2$ is exponentially suppressed in large $\bar h$, schematically as
\begin{align}
\int_2^\infty \frac{dz}{z} z^{\Delta_\phi-\bar h -m} f(z) \sim 2^{-\bar h}\,.
\end{align} Therefore, we can work with the $t$-channel OPE, integrate it in the region of its validity (from $z=1$ to $z_{\text{max}} = 2$), and obtain an all-orders expansion in $1/\bar h$, with undetermined exponentially-suppressed corrections.
\end{itemize}

Now that we have oriented ourselves, let's calculate the contributions to $a(\Delta,J)$ from a single primary operator $\mathcal{O} \in \phi\times \phi$ in the $t$-channel OPE in its full glory. We take the terms in the $t$-channel OPE corresponding to $\cO$, and invert them to the $s$-channel via the inversion formula. We use $a^{(\cO)}(\Delta, J)$ to denote the contribution to $a(\Delta, J)$ from the inversion of the contribution of $\cO$ in the $t$-channel. Then we find
\begin{align}
a^{(\cal O)}(\Delta,J) &\approx (1+(-1)^J) K_J \int_0^1 \frac{d\bar z}{\bar z} \int_1^{z_{\text{max}}} \frac{dz}{z} \sum_{m=0}^\infty q_m(J) z^{\Delta_\phi-\bar h-m}\bar z^{\Delta_\phi-h+m} \nonumber
\\ & \qquad \times {\rm Disc}\left[a_{\cal O} \sum_{k=0}^{\ell_{\cal O}} p_k(\ell_\cO) (1-z)^{h_{\mathcal{O}}-\Delta_\phi +k} (1-\bar z)^{\bar h_{\mathcal{O}} -\Delta_\phi -k}
\right]
\nn\\ &=  a_{\cal O} (1+(-1)^J) K_J   \sum_{m=0}^\infty \sum_{k=0}^{\ell_{\cal O}} q_m(J) p_k(\ell_\cO) \frac{\Gamma(1+\bar h_\cO - \Delta_\phi-k) \Gamma(\Delta_\phi+m-h) }{\Gamma(\bar h_\cO -h +1-k+m)} \nonumber
\\ & \qquad \times   2\pi S_{h_\cO-\Delta_\phi+k,\Delta_\phi-m}(\bar h)\,. \label{eq:single t-chan op inversion}
\end{align}
In general, we will think of $a(\Delta, J)$ as a sum of such $a^{(\cO)}(\Delta, J)$, up to some finite spin $\ell_\mathrm{max}$, plus contributions from sums over infinite families of operators with unbounded spin, so
\be 
a(\Delta, J)  \supset \sum_{\cO \in \phi \times \phi, \,\ell < \ell_{\text{max}}} a^{(\cO)}(\Delta, J)\,.
\ee
The reasoning behind separating out the sums to infinite spin will become apparent in section~\ref{sec:self-corrections}. We have defined the function 
\begin{align}
S_{c,\Delta}(\bar h) &= \frac{\sin(-\pi c)}{\pi} \int_1^{z_\text{max}} \frac{dz}{z} z^{\Delta-\bar h} (z-1)^c 
\nn\\ &= \frac{1}{\Gamma(-c)} \frac{\Gamma(\bar h-\Delta-c)}{\Gamma(\bar h-\Delta+1)} - \frac{1}{\Gamma(-c)\Gamma(1+c)} B_{1/z_\text{max}}(\bar h-\Delta-c,1+c)\,. \label{eq:def S}
\end{align} Here $B_{1/z_\text{max}}(\bar h-\Delta-c,1+c)$ is the incomplete beta function, which decays as $z_\text{max}^{-\bar h}$ at large $\bar h$. 

The factors $\Gamma(\Delta_\phi+m-h)$ in the numerator in \eqref{eq:single t-chan op inversion} give poles at $h = \Delta_\phi+n$ for $n\in \mathbb{Z}_{\ge 0}$. Na\"ively, these are the poles corresponding to the $[\phi\phi]_{n}$ families, at the na\"ive dimensions $\Delta = 2\Delta_\phi+2n+J$, without anomalous dimensions. However,
the correct $a(\Delta,J)$ has poles in $\Delta$ at the exact dimensions, including anomalous dimensions, so our na\"ive poles are shifted to their correct values,
\begin{align}
\frac{1}{\Delta_\phi +n-h} \rightarrow \frac{1}{\Delta_\phi + n + \delta_n(\bar h) -h}.
\end{align} This has a subtle but important effect on the thermal coefficients. When one takes the residue, there is an extra factor $d\bar h/dJ$ that depends on the derivative of $\delta_n(\bar h)$, since
\begin{align}
\Res\limits_{\Delta = 2\Delta_\phi +2n+J +2\delta_n(\bar h)}\frac{1}{\Delta_\phi + n + \delta_n(\bar h) -h} = -2 \frac{1}{1-\delta'_n(\bar h)} = -2 \frac{d\bar h}{dJ},
\end{align} as opposed to
\begin{align}
\Res\limits_{\Delta = 2\Delta_\phi +2n+J }\frac{1}{\Delta_\phi + n -h} = -2.
\end{align}
Note that $d\bar h/dJ = 1$ when anomalous dimensions vanish. In section~\ref{sec:pole-locations}, we will provide a consistency check that the poles are indeed shifted to their correct locations. In our discussion of sums over families below, we include $\frac{d\bar h}{dJ}$ for two reasons: firstly, it greatly simplifies the analysis of the asymptotics of such sums; secondly, we have in mind a situation where the anomalous dimensions $\de_n(\bar h)$ are known through other means (e.g.\ the vacuum four-point function bootstrap), and we would like to use that information in the thermal bootstrap.

Finally, evaluating the residues of $a^{(\cO)}(\Delta,J)$ at the $[\f\f]_n$ poles, we get the contribution of $\mathcal{O}$ to the thermal coefficients of the $[\phi\phi]_n$ family,
\begin{align}
& a_{[\phi\phi]_{n}}^{(\cO)} ( J) = -\Res\limits_{\Delta = 2\Delta_\phi +2n+J} a^{(\cal O)}(\Delta,J)
\nn\\ &\quad = a_\cO (1+(-1)^J) 4\pi K_J \frac{d\bar h}{dJ} \sum_{r=0}^{n} \sum_{k=0}^{\ell_{\cal O}} q_r(J) p_k(\ell_\cO) (-1)^{n-r} \binom{\bar h_\cO-\Delta_\phi -k}{n-r} 
S_{h_\cO-\Delta_\phi+k,\Delta_\phi-r}(\bar h). \label{eq:residue of single t-chan op inversion}
\end{align} Note that $\bar h$ is implicitly defined as a function of $J$ by $\bar h = \Delta_\f + n + \delta(\bar h) +J$.
As we have emphasized above, properties of using the OPE with the inversion formula, 
these contributions are naturally organized as power-law corrections in large $\bar h$. The function $S_{c,\Delta}(\bar h)$ behaves as
\begin{align}
S_{c, \Delta}(\bar h) = \frac{1}{\Gamma(-c)} \frac{1}{\bar h^{c+1}} + \cO\left(\frac{1}{\bar h^{c+2}}\right) \label{eq:S leading asymptotics}
\end{align} at large $\bar h$, and $c_m(J)$ behaves as a constant to leading order in $1/\bar h$.
So a given term in \eqref{eq:residue of single t-chan op inversion} starts at order $\bar h^{-(h_\cO-\Delta_\f +k+1)}$.
Thus, we see that the contribution of an operator $\cO$ in the $t$-channel behaves at a rate controlled by its twist. Concretely, the leading contributions in $1/\bar h$ are given by the $k=0$ term of the sum in \eqref{eq:residue of single t-chan op inversion},
\begin{align}
a_{[\phi\phi]_{n}}^{(\cO)} ( J) & = a_\cO (1+(-1)^J) \frac{ K_J}{K_{\ell_\cO}} \frac{d\bar h}{dJ} \sum_{r=0}^{n} q_r(J) (-1)^{n-r} \binom{\bar h_\cO-\Delta_\phi}{n-r} S_{h_\cO-\Delta_\phi,\Delta_\phi-r}(\bar h), \label{eq:leading double twist 1pt function}
\end{align} 
For the leading double-twist family $[\phi\phi]_0$, this further simplifies to
\begin{align}
a_{[\phi\phi]_{0}}^{(\cO)} ( J) & = a_\cO (1+(-1)^J) \frac{ K_J}{K_{\ell_\cO}} \frac{d\bar h}{dJ}S_{h_\cO-\Delta_\phi,\Delta_\phi}(\bar h) \label{eq:leading [phiphi]_0 1pt function}
\nn\\ & \sim  a_\cO (1+(-1)^J) \frac{ K_J}{K_{\ell_\cO}} \frac{1}{\bar h^{h_\cO-\Delta_\phi+1}} + \dots.
\end{align} In writing the last line, we have assumed that $\delta_{[\f\f]_0}(\bar h)$ grows slower than $\bar h$ as $\bar h \rar \oo$.

To help understand the examples that follow, let us introduce a diagrammatic language that  helps keep track of terms in large-spin perturbation theory of thermal data. Our diagrams can be thought of as analogs of the four-point function large-spin diagrams for the thermal case. We do not have a rigorous definition of these diagrams or a complete set of rules for using them. Nevertheless, they will help organize the discussion.\footnote{Large-spin diagrams for four-point functions can be understood as physical processes in the massive 2d effective theory defined in \cite{Alday:2007mf}. It would be nice to develop a similar understanding of the diagrams here. For now, our diagrams are simply mnemonic devices.}

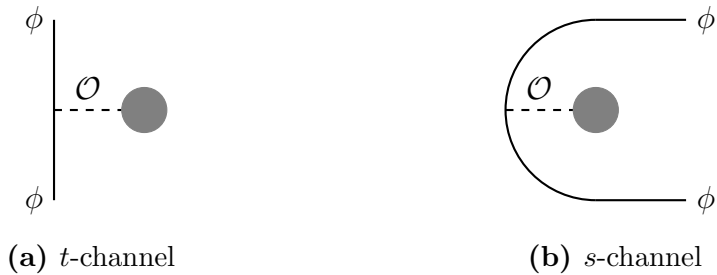
\begin{figure}[htb]
\centering
\begin{subfigure}[t]{.4\textwidth}
\centering
\begin{tikzpicture}[xscale=0.6,yscale=0.6]
\draw[thick] (0,0) -- (0,4);
\draw[thick, dashed] (0,2) -- (1.5,2);
\node[left] at (0,0) {$\phi$};
\node[left] at (0,4) {$\phi$};
\node[above] at (.75,2) {$\cO$};
\filldraw[gray] (2,2) circle (.5);
\end{tikzpicture}
\caption{$t$-channel}
\label{fig:t-channel phi phi O}
\end{subfigure}
~
\begin{subfigure}[t]{.4\textwidth}
\centering
\begin{tikzpicture}[xscale=0.6,yscale=0.6]
\draw[thick] (2,0) arc (270:90:2);
\draw[thick] (2,0) -- (4,0);
\draw[thick] (2,4) -- (4,4);
\draw[thick, dashed] (0,2) -- (1.5,2);
\node[right] at (4,0) {$\phi$};
\node[right] at (4,4) {$\phi$};
\node[above] at (.75,2) {$\cO$};
\filldraw[gray] (2,2) circle (.5);
\end{tikzpicture}
\caption{$s$-channel}
\label{fig:s-channel phi phi O}
\end{subfigure}
\caption{An illustration of the relation between $s$- and $t$-channels in the $\langle \f\f\rangle_\beta$ correlator. The two channels are related by moving the external operators around the thermal circle (gray). A single term in the $t$-channel OPE $\cO\in \phi\times \phi$ inverts to the sum over the $[\f\f]_n$ families in the $s$-channel. Alternatively, the sum over the $[\f\f]_n$ families in the $s$-channel reproduces the $\cO$ term in the $t$-channel.}
\label{fig:phi phi O crossing}
\end{figure}

For example, we can understand the fact that $\cO\in \f\times \f$ in the $t$-channel OPE inverts to give contributions to $a_{[\f\f]_n}$ proportional to $a_\cO$ via the diagrams in figure~\ref{fig:phi phi O crossing}. Let's start with the $t$-channel diagram in figure~\ref{fig:t-channel phi phi O}. We should read this diagram from left to right as two $\f$ operators approach each other on one side of the thermal circle (corresponding to the $t$-channel), and fuse into $\cO$, which in turn gets an expectation value. The diagram illustrates that this process should be proportional to the three-point coefficient $f_{\f\f\cO}$ and to the one-point function $b_\cO$, which is indeed the case by the definition of $a_\cO$.

Now, let's relate this process to the $s$-channel. The diagrammatic rule relating the $s$- and $t$-channels is given by taking the two external operators to the other side of the thermal circle around opposite sides. This converts the process in figure~\ref{fig:t-channel phi phi O} to the process in figure~\ref{fig:s-channel phi phi O}. Reading the resulting process from right to left, we interpret it as two external $\f$'s fusing into operators in the $[\f\f]_n$ families, which get expectation values proportional to $a_\cO$.

Summarizing, we reiterate that the thermal coefficients of families of operators are organized into large-spin expansions, with the operators $\cO$ in the OPE contributing perturbatively at order determined by their twist. Since the unit operator has the lowest twist in any unitary theory, it gives the leading contribution for large-spin members of double-twist families. A second important contribution comes from the stress tensor $\cO=T_{\mu\nu}$, which gives a universal contribution proportional to the free energy density. These two universal contributions were written in (\ref{eq:universalcontribution}) in the introduction. Furthermore, we also see that the contribution of a given $\cO$ is linear in its one-point function. This greatly simplifies perturbatively solving for the one-point functions, especially when one considers the corrections from sums of families, which we shall explore below.

\subsection{Contributions of double-twist families: resumming, asymptotics, and other families}
\label{sec:self-corrections}

We have seen that inverting any single operator $\cO$ in the $\phi\times \phi$ OPE gives contributions to the one-point functions of the $[\phi\phi]_n$ families, and only these families.  But many other operators appear in the $\phi\times \phi$ OPE, and the inversion formula must pick up their existence. How can we extract their one-point functions from $\langle \phi \phi \rangle_\beta$? 

As seen in section~\ref{sec:leading-double-twist-pert-theory}, any finite number of terms in the $t$-channel OPE have ``regular'' $\bar z$ expansions around $\bar z \sim 0$, with integer $\bar z$ powers obtained from the Taylor series of some collection of terms of the form $(1-\bar z)^c$.
Inverting such terms  will only give poles at $h = \Delta_\phi + n$, which thus correspond to the  $[\phi\phi]_n$ families. So, to obtain the necessary poles at other locations, we need to find terms in the $t$-channel expansion that behave as $\bar z^c$ with $c \notin \mathbb{Z}_{\ge 0}$ near $\bar z \sim 0$. Such terms invert to poles at $h= \Delta_\phi+c+n$, and correspond to different families determined by $c$. We will call such terms ``singular'', in analogy with the Casimir-singular definition of \cite{Alday:2007mf, Simmons-Duffin:2016wlq}. Singular terms are characterized by having discontinuity around $\bar z \sim 0$, which means they would be picked up by the Lorentzian inversion formula that takes the two-point function and inverts it to the $t$-channel. The only way we can obtain singular terms is from tails of the infinite sums of families in the $t$-channel OPE, which, when summed up, will have different $\bar z$ behavior compared to any finite number of terms. A related motivation for understanding this problem is to compute the contributions of the $[\phi\phi]_n$ families to their own one-point functions. 

We will now explain how to systematically compute the contributions of double-twist families to the one-point functions of operators that appear in the $\phi\times \phi$ OPE. To begin, let's focus on the contribution from summing the tail of the particular double-twist family $[\phi\phi]_0$. 
Once again, we start with the $t$-channel OPE expansion, which we now try to understand in the double-lightcone limit $(z,\bar z) \sim (1,0)$. 

Let $\{\mathcal{O}\}$ be a set of operators in the $\phi\times \phi$ OPE with low twist.
Inverting their terms in the $t$-channel via section~\ref{sec:leading-double-twist-pert-theory}, we obtain from \eqref{eq:leading [phiphi]_0 1pt function} the leading $1/{\bar h}$ behavior of the 1-point functions of the $[\phi\phi]_0$ family,
\begin{align}
a_{[\phi\phi]_0} (J)  \sim \sum_{\mathcal{O}} \frac{a_{\mathcal{O}}}{K_{\ell_{\mathcal{O}}}} (1+ (-1)^J) K_J \frac{d\bar h}{dJ} S_{h_\mathcal{O} - \Delta_\phi,\Delta_\phi}(\bar h). \label{eq:leading-double-twist-one-pt-function}
\end{align} Now, let's insert this expression for $a_{[\phi\phi]_0}$ back into the $t$-channel OPE, and consider the $t$-channel sum over the $[\phi\phi]_0$ family
\begin{equation}
\begin{split}
\sum_{[\phi\phi]_{0,J} \in [\phi\phi]_0} &\frac{a_{[\phi\phi]_0}( J)}{4\pi K_J}  (1-z)^{h-\Delta_\phi} (1-\bar z)^{\bar h-\Delta_\phi} 
\\ &\sim \sum_{[\phi\phi]_0} \sum_{\mathcal{O}} \frac{a_{\mathcal{O}}}{4\pi K_{\ell_{\cal O}}} (1+(-1)^J) \frac{d\bar h}{dJ}  S_{h_{\cal O}-\Delta_\phi,\Delta_\phi} (\bar h) (1-z)^{h-\Delta_\phi} (1-\bar z)^{\bar h-\Delta_\phi} + \dots. \label{eq:sum-over-0th-family}
\end{split}
\end{equation}
We wish to apply the inversion formula to this sum. If we na\"ively invert each term in the sum, and then sum over the family, we notice that the sum over the contributions of each individual member of the family diverges. In other words, the inversion formula and the infinite sum over the family do not commute. So, we have to sum over the family first before inverting to the $s$-channel. This is in line with our anticipation that the poles for other families must arise from the tails of the sums over infinite families, such as $[\phi\phi]_0$. If the $t$-channel sum and the inversion integral commuted, we would only ever get poles for the $[\phi\phi]_n$ families from the inversion formula.

\subsubsection{Analytic and numerical formulae for sums over families}

\label{sec:numerical-method-for-performing-asymp-sum}

Let's analyze the $t$-channel sum over a double-twist family more carefully, in the spirit of \cite{Simmons-Duffin:2016wlq}. Consider the sum over a particular term in the $S_{c,\Delta}(\bar h)$ expansion of one-point functions of an arbitrary double-twist family,\footnote{For $h_e$ and $h_f$, the indices $e$ and $f$ stand for ``external'' and ``family''. For the two-point function of two identical scalars $\f$, $h_e = \Delta_\f$.}
\begin{align}	
\sum_{\bar h = h_f + \ell + \delta(\bar h)} \frac{d \bar h}{d\ell} S_{c,\Delta}(\bar h) (1-z)^{h_f+\delta(\bar h)-h_e} (1-\bar z)^{\bar h-h_e}.
\end{align}
Here, $\bar h = h_f+\ell+\delta(\bar h)$ runs over the family with anomalous dimensions $\delta(\bar h)$, and $h(\bar h) = h_f+\delta(\bar h)$ are the half-twists of the operators in the family. We have switched to denoting spin by $\ell$ for sums over families, to avoid conflict with applying the inversion formula to these sums later on. We have left out the $(-1)^\ell$ factor for now; we will return to it later. In general, this is a difficult sum to evaluate, and we don't yet know of an exact treatment.
However, since anomalous dimensions $\delta(\bar h)$ for the families of interest are small for large $\bar h$, we can work order by order in $\delta(\bar h)$. Concretely, we can split the sum over $\bar h$ as
\begin{equation}
\sum_{\bar h} = \sum_{\bar h < {\bar h_0} + \delta({\bar h_0})} + \sum_{\substack{\bar h = {\bar h_0} + \ell + \delta(\bar h) \\ \ell=0,1,\dots}},
\end{equation} for some large enough ${\bar h_0} = h_f + \ell_0$ such that the anomalous dimensions $\delta(\bar h)$ are sufficiently small for $\bar h>\bar h_0$, and expand the infinite sum piece in small $\delta(\bar h) \log(1-z)$,
\begin{align}
\sum_{\bar h = \bar h_0 + \ell + \delta(\bar h)} \frac{d \bar h}{d\ell} S_{c,\Delta} (\bar h) (1-\bar z)^{\bar h-h_e} \sum_{m=0}^\infty \frac{\delta(\bar h)^m}{m!} \log^m (1-z) (1-z)^{h_f-h_e}. \label{eq:double-twist-sum-delta-expansion}
\end{align} This expansion is valid in a regime $  e^{-1/|\delta(\bar h)|} < |1-z| < e^{1/|\delta(\bar h)|}$, which is near the double-lightcone limit for small anomalous dimensions.
Now, the dependence on $z$ --- which controls the discontinuity in the inversion formula --- can be factored out of the $\bar h$ sum. Recall that the anomalous dimensions $\delta(\bar h)$ are themselves analytic functions of $\bar h$, and they can be computed perturbatively in a large-$\bar h$ expansion \cite{Simmons-Duffin:2016wlq,Fitzpatrick:2012yx,Komargodski:2012ek,Alday:2015eya,Alday:2015ota,Alday:2015ewa,Alday:2016njk}. For example, the anomalous dimensions of a family $[\f\f]_0$ can be expanded at large $\bar h$, and include terms such as
\begin{align}
\delta(\bar h)  &\supset \sum_{\cO \in \phi\times \phi} \frac{\delta^{(\cO)}}{\bar h^{2h_\cO}} \, . \label{eq:large-spin exp of anomalous dimensions}
\end{align}

How can we compute the sums over $\bar h$?
With a more general treatment in mind, let's consider the sum 
\begin{align}
\sum_{\bar h = \bar h_0 + \ell + \delta(\bar h)} \frac{d \bar h}{d\ell} f(\bar h) (1-\bar z)^{\bar h - h_e},
\end{align}
with a summand $f(\bar h)$ which grows at most as a power law at large $\bar h$. The summands of interest for us are of the form\footnote{We will also be interested in sums including derivatives $\partial_{c}^m S_{c,\Delta}(\bar h)$, as we will discuss later.}
\be 
f(\bar h) = \delta^m(\bar h) S_{c,\Delta}(\bar h)\,. \label{eq:summand of interest}
\ee If we tried to expand in small $\bar z$ and compute the $\bar h$ sum order by order in $\bar z$, we see that for large enough powers of $\bar z$ we get divergent sums in $\bar h$. In fact, this is to be expected. By the existence of the inversion formula, we know that such sums must have asymptotic pieces that reproduce the singular terms $\bar z^a$, which could not possibly be obtained from expanding in $\bar z$ first (which produces only integer powers of $\bar z$).
Thus, we expect the result of such a sum to be of the form
\begin{align}
\sum_{\bar h} \frac{d \bar h}{d\ell} f(\bar h) (1-\bar z)^{\bar h - h_e} = \sum_{a\in A} c_a \bar z^a + \sum_{k=0}^{\infty} \alpha_k \bar z^k,
\end{align} with $A\subset \mathbb{R}\backslash \mathbb{Z}_{\ge0}$ some set of numbers which are not non-negative integers that we have to determine. (We can also have $\bar z^a \log^m \bar z$ terms that we will write as $\frac{\ptl^m}{\ptl a^m} \bar z^a$.) Our task is reduced to computing the coefficients $c_a$ and $\alpha_k$.
To do this, we will separate the sum into asymptotic parts, which reproduce the $\bar z^a$ terms, and leftover regular parts that are convergent sums in $\bar h$, with which we can compute the $\alpha_k$ coefficients.

First, we determine the large-$\bar h$ asymptotics of $f(\bar h)$ in terms of the known functions $S_{a,\Delta}(\bar h)$,
\begin{align}
f(\bar h) \sim \sum_{a\in A} c_{a,\Delta}[f] S_{a,\Delta}(\bar h). \label{eq:S asymptotics of summand}
\end{align} Note that the set $A$ is determined by the asymptotics of $f(\bar h)$, but the expansion can be written for any choice of $\Delta$. For summands of interest like in \eqref{eq:summand of interest}, the asymptotic expansion is determined algorithmically from the large-$\bar h$ expansions of $S_{c,\Delta}(\bar h)$ and of $\delta(\bar h)$ \emph{\`a la} \eqref{eq:large-spin exp of anomalous dimensions}. Once the asymptotics in \eqref{eq:S asymptotics of summand} are obtained, we can compute the singular terms from the tails of the sums over the asymptotics, by using the crucial identity of the integer-spaced sum
\begin{align}
\sum_{\substack{\bar h={\bar h_0} + \ell \\ \ell=0,1,\dots}} S_{c,\Delta}(\bar h) (1-\bar z)^{\bar h} &= (1-\bar z)^{{\bar h_0}} S_{c,\Delta}({\bar h_0}) \PFQ{2}{1}{1,{\bar h_0}-\Delta-c}{{\bar h_0}-\Delta+1}{1-\bar z}
\nn\\ &= \bar z^c (1-\bar z)^\Delta - S_{c-1,\Delta+1} ({\bar h_0}) (1-\bar z)^{{\bar h_0}} \PFQ{2}{1}{1,{\bar h_0}-\Delta-c}{-c+1}{\bar z}. \label{eq:integer-spaced-sum}
\end{align}
Note that the first term is singular and the second term, proportional to ${}_2F_1(\cdots,\bar z)$ is regular.

We claim that the noninteger-spaced sum over $S_{c,\Delta}(\bar h)$ (with $\bar h$ determined by the anomalous dimensions $\delta(\bar h)$) has the same singular piece as the integer-spaced sum,
\begin{align}
\sum_{\bar h = {\bar h_0} +\ell + \delta(\bar h)} \frac{d \bar h}{d\ell} S_{c,\Delta}(\bar h) (1-\bar z)^{\bar h} = \bar z^c (1-\bar z)^\Delta + \text{regular}.
\end{align} This can be verified by an argument due to \cite{Simmons-Duffin:2016wlq} applied to the present case. We convert the sum to a contour integral via Cauchy's residue theorem,\footnote{This is also known as the Sommerfeld-Watson transform.}
\begin{align}
\sum_{\ell=0}^\infty &\frac{d \bar h}{d\ell} S_{c,\Delta}(\bar h) (1-\bar z)^{\bar h}
= -\oint_\gamma \frac{d\bar h}{2\pi i} \pi \cot(\pi(\bar h-{\bar h_0}-\delta(\bar h)))  
S_{c,\Delta}(\bar h) (1-\bar z)^{\bar h},
\end{align} where $\gamma$ is a contour along the real axis that picks up the desired poles. We can deform the contour to one that runs parallel to the imaginary axis, plus arcs at infinity. The singular terms come from the asymptotics of this integral. As long as $\delta(\bar h)$ grows slower than $\bar h$ as $\bar h\rightarrow \pm i \oo$, the asymptotic region of the integral approaches a $\delta(\bar h)$-independent constant exponentially quickly, since 
\begin{align}
\pi \cot(\pi(\bar h - {\bar h_0} -\delta(\bar h))) \rightarrow \mp 1 + O(e^{\mp 2s}) \quad \text{as } \bar h \rightarrow \pm i s.
\end{align} Therefore, the singular pieces are independent of $\delta(\bar h)$.
To be even more concrete, we can subtract the contour integral versions of the noninteger- and integer-spaced sums, and notice that the asymptotics vanish, or that if we expand the difference in small $\bar z$, the integrals in $\bar h$ are convergent term by term so the singular terms must have canceled. 
Thus, we see that 
\begin{align}
\sum_{\bar h} \frac{d \bar h}{d\ell} f(\bar h) (1-\bar z)^{\bar h-h_e} = \sum_{a\in A} c_{a,\Delta}[f]\, \bar z^a (1-\bar z)^{\Delta-h_e} + \text{regular}.
\end{align} Note that we can always choose $\Delta = h_e$ in the asymptotic expansion in \eqref{eq:S asymptotics of summand} to simplify the organization of the singular terms,
\begin{align}
\sum_{\bar h} \frac{d \bar h}{d\ell} f(\bar h) (1-\bar z)^{\bar h-h_e} = \sum_{a\in A} c_{a,h_e}[f]\, \bar z^a + \text{regular}.
\end{align} 

We are left with computing the coefficients $\alpha_k$ of the regular terms. Since we have extracted the asymptotics, we can write convergent expressions for $\alpha_k$ by subtracting the asymptotics and expanding in small $\bar z$. This will be best done by once again converting the sum over $\bar h$ into a contour integral in the complex $\bar h$ plane, via Cauchy's theorem. We want to write a contour integral of the form
\begin{align}
\sum_{\bar h= {\bar h_0} +\ell + \delta(\bar h)} & \frac{d \bar h}{d\ell} f(\bar h)  (1-\bar z)^{\bar h - h_e} = \sum_{\bar h= {\bar h_0} +\ell + \delta(\bar h)} \frac{d \bar h}{d\ell} f(\bar h) \sum_{k=0}^\infty \genfrac{(}{)}{0pt}{}{\bar h - h_e}{k} (-1)^k \bar z^k
\nn\\ &= -\oint_{h_c -i\infty}^{h_c + i\infty} \frac{d\bar h}{2\pi i} \pi \cot(\pi(\bar h-{\bar h_0}-\delta(\bar h))) f(\bar h) \sum_{k=0}^\infty \genfrac{(}{)}{0pt}{}{\bar h - h_e}{k} (-1)^k \bar z^k,
\end{align} where $h_c = {\bar h_0}+\delta({\bar h_0}) - \epsilon$, for some small $\epsilon>0$. This contour integral will equal the sum only if $f(\bar h)$ decays fast enough on the arcs at infinity so that we may drop them, and if $f(\bar h)$ does not have any simple poles for ${\rm Re}~ \bar h \ge h_c$. The second condition is easily remedied in case $f(\bar h)$ does have poles, by simply removing the residues coming from those poles. The first condition is related to the more important issue that for large enough $k$, the $\bar h$ growth is divergent, and 
the sum over $k$ and the contour integral do not commute. However, we can regulate the integral by subtracting the divergent asymptotics in the form of the integer-spaced sum until we get a convergent integral (for which the arcs vanish as well), and add back the known result of the integer-spaced sum. This gives the following formula
\begin{align}
\alpha_k [p,\delta,h_e]({\bar h_0}) &= -\oint_{h_c -i\infty}^{h_c + i\infty} \frac{d\bar h}{2\pi i}  \genfrac{(}{)}{0pt}{}{\bar h - h_e}{k} (-1)^k \nonumber
\\ & \qquad \times \left(\pi \cot(\pi(\bar h-{\bar h_0}-\delta(\bar h))) ~ f(\bar h) - \pi \cot(\pi(\bar h -{\bar h_0})) \sum_{\substack{a\in A \\ a<K}} c_{a,\Delta} S_{a,\Delta}(\bar h) \right) \nonumber
\\ & \quad + \sum_{\substack{a\in A \\ a<K}} c_{a,\Delta} \left(r_k(a,\Delta,h_e,{\bar h_0}) +s_k(a,\Delta,h_e,{\bar h_0})\right). \label{eq:def:alpha_k}
\end{align}
Here, $K$ should be at least $k$, but larger $K$ gives a faster converging integral.
In the last line, we have added back terms with $r_k$, which is the coefficient of $\bar z^k$ for the integer spaced sum in \eqref{eq:integer-spaced-sum},
\begin{align}
r_k(a,\Delta,h_e,{\bar h_0}) &= \left.- S_{a-1,\Delta+1} ({\bar h_0}) (1-\bar z)^{{\bar h_0}-h_e} \PFQ{2}{1}{1,{\bar h_0}-\Delta-a}{-a+1}{\bar z} \right|_{\bar z^k}
\nn\\ &= - S_{a-1,\Delta+1} ({\bar h_0}) \sum_{m=0}^k (-1)^m \genfrac{(}{)}{0pt}{}{{\bar h_0} - h_e}{m} \frac{({\bar h_0}-\Delta-a)_{k-m}}{(-a+1)_{k-m}} \label{eq:def:regular-terms}
\end{align} and $s_k$, which is the contribution of spurious poles (coming from the asymptotics $S_{a,\Delta}(\bar h)$ we subtracted) that are picked up by the contour when $h_c -\Delta - a \le 0$,
\begin{align}
&s_k(a,\Delta,h_e,{\bar h_0}) = \sum_{n=0}^{\lfloor a+\Delta-h_c\rfloor} \operatorname*{Res}\limits_{\bar h= a+\Delta-n} \genfrac{(}{)}{0pt}{}{\bar h - h_e}{k} (-1)^k \pi \cot (\pi(\bar h -{\bar h_0}) )S_{a,\Delta}(\bar h)
\nn\\
&\qquad = \sum_{n=0}^{\lfloor a+\Delta-h_c\rfloor} \genfrac{(}{)}{0pt}{}{a+\Delta-n - h_e}{k} (-1)^k \pi \cot (\pi(a+\Delta-n -{\bar h_0})) \frac{(-1)^n}{n! \Gamma(-a) \Gamma(a-n+1)}.\label{eq:def:spurious-poles}
\end{align} The contour integral can be integrated numerically in \textit{Mathematica} to high precision.

Finally, to finish our discussion, let's consider the alternating sum,
\begin{align}
\sum_{\bar h} (-1)^\ell f(\bar h) (1-\bar z)^{\bar h-h_e}.
\end{align} This sum is convergent order by order in the $\bar z$ expansion, so one does not need to subtract off asymptotics. This sum is given by
\begin{align}
\sum_{\bar h} (-1)^\ell \frac{d \bar h}{d\ell} f(\bar h) (1-\bar z)^{\bar h-h_e} = \sum_{k} \alpha^{-}_k[f,\delta,h_e]({\bar h_0}) \bar z^k,
\end{align} where the coefficients $\alpha_k^-$ are given by the contour integral with the replacement $\cot \rightarrow \csc$,
\begin{align}
\alpha^{-}_k[f,\delta,h_e]({\bar h_0}) = -\oint_{h_c -i\infty}^{h_c + i\infty} \frac{d\bar h}{2\pi i}  \genfrac{(}{)}{0pt}{}{\bar h - h_e}{k} (-1)^k \pi \csc(\pi(\bar h-{\bar h_0}-\delta(\bar h))) ~ f(\bar h).\label{eq:def:alpha-minus}
\end{align}
Collecting our calculations, the full sum over operators with even spin is given by
\begin{align}
\sum_{\bar h = {\bar h_0} +\ell + \delta(\bar h)} (1+(-1)^\ell) \frac{d \bar h}{d\ell} f(\bar h) (1-\bar z)^{\bar h- h_e} = \sum_{a\in A} c_{a,\Delta}[f] \bar z^a(1-\bar z)^{\Delta - h_e} + \sum_{k=0}^{\infty} \alpha_k^{\rm even}[f,\delta,h_e]({\bar h_0}) \bar z^k,
\end{align} where 
\begin{align}
\alpha_k^{\rm even}[f,\delta,h_e]({\bar h_0}) =  \alpha_k [f,\delta,h_e]({\bar h_0}) + \alpha_k^{-}[f,\delta,h_e]({\bar h_0}). \label{eq:def:alpha-even}
\end{align}

\subsubsection{Corrections to one-point functions from double-twist families}
\label{sec:double-twist-contributions}

Armed with the technology to compute the sums over double-twist families, we return to understanding their contributions to one-point functions of operators. Let's recall the $t$-channel sum over $[\phi\phi]_0$ in \eqref{eq:sum-over-0th-family}, and expand it in $\delta \log (1-z)$ as in \eqref{eq:double-twist-sum-delta-expansion},
\begin{align}
\sum_{\cal O} \frac{a_{\cal O}}{4\pi K_{\ell_{\cal O}}} \sum_{\bar h} (1+(-1)^\ell) \frac{d \bar h}{d\ell} S_{h_{\cal O}-\Delta_\phi,\Delta_\phi} (\bar h) (1-\bar z)^{\bar h -\Delta_\phi} \sum_{m=0}^\infty \frac{\delta(\bar h)^m}{m!} \log^m (1-z) (1-z)^{h_f-\Delta_\phi}.
\end{align}
Here, the sum is over the operators $[\f\f]_{0,\ell}$ with $\bar h = {\bar h_0} + \ell + \delta_{[\f\f]_0}(\bar h)$, where ${\bar h_0}$ is the $\bar h$ of the lowest spin member of the family where we started the sum. For each $\cal O$, summing over $\bar h$ yields
\begin{equation}
\begin{split}
\sum_{m=0}^\infty &\left( \sum_{a} c_{a,\Delta_\f}\! \left[\frac{\delta^{m}}{m!} S_{h_{\cal O}-\Delta_\phi,\Delta_\phi} \right] \bar z^a  + \sum_{k=0}^\infty \alpha_k^{\rm even}\! \left[\frac{\delta^{m}}{m!} S_{h_{\cal O}-\Delta_\phi,\Delta_\phi},\delta, \Delta_\phi\right]\!({\bar h_0}) \bar z^k\right) \\ & \times \log^m(1-z) (1-z)^{h_f-\Delta_\phi}. \label{eq:regulated-double-twist-sum}
\end{split}
\end{equation} For $[\phi\phi]_0$, $h_f = \Delta_\phi$, but we've kept it general here to demonstrate the general structure. Now, let's invert this piece of the two-point function to the $s$-channel. The integer powers $\bar z^k$ invert to poles for $[\phi\phi]_n$, giving contributions to the one-point functions of these families, including $[\phi\phi]_0$ itself. The contributions are controlled by \begin{equation}
    {\rm Disc}[\log^m (1-z) (1-z)^{h_f-\Delta_\phi}] =\partial_{h_f}^m {\rm Disc}[(1-z)^{h_f-\Delta_\phi}],
\end{equation}
which inverts to a term 
\begin{align}
S^{(m)}_{h_f-\Delta_\phi}(\bar h) = \partial_{h_f}^m S_{h_f-\Delta_\phi} (\bar h)
\end{align}
in the large-$\bar h$ expansion of thermal coefficients. 
For example, including the self-corrections of $[\phi\phi]_0$ to leading order in large spin yields
\begin{equation}
\begin{split}
a_{[\phi\phi]_0}(J) &= (1+(-1)^J) K_J \frac{d\bar h}{dJ} \\ &\quad \times \sum_{\cal O} \frac{a_{\cal O}}{K_{\ell_{\cal O}}} \left( S_{h_{\cal O}-\Delta_\phi,\Delta_\phi}(\bar h) + \sum_{m=0}^{\infty} \alpha^{\rm even}_0\left[\frac{\delta^m}{m!} S_{h_{\cal O}-\Delta_\phi,\Delta_\phi},\delta,\Delta_\phi\right]\!({\bar h_0}) S^{(m)}_{0,\Delta_\phi}(\bar h)\right). \label{eq:first-self-correction}
\end{split}
\end{equation}
We should remember that if we started the sum at some high ${\bar h_0}$, we should individually add the contributions of the low-lying members of the family that were excluded from the sum.

Once we have computed the self-corrections as in \eqref{eq:first-self-correction}, there is nothing that stops us from iterating this procedure and computing the self-corrections from the new, once-self-corrected one-point functions. Instead of iterating indefinitely, we can solve for the fixed point of these self-corrections with a little cleverness. We have provided a method for this in appendix~\ref{sec:self-corr-fixed-pt}.

The singular terms $\bar z^a$ in \eqref{eq:regulated-double-twist-sum} give poles at $h=\Delta_\phi + a + n$. We expect that these poles correspond to other families of operators with the given na\"ive twist --- we will soon explore which families. Let's denote these families by $[\Delta+a]_n$ for now. We see that the $[\f\f]_0$ family contributes 
\begin{align}
a_{[\Delta+a]_n}^{([\f\f]_0)} (J)= (1+(-1)^J) K_J \frac{d\bar h}{dJ} \sum_{\cal O} \frac{a_{\cal O}}{K_{\ell_{\cal O}}} \sum_{m} c_{a,\Delta_\f}\left[\frac{\delta^{m}}{m!} S_{h_{\cal O}-\Delta_\phi,\Delta_\phi} \right] S^{(m)}_{0,\Delta_\phi}(\bar h)
\label{eq:other-families}
\end{align} to the one-point functions of the $[\Delta+a]_n$ families. The sum is over $m$ such that $S_{a,\Delta_\f}$ appears in the asymptotic expansion of $\frac{\delta^{m}}{m!} S_{h_{\cal O}-\Delta_\phi,\Delta_\phi}$. Of course, this is rather schematic, since for interacting CFTs, the spectrum of families of higher-twist operators is very complicated, with large anomalous dimensions and mixing among families. Regardless, these contributions are present asymptotically in large $J$.

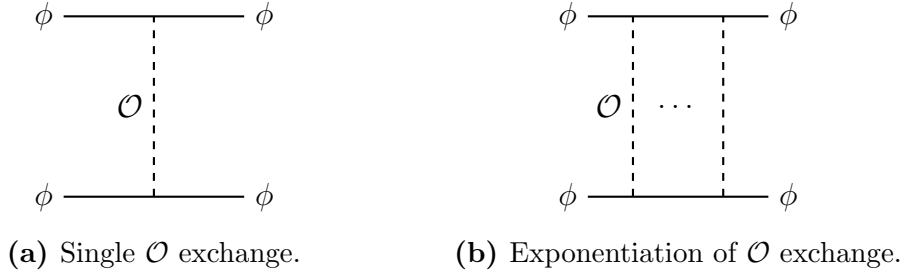
\begin{figure}[tbh]
\centering
\begin{subfigure}[t]{.4\textwidth}
\centering
\begin{tikzpicture}[xscale=0.6,yscale=0.6]
\draw[thick] (0,0) -- (4,0);
\draw[thick] (0,4) -- (4,4);
\draw[thick, dashed] (2,0) -- (2,4);
\node[left] at (0,0) {$\phi$};
\node[right] at (4,0) {$\phi$};
\node[left] at (0,4) {$\phi$};
\node[right] at (4,4) {$\phi$};
\node[left] at (2,2) {$\cO$};
\end{tikzpicture}
\caption{Single $\cO$ exchange.}
\label{fig:single O exchange to delta}
\end{subfigure}
~
\begin{subfigure}[t]{.4\textwidth}
\centering
\begin{tikzpicture}[xscale=0.6,yscale=0.6]
\draw[thick] (0,0) -- (4,0);
\draw[thick] (0,4) -- (4,4);
\draw[thick, dashed] (1,0) -- (1,4);
\draw[thick, dashed] (3,0) -- (3,4);
\node[left] at (0,0) {$\phi$};
\node[right] at (4,0) {$\phi$};
\node[left] at (0,4) {$\phi$};
\node[right] at (4,4) {$\phi$};
\node[left] at (1,2) {$\cO$};
\node at (2,2) {\dots};
\end{tikzpicture}
\caption{Exponentiation of $\cO$ exchange.}
\label{fig:exponentiation of O exchange to delta}
\end{subfigure}
\caption{Large-spin diagrams illustrating the contribution to the anomalous dimensions of $[\f\f]_n$ from the exchange of $\cO$.}
\label{fig:double-twist anomalous dimension}
\end{figure}

\begin{figure}[htb]
\centering
\begin{subfigure}[t]{.4\textwidth}
\centering
\begin{tikzpicture}[xscale=0.6,yscale=0.6]
\draw[thick] (0,0) -- (3,0);
\draw[thick] (0,4) -- (3,4);
\draw[thick, dashed] (1,0) -- (1,4);
\node[left] at (0,0) {$\phi$};
\node[left] at (0,4) {$\phi$};
\node[left] at (1,2) {$\cO$};
\filldraw[gray] (3,2) circle (.5);
\draw[thick] (3,0) arc (-90:90:2);
\end{tikzpicture}
\caption{$t$-channel}
\label{fig:t-channel phi phi with delta}
\end{subfigure}
~
\begin{subfigure}[t]{.4\textwidth}
\centering
\begin{tikzpicture}[xscale=0.6,yscale=0.6]
\draw[thick,dashed] (2,0) arc (270:90:2);
\draw[thick,dashed] (2,0) -- (4,0);
\draw[thick,dashed] (2,4) -- (4,4);
\draw[thick] (4,0) -- (5,0);
\draw[thick] (4,4) -- (5,4);
\draw[thick] (4,0) -- (4,4);
\node[right] at (5,0) {$\phi$};
\node[right] at (5,4) {$\phi$};
\node[left] at (0,2) {$\cO$};
\filldraw[gray] (2,2) circle (.5);
\end{tikzpicture}
\caption{$s$-channel}
\label{fig:s-channel phi phi with delta}
\end{subfigure}
\caption{The $t$-channel sum over the asymptotic parts represented by the diagram on the left inverts to the $s$-channel process on the right. Accordingly, the inversion should produce poles for the $[\cO\cO]_m$ families. The diagram on the right can itself be deciphered by reading it from right to left; first the external $\f$ operators form $[\f\f]_n$, which mixes with $[\cO\cO]_m$ via exchange of a $\f$, then the $[\cO\cO]_m$ receive expectation values proportional to $ b_{\mathbf{1}}$.}
\label{fig:phi phi with delta}
\end{figure}
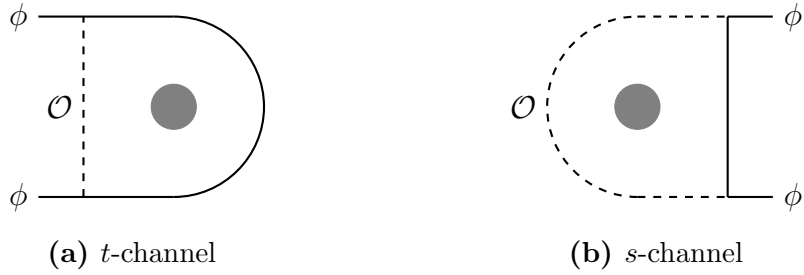

In general, can we say which other families of operators appear in the asymptotics of the sum of a given family, and therefore receive contributions via \eqref{eq:regulated-double-twist-sum}? The large-spin expansion of the anomalous dimensions and OPE coefficients allows us to answer this question. Suppose $\cO$ is an operator in the $\f\times \f$ OPE. Then, $\cO$ corrects the anomalous dimensions of the $[\f\f]_0$ family, via the large-spin diagram in figure \ref{fig:double-twist anomalous dimension} \cite{ Simmons-Duffin:2016wlq, Alday:2007mf}. Consequently, there is a term in the asymptotic expansion of $\delta_{[\f\f]_0}$ that goes like \begin{align}
\delta_{[\f\f]_0}(\bar h) 
\sim \delta_{[\f\f]_0}^{(\cO)} \bar h^{-2h_\cO} +\dots
\end{align} where $\delta_{[\f\f]_0}^{(\cO)}$ is some coefficient. Now, imagine the contribution of the identity operator to the $[\f\f]_0$ thermal coefficients, which goes like $S_{-\Delta_\phi,\Delta_\phi}(\bar h)$ to leading order. Therefore, the sum over the $[\f\f]_0$ family to first order in $\delta_{[\f\f]_0}$ contains the asymptotic term
\begin{align}
\delta_{[\f\f]_0}(\bar h) S_{-\Delta_\phi,\Delta_\phi}(\bar h) \sim \delta_{[\f\f]_0}^{(\cO)} S_{2h_\cO-\Delta_\phi,\Delta_\phi}(\bar h) +\dots \label{eq:asymptotics for OO pole}
\end{align} This asymptotic piece corresponds to the diagram depicted in figure \ref{fig:decomposition of phi phi with delta}.
In the $t$-channel sum over the $[\f\f]_0$ family, this term produces the singular term $\bar z^{2h_\cO-\Delta_\phi}$, which inverts to poles at $h=2h_\cO +n$, na\"ively corresponding to the families $[\cO\cO]_n$. 
For example, the residue for $[\cO\cO]_0$ from this contribution is
\begin{align}
a_{[\cO\cO]_0}^{([\f\f]_0)} (J) = (1+(-1)^J) 4\pi K_J \frac{d\bar h}{dJ} \delta_{[\f\f]_0}^{(\cO)} S_{0,\Delta_\f}^{(1)}(\bar h). \label{eq:OO leading thermal coefficients}
\end{align} 
Thus, we see that the $[\f\f]_0$ family contributes to the $[\cO\cO]_n$ families through its anomalous dimension! 
Similar arguments apply to the $[\f\f]_n$ families. 
We could have guessed that we should obtain poles for the $[\cO\cO]_n$ families by crossing the diagram in figure \ref{fig:decomposition of phi phi with delta} to the $s$-channel. As demonstrated in figure \ref{fig:phi phi with delta}, the resulting $s$-channel process is proportional to $b_{[\cO\cO]_n}$, so the inversion to the $s$-channel must have produced poles for $[\cO\cO]_n$. We therefore see that the intuition from the diagrams agree with concrete calculations!
One can also check that the expression for $b_{[\cO\cO]_0}$ obtained from the $\langle \f\f\rangle_\beta$ correlator agrees with the expression obtained from the $\langle \cO\cO \rangle_\beta$ correlator to leading order in the large-$\bar h$ expansion.

\begin{figure}[tbh]
\centering
\begin{tikzpicture}[xscale=0.6,yscale=0.6]
\draw[thick] (0,0) -- (2,0);
\draw[thick] (0,4) -- (2,4);
\draw[thick, dashed] (1,0) -- (1,4);
\node[left] at (0,0) {$\phi$};
\node[left] at (0,4) {$\phi$};
\node[left] at (1,2) {$\cO$};
\node at (3,2) {$\times$};
\node at (3,0) {$\phi$};
\node at (3,4) {$\phi$};
\filldraw[gray] (5,2) circle (.5);
\draw[thick] (4,0) -- (5,0);
\draw[thick] (4,4) -- (5,4);
\draw[thick] (5,0) arc (-90:90:2);
\node at (8,2) {$=$};
\draw[thick] (9,0) -- (12,0);
\draw[thick] (9,4) -- (12,4);
\draw[thick, dashed] (10,0) -- (10,4);
\node[left] at (9,0) {$\phi$};
\node[left] at (9,4) {$\phi$};
\node[left] at (10,2) {$\cO$};
\filldraw[gray] (12,2) circle (.5);
\draw[thick] (12,0) arc (-90:90:2);
\end{tikzpicture}
\caption{The diagram on the right can be thought of as the product of the two subdiagrams. Reading it from left to right, it's comprised of the contribution of $\cO$ to the anomalous dimensions of $[\f\f]_n$, and the thermal coefficients of the $[\f\f]_n$ families proportional to $b_{\mathbf{1}}$. Accordingly, it should be interpreted as the asymptotic piece with $\delta^{(\cO)}_{[\f\f]_n}(\bar h)\times a_{[\f\f]_n}^{(\mathbf{1})}(\bar h)$ in the sum over $[\f\f]_0$.}
\label{fig:decomposition of phi phi with delta}
\end{figure}
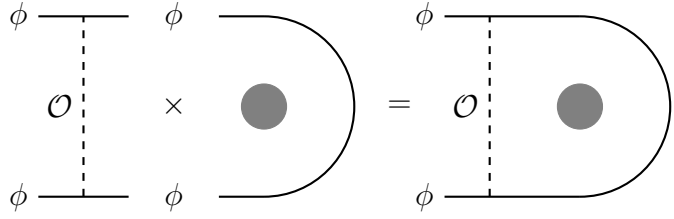

\begin{figure}[htb]
\centering
\begin{subfigure}[t]{.4\textwidth}
\centering
\begin{tikzpicture}[xscale=0.6,yscale=0.6]
\draw[thick] (0,0) -- (3,0);
\draw[thick] (0,4) -- (3,4);
\draw[thick, dashed] (1,0) -- (1,4);
\node[left] at (0,0) {$\phi$};
\node[left] at (0,4) {$\phi$};
\node[left] at (1,2) {$\cO$};
\filldraw[gray] (3,2) circle (.5);
\draw[thick] (3,0) arc (-90:90:2);
\draw[thick, dotted] (3.5,2) -- (5,2);
\node[above] at (4.25,2) {$\cO'$};
\end{tikzpicture}
\caption{$t$-channel}
\label{fig:t-channel aOprime phi phi with delta}
\end{subfigure}
~
\begin{subfigure}[t]{.4\textwidth}
\centering
\begin{tikzpicture}[xscale=0.6,yscale=0.6]
\draw[thick,dashed] (2,0) arc (270:90:2);
\draw[thick,dashed] (2,0) -- (4,0);
\draw[thick,dashed] (2,4) -- (4,4);
\draw[thick] (4,0) -- (5,0);
\draw[thick] (4,4) -- (5,4);
\draw[thick] (4,0) -- (4,4);
\node[right] at (5,0) {$\phi$};
\node[right] at (5,4) {$\phi$};
\node[left] at (0,2) {$\cO$};
\filldraw[gray] (2,2) circle (.5);
\draw[thick, dotted] (2.5,2) -- (4,2);
\node[above] at (3.25,2) {$\cO'$};
\end{tikzpicture}
\caption{$s$-channel}
\label{fig:s-channel aOprime phi phi with delta}
\end{subfigure}
\caption{The $t$-channel diagram denotes a sum over the asymptotics $\delta_{[\f\f]_n}^{(\cO)}(\bar h) \times a_{[\f\f]_n}^{(\cO')}(\bar h)$. This inverts to poles for the $[\cO\cO\cO']_m$ families in the $s$-channel. When $\cO'=\mathbf{1}$ we omit the line by convention and recover figure \ref{fig:phi phi with delta}.}
\label{fig:aOprime phi phi with delta}
\end{figure}
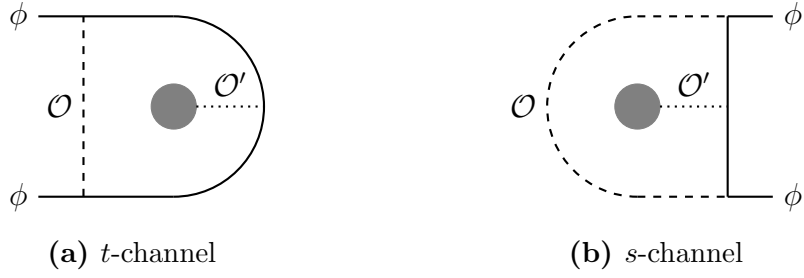

In fact, the situation is much more general. For example, we can consider other terms in the asymptotic expansion of $a_{[\f\f]_0}$, such as $S_{h_{\cO'}-\Delta_\f,\Delta_\f}(\bar h)$ coming from some other operator $\cO'$. Then, the sum over $[\f\f]_0$ to first order in $\delta_{[\f\f]_0}$ produces the singular term $\bar z^{2h_\cO + h_\cO' -\Delta_\f}$, na\"ively corresponding to multi-twist families $[\cO\cO\cO']$. The diagrams for the sum over this asymptotic piece and the corresponding $s$-channel process are given in figure \ref{fig:aOprime phi phi with delta}.
We could also work to higher order in the anomalous dimensions, and obtain poles for multi-twist families, and so on.

However, we are unsure what the precise rules are for which diagrams are allowed, and how to interpret them in general. We will leave deriving these diagrams from physical arguments and further generalizing them to a future project.

\subsubsection{Corrections to pole locations}
\label{sec:pole-locations}

We are also in a position to address the issue of na\"ive versus true locations of poles of $a(\Delta,J)$, raised in section~\ref{sec:leading-double-twist-pert-theory}. Corrections to the locations of poles essentially arise from the asymptotics of the sums over the terms $S_{c,\Delta}^{(m)}(\bar h)$ with $m>0$. By taking derivatives of the integer-spaced sum in \eqref{eq:integer-spaced-sum}, 
\begin{align}
\sum_{\substack{\bar h={\bar h_0} + \ell \\ \ell=0,1,\dots}} S_{c,\Delta}^{(m)}(\bar h) (1-\bar z)^{\bar h} &=  \bar z^c (1-\bar z)^\Delta \log^{m}\bar z +\text{regular}, \label{eq:integer-spaced-sum-derivative}
\end{align} we see that sums over the asymptotics $S_{c,\Delta}^{(m)}(\bar h)$ produce $\log^m\bar z$ terms. Such $\log^m\bar z$ terms in $g(z,\bar z)$ in turn shift the location of the poles of $a(\Delta,J)$ obtained from the $\bar z$ integral in the inversion formula. For a nice way to see this, let's define the generating function 
\begin{align}
\tilde g_n (\bar z,\bar h) = \int_{1}^{z_{\text{max}}}\frac{dz}{z} q_n(J)z^{\Delta_\f-\bar h-n}\Disc[g(z,\bar z)].
\end{align}
The thermal data $a(\Delta,J)$ is obtained from $\tilde g_n (\bar z,\bar h)$ by doing the remaining $\bar z$ integral in the inversion formula,
\begin{align}
a(\Delta,J) = (1+(-1)^J) K_J \int_0^1 \frac{d\bar z}{\bar z} \sum_{n=0}^\infty \bar z^{\Delta_\f-h+n} \tilde g_n(\bar z,\bar h). 
\end{align}
Now, consider the role of terms in $\tilde g_n(\bar z,\bar h)$ of the form
\begin{align}
\sum_{m=0}^\infty f_m(\bar h) \, \bar z^{c}  \log^m \bar z.
\end{align} The claim is that such terms resum to $f(\bar h)\, \bar z^{c + \delta(\bar h)}$, thus changing the location of poles as a function of $\bar h$, i.e.\ introducing anomalous dimensions!

Let's try to see this concretely. For example, how can we see the anomalous dimensions of $[\f\f]_0$ arise? The anomalous dimensions must arise from sums of infinite families, yet which families? The anomalous dimension of $[\f\f]_0$ contributes to other data in the theory, like in figures \ref{fig:phi phi with delta} and \ref{fig:aOprime phi phi with delta}. What is the data in the theory that gives rise to the anomalous dimensions? Of course, there is nothing special about interpreting figures \ref{fig:phi phi with delta} and \ref{fig:aOprime phi phi with delta} as sums over $[\f\f]_0$ in the $t$-channel, which invert to poles for the $[\cO\cO]$ and $[\cO\cO\cO']$ families. Rather, these diagrams are supposed to be crossing symmetric! So, we can flip $s$- and $t$-channels in these diagrams, sum over the $[\cO\cO]$ and $[\cO\cO\cO']$ families in the $t$-channel, and hopefully obtain the expected corrections to the $[\f\f]$ anomalous dimensions when inverted to the $s$-channel. 

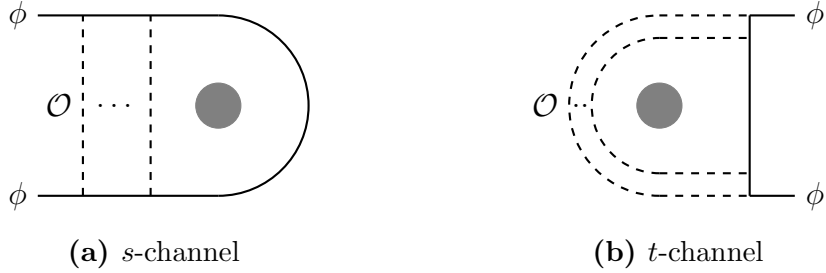
\begin{figure}[htb]
\centering
\begin{subfigure}[t]{.4\textwidth}
\centering
\begin{tikzpicture}[xscale=0.6,yscale=0.6]
\draw[thick] (0,0) -- (4,0);
\draw[thick] (0,4) -- (4,4);
\draw[thick, dashed] (1,0) -- (1,4);
\draw[thick, dashed] (2.5,0) -- (2.5,4);
\node at (1.75,2) {\dots};
\node[left] at (0,0) {$\phi$};
\node[left] at (0,4) {$\phi$};
\node[left] at (1,2) {$\cO$};
\filldraw[gray] (4,2) circle (.5);
\draw[thick] (4,0) arc (-90:90:2);
\end{tikzpicture}
\caption{$s$-channel}
\label{fig:s-channel phi phi with delta exponentiated}
\end{subfigure}
~
\begin{subfigure}[t]{.4\textwidth}
\centering
\begin{tikzpicture}[xscale=0.6,yscale=0.6]
\draw[thick,dashed] (2,0) arc (270:90:2);
\draw[thick,dashed] (2,0.5) arc (270:90:1.5);
\draw[thick,dashed] (2,0) -- (4,0);
\draw[thick,dashed] (2,4) -- (4,4);
\node at (.25,2) {..};
\draw[thick,dashed] (2,.5) -- (4,0.5);
\draw[thick,dashed] (2,3.5) -- (4,3.5);
\draw[thick] (4,0) -- (5,0);
\draw[thick] (4,4) -- (5,4);
\draw[thick] (4,0) -- (4,4);
\node[right] at (5,0) {$\phi$};
\node[right] at (5,4) {$\phi$};
\node[left] at (0,2) {$\cO$};
\filldraw[gray] (2,2) circle (.5);
\end{tikzpicture}
\caption{$t$-channel}
\label{fig:t-channel sum over multi-twist reproducing exponentiated phi phi delta}
\end{subfigure}
\caption{Higher order terms $\delta_{[\f\f]}^m \log^m \bar z$ that sum up to shift the $[\f\f]$ poles in $a(\Delta,J)$ are produced by sums over multi-twist families in the $t$-channel.}
\label{fig:phi phi with delta exponentiated}
\end{figure}

Let's start with the simpler process in figure \ref{fig:phi phi with delta}, but now with the $[\cO\cO]_0$ family running in the $t$-channel. Recall that the $t$-channel sum over the $[\f\f]_0$ family inverted to poles for the $[\cO\cO]_0$ family (through the asymptotics in \eqref{eq:asymptotics for OO pole}) with residue $a_{[\cO\cO]_0}^{([\f\f]_0)} (J)$ given in \eqref{eq:OO leading thermal coefficients}. 
The $t$-channel sum over the $[\cO\cO]_0$ family looks like
\begin{align}
\sum_{[\cO\cO]_0} \frac{a_{[\cO\cO]_0}(J)}{4\pi K_J} (1-z)^{h-\Delta_\f} (1-\bar z)^{\bar h-\Delta_\f}
\end{align} to leading order in $(1-z)$. Focusing on the term $a_{[\cO\cO]_0}^{([\f\f]_0)} (J)$ of $a_{[\cO\cO]_0}(J)$, we have the sum
\begin{align}
\sum_{[\cO\cO]_0} (1+(-1)^J) \frac{d\bar h}{dJ} \delta_{[\f\f]_0}^{(\cO)} S_{0,\Delta_\f}^{(1)}(\bar h) (1-z)^{h-\Delta_\f} (1-\bar z)^{\bar h-\Delta_\f}.
\end{align} Let's assume that $\cO \ne \f$, so $\Delta_\cO \ne \Delta_\f$. Then, expanding to leading (constant) order in $\delta_{[\cO\cO]_0}(\bar h)$, the sum becomes
\begin{align}
\sum_{[\cO\cO]_0} (1+(-1)^J) \frac{d\bar h}{dJ} \delta_{[\f\f]_0}^{(\cO)} S_{0,\Delta_\f}^{(1)}(\bar h) (1-z)^{\Delta_\cO-\Delta_\f} (1-\bar z)^{\bar h-\Delta_\f} = \delta_{[\f\f]_0}^{(\cO)} \log\bar z (1-z)^{\Delta_\cO-\Delta_\f} +\dots.
\end{align} If $\cO = \f$ is in the $\f\times\f$ OPE, we should consider the sum over $\delta_{[\f\f]_0}(\bar h) a_{[\f\f]_0}(J)$, 
\begin{align}
\sum_{[\f\f]_0} (1+(-1)^J) &\frac{d\bar h}{dJ} \delta_{[\f\f]_0}^{(\f)}\frac{1}{\bar h^{\Delta_\f}} S_{-\Delta_\f,\Delta_\f}(\bar h) \log (1-z) (1-\bar z)^{\bar h-\Delta_\f} \nonumber \\ &= \sum_{[\f\f]_0} (1+(-1)^J) \frac{d\bar h}{dJ} \frac{-\delta^{(\f)}_{[\f\f]_0}}{\Gamma(\Delta_\f)} S_{0,\Delta_\f}^{(1)}(\bar h) \log (1-z) (1-\bar z)^{\bar h-\Delta_\f} +\dots \nn
\\ &= -\frac{\delta^{(\f)}_{[\f\f]_0}}{\Gamma(\Delta_\f)} \log \bar z \log(1-z) +\dots.
\end{align}
Note that we have we have used the asymptotic expansion 
\begin{align}
    \frac{1}{\bar h^a}S_{-a,\Delta}(\bar h) = -\frac{1}{\Gamma(a)}S^{(1)}_{0,\Delta}(\bar h) + \dots. \label{eq:degenerate-asymptotic-expansion}
\end{align} 
Doing the $z$ integral and summing over $\cO$, we obtain the contribution to $\tilde g_0(\bar z,\bar h)$
\begin{align}
\sum_{\substack{\cO \in \f\times \f \\ \cO \ne \f}}\delta_{[\f\f]_0}^{(\cO)} \log\bar z \, S_{\Delta_\cO-\Delta_\f,\Delta_\f}(\bar h) -\frac{\delta^{(\f)}_{[\f\f]_0}}{\Gamma(\Delta_\f)} \log \bar z \, S^{(1)}_{0,\Delta_\f}(\bar h).
\end{align}
Now, let's combine this with the contribution of the unit operator to $\tilde g_0(\bar z,\bar h)$,
\begin{align}
\tilde g_0(\bar z,\bar h) &= S_{-\Delta_\f,\Delta_\f}(\bar h) + \log\bar z\left(\sum_{\substack{\cO \in \f\times \f \\ \cO \ne \f}} \delta_{[\f\f]_0}^{(\cO)} S_{\Delta_\cO-\Delta_\f,\Delta_\f}(\bar h) -\frac{\delta^{(\f)}_{[\f\f]_0}}{\Gamma(\Delta_\f)} S^{(1)}_{0,\Delta_\f}(\bar h)\right)+ \dots 
\\ &= S_{-\Delta_\f,\Delta_\f}(\bar h) \left(1+ \log\bar z \sum_{\cO\in \f\times \f} \delta_{[\f\f]_0}^{(\cO)} \frac{1}{\bar h^{\Delta_\cO}} +\dots\right) +\dots.
\end{align} In the last step, we reverted the asymptotic expansion in \eqref{eq:degenerate-asymptotic-expansion} to separate the contribution to the pole location from the residue. This looks like the first few terms in the expansion of $S_{-\Delta_\f,\Delta_\f}(\bar h) \bar z^{\delta_{[\f\f]_0}(\bar h)}$ in small $\delta_{[\f\f]_0}(\bar h)\log \bar z$ and in large $\bar h$. The higher powers of $\log\bar z$ come from the exponentiation of the anomalous dimension, which arise from sums over multi-twist families, such as in the diagram depicted in figure \ref{fig:phi phi with delta exponentiated}. Essentially, any contribution to the anomalous dimension can be recovered by embedding the corresponding four-point function large-spin diagram and performing the thermal crossing operation. Thus, we see the beginnings of a self-consistent story of how the anomalous dimensions are incorporated into the large-spin perturbation theory of the thermal data. We can do the exact analysis with other asymptotics of $a_{[\f\f]_0}(J)$, by reversing the diagram in figure \ref{fig:aOprime phi phi with delta} and thinking of the sum over $[\cO\cO\cO']$. Repeating our analysis above line by line, we'll start to recover the $\cO$ corrections to the anomalous dimensions for the poles proportional to $a_{\cO'}$.

\begin{figure}[htb]
\centering
\begin{subfigure}[t]{.4\textwidth}
\centering
\begin{tikzpicture}[xscale=0.6,yscale=0.6]
\draw[thick] (0,0) -- (1,0);
\draw[thick] (0,4) -- (1,4);
\draw[thick,dashed] (1,0) -- (4,0);
\draw[thick,dashed] (1,4) -- (4,4);
\draw[thick] (1,0) -- (1,4);
\draw[thick, dotted] (2,0) -- (2,4);
\node[right] at (2,2) {$\cO'$};
\node[left] at (0,0) {$\phi$};
\node[left] at (0,4) {$\phi$};
\filldraw[gray] (4,2) circle (.5);
\draw[thick,dashed] (4,0) arc (-90:90:2);
\node[right] at (6,2) {$\cO$};
\end{tikzpicture}
\caption{$s$-channel}
\label{fig:s-channel OO pole with delta}
\end{subfigure}
~
\begin{subfigure}[t]{.4\textwidth}
\centering
\begin{tikzpicture}[xscale=0.6,yscale=0.6]
\draw[thick] (2,0) arc (270:90:2);
\draw[thick,dotted] (2,0.5) arc (270:90:1.5);
\draw[thick] (2,0) -- (4,0);
\draw[thick] (2,4) -- (4,4);
\draw[thick,dotted] (2,.5) -- (4,0.5);
\draw[thick,dotted] (2,3.5) -- (4,3.5);
\draw[thick] (4,0) -- (5,0);
\draw[thick] (4,4) -- (5,4);
\draw[thick,dashed] (4,0) -- (4,4);
\node[right] at (5,0) {$\phi$};
\node[right] at (5,4) {$\phi$};
\node[right] at (4,2) {$\cO$};
\node[below] at (3,3.5) {$\cO'$};
\filldraw[gray] (2,2) circle (.5);
\end{tikzpicture}
\caption{$t$-channel}
\label{fig:t-channel sum over multi-twist reproducing OO pole with delta}
\end{subfigure}
\caption{Poles in $a(\Delta,J)$ for other double-twist families $[\cO\cO]$ in $\langle\f\f\rangle$ shift by anomalous dimensions through sums over multi-twist families $[\f\f\cO'\cO']$ in the $t$-channel.}
\label{fig:OO pole with delta}
\end{figure}
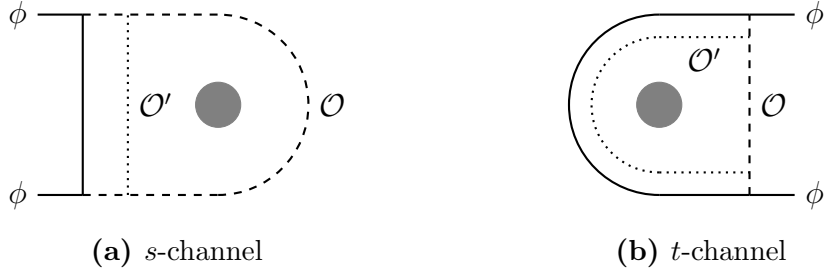

We can also see how the anomalous dimensions of families other than $[\f\f]_0$ arise as well. Let's consider a double-twist family $[\cO\cO]$ for some $\cO\in\f\times\f$ with $\cO\ne \f$, and think about how their anomalous dimensions appear to correct their pole locations in $a(\Delta,J)$. Applying our thinking above, we can first find which families $\delta_{[\cO\cO]}$ contribute to, and then reverse the process. This leads us to the process illustrated in figure \ref{fig:OO pole with delta}. Let's see if the diagram indeed checks out. For simplicity, let's consider the $[\cO\cO]_0$ family, and suppose $\cO'\in\cO\times \cO$, so
\begin{align}
    \delta_{[\cO\cO]_0}(\bar h) \supset \delta_{[\cO\cO]_0}^{(\cO')} \frac{1}{\bar h^{2h_{\cO'}}}.
\end{align} 
Using the leading expression for $a_{[\cO\cO]_0}$ computed in \eqref{eq:OO leading thermal coefficients}, we see that the $t$-channel sum over $[\cO\cO]_0$ to first order in $\delta_{[\cO\cO]_0}$ produces the singular term 
\begin{equation}
\begin{split}
\sum_{[\cO\cO]_0} &(1+(-1)^J) a_{[\cO\cO]_0}(J) \delta_{[\cO\cO]_0}(\bar h) (1-z)^{\Delta_\cO -\Delta_\f}\log(1-z) (1-\bar z)^{\bar h-\Delta_\f} 
\\ &\quad \supset -\Gamma(-2h_{\cO'}) \delta_{[\f\f]_0}^{(\cO)} \delta_{[\cO\cO]_0}^{(\cO')}  \bar z^{2h_{\cO'}} (1-z)^{\Delta_\cO -\Delta_\f}\log(1-z),
\end{split}
\end{equation} where we have used the asymptotic expansion\footnote{For positive integer values of $a$, this asymptotic expansion is slightly modified, and one needs to be more careful with the analysis that follows.}
\begin{equation}
    \frac{1}{\bar h^a} S_{0,\Delta}^{(1)}(\bar h) = -\Gamma(-a)S_{a,\Delta}(\bar h) + \dots, \label{eq:asymptotic expansion of Q and Sm[1](0)}
\end{equation} which can be obtained from the asymptotic expansion in \eqref{eq:degenerate-asymptotic-expansion}.
Such a term inverts to poles for the multi-twist families $[\f\f\cO'\cO']_n$, as we expected from the diagram. The lowest-twist family has the leading residue
\begin{align}
    a_{[\f\f\cO'\cO']_0}(J) \supset (1+(-1)^J) 4\pi K_J \frac{d\bar h}{dJ} \delta_{[\f\f]_0}^{(\cO)} (-\Gamma(-2h_{\cO'})) \delta_{[\cO\cO]_0}^{(\cO')} S^{(1)}_{\Delta_\cO-\Delta_\f,\Delta_\f}(\bar h).
\end{align} Now, let's reverse the diagram, which tells us to sum over $[\f\f\cO'\cO']_0$ in the $t$-channel to constant order in their anomalous dimensions. Performing this sum yields the singular terms
\begin{align}
    -\Gamma(-2h_{\cO'})\delta_{[\f\f]_0}^{(\cO)} \delta_{[\cO\cO]_0}^{(\cO')}  \bar z^{\Delta_\cO-\Delta_\f} \log \bar z (1-z)^{2h_{\cO'}} + \dots
\end{align} which contribute the expected first-order shift to the $[\cO\cO]_0$ poles, since
\begin{align}
    \tilde g_0(\bar z,\bar h) &\supset \delta_{[\f\f]_0}^{(\cO)} \bar z^{\Delta_\cO - \Delta_\f} \left( S^{(1)}_{0,\Delta_\f}(\bar h) -  \log\bar z \sum_{\cO'\in \cO\times \cO} \Gamma(-2h_{\cO'})\delta_{[\cO\cO]_0}^{(\cO')}  S_{2h_{\cO'},\Delta_\f}(\bar h) +\dots \right) 
    \\ &=  \delta_{[\f\f]_0}^{(\cO)} \bar z^{\Delta_\cO - \Delta_\f} S^{(1)}_{0,\Delta_\f}(\bar h) \left( 1+ \log\bar z \sum_{\cO'\in \cO\times \cO} \delta_{[\cO\cO]_0}^{(\cO')}\frac{1}{h^{2h_{\cO'}}} +\dots\right).
\end{align} 
We see that we begin recovering the correct pole locations of the $[\cO\cO]_0$ family in the inversion of the $\langle \f\f\rangle_\beta$ correlator. Once again, the higher-order corrections come from diagrams with exponentiated anomalous dimensions analogous to figure \ref{fig:phi phi with delta exponentiated}.

\subsection{Case study: $\la[\s\s]_0\ra_\b$ in the 3d Ising model}
\label{sec:3d-Ising-large-spin-pert-theory}

Our primary example for applying the above technology is the 3d Ising CFT. At this point, much is known both analytically and numerically about the spectrum and OPE data of the 3d Ising CFT. This abundance of data makes the 3d Ising CFT a natural and ideal candidate for studying thermal correlators. In \cite{Simmons-Duffin:2016wlq}, the low-twist spectrum of the 3d Ising CFT has been computed via the lightcone four-point function bootstrap. Especially relevant to our analysis here is the analytic computation for the anomalous dimensions and OPE coefficients of the most important double-twist family, $[\s \s]_0$, which has the lowest twist trajectory. Taking the spectrum and OPE data as input, we will apply the thermal bootstrap to study the thermal coefficients of the $[\s\s]_0$ family.

The most natural way to get a handle on the $[\s\s]_0$ family is by studying the thermal correlator $\langle \s\s\rangle_\beta$. Let's remind ourselves about the relevant low-twist spectrum of the 3d Ising CFT. The first few lowest-twist primary operators in the $\s\times\s$ OPE are
\begin{align}
\s\times \s = \mathbf{1} + \epsilon + T +\sum_{\ell=4,6,\dots} [\s\s]_{0,\ell}+\dots.
\end{align} 
Our strategy will be to determine the thermal coefficients of $[\s\s]_0$ in terms of $b_\e$ and $b_T$, which we treat as unknowns. 
While we do not determine the values of $b_\epsilon$ and $b_T$ here, our work paves the way towards it.  
In a future paper \cite{Iliesiu:2018}, we will show how information about the low-twist families can be used in conjunction with the KMS condition in the Euclidean regime to ``tie the knot'' on the thermal bootstrap and estimate some thermal coefficients in the theory. 

To numerically study the thermal coefficients in the $[\s\s]_0$ family, we use the scaling dimensions of $\s$ and $\e$, obtained from the numerical bootstrap study \cite{Kos:2016ysd}
\begin{align}
\Delta_\s = .5181489(10), \qquad \Delta_\e = 1.412625(10)\,.
\end{align} 
Using our result \eqref{eq:leading [phiphi]_0 1pt function}, together with these numerical values, we compute the leading contributions to the $[\s\s]_0$ one-point functions,
\begin{align}
a_{[\s\s]_0}(J) = \sum_{\cO = \mathbf{1},\e,T} a_\cO (1+(-1)^J) \frac{ K_J}{K_{\ell_\cO}} \pdr{\bar h}{J}S_{h_\cO-\Delta_\s,\Delta_\s}(\bar h). \label{eq:[ss]_0 leading one point}
\end{align} 
To emphasize the utility of this result we can write the large-spin expansion of the thermal coefficients
\begin{equation}
\begin{split}
a_{[\s\s]_0}(J) = (1+(-1)^J) \bigg[&\frac{1}{J^{\frac{1}{2}-\Delta_\s}} \left(1.0354 + 0.000171 \frac{1}J + \cO\left(\frac{1}{J^2}\right)\right)\\ &+ \frac{a_T}{J^{1-\Delta_\sigma}}\left(0.01218 + 0.001414 \frac{1}{J} + \cO\left(\frac{1}{J^2}\right)\right)\\ &+ \frac{a_\e}{J^{\frac{1}2 + \frac{\Delta_\e}2 - \Delta_\s}}\left(-0.28971 - 0.06859 \frac{1}J - \cO\left(\frac{1}{J^2}\right)\right)
\\& + \dots  \bigg]\,,
\end{split}
\end{equation}
where terms on each line come from the unit operator, the stress tensor and the $\e$ operator respectively. The final ``$\dots$" include contributions of other operators that are either suppressed in the $1/J$ expansion or with coefficients small enough that they can be neglected for reasonable values of the spin.

To go beyond asymptotically large spin and estimate thermal coefficients for operators with small spin, we should include higher-order corrections in $1/J$. 
The next contributions come from the $[\s\s]_0$ family themselves.
Thus, we need to sum over the $[\s\s]_0$ family next. We use the leading expressions in the large-spin expansion \eqref{eq:large-spin exp of anomalous dimensions} of the anomalous dimensions of $[\s\s]_0$, which were computed in \cite{Simmons-Duffin:2016wlq} as 
\begin{align}
\delta_{[\s\s]_0}(\bar h)
\sim -0.001422 \frac{1}{\bar h}-0.04627 \frac{1}{\bar h^{\Delta_\e}} +\dots\,.
\end{align} Upon first iteration, when considering the corrections from $[\s\s]_0$ to itself only once, the corrected thermal coefficient is given by (\ref{eq:first-self-correction}),
\begin{align}
a_{[\s\s]_0}(J ) = &\sum_{\cO = \mathbf{1},\e,T} a_\cO (1+(-1)^J) \frac{K_J}{K_{\ell_\cO}} \frac{d\bar h}{dJ} \nonumber
\\ & \times \left(S_{h_\cO-\Delta_\s,\Delta_\s}(\bar h) + \sum_{m=0}^{\infty}\alpha_0^{\text{even}}\! \left[\frac{\delta^m_{[\s\s]_0}}{m!} S_{h_\cO-\Delta_\s,\Delta_\s},\delta_{[\s\s]_0}, \Delta_\s \right]\!(2h_\s+4) \, S^{(m)}_{0,\Delta_\s}(\bar h) \right). \label{eq:[ss]_0 self corrected}
 \end{align} 
 We can compute the fixed point of the self corrections above using appendix~\ref{sec:self-corr-fixed-pt}, with \eqref{eq:[ss]_0 leading one point} as input. It turns out that the self-corrections of operators in the $[\s\s]_0$ family is given by convergent sums over operators in the $[\s\s]_0$ family, so one can also evaluate the sums numerically by choosing a large spin cut-off. By recursively repeating this numerical process the results converge to the fixed point determined analytically \emph{\`{a} la} appendix~\ref{sec:self-corr-fixed-pt}.\footnote{We find that for small values of the spin the contribution of the stress-energy tensor is the most affected by self-corrections, with a $20\%$ correction for $J=4$.}

To be concrete, the table below shows a few examples for the values of the thermal coefficients $a_{[\s\s]_{0, \ell}}$ and for the thermal one-point functions  $b_{[\s\s]_{0, \ell}}$:

\begin{table}[h!]
\begin{center}
\begin{tabular}{ |c|c|c| } 
 \hline
 $\ell$ & $a_{[\s\s]_{0, \ell}}$ & $b_{[\s\s]_{0, \ell}}/\sqrt{c_{[\s\s]_{0, \ell}}}$ \\ \hline\hline
 4 & $2.1113 - 0.2163 a_\epsilon + 0.0102 a_T$ & $33.431 - 3.4255 a_\epsilon + 0.16182 a_T$ \\ \hline
 6 & $2.1483 - 0.1724 a_\epsilon + 0.0092 a_T$ & $246.29 - 19.773 a_\epsilon+ 1.0500 a_T$  \\ \hline
 8 & $2.1628 - 0.1428 a_\epsilon + 0.0083 a_T$ & $1844.1 - 121.72 a_\epsilon + 7.0586 a_T$ \\ \hline
 10 & $2.1714 - 0.1223 a_\epsilon + 0.0076 a_T$ & $1.3982 \times 10^4 - 787.68 a_\epsilon + 48.839 a_T$  \\ \hline
\end{tabular}
\end{center}
\caption{The thermal coefficients and one-point functions for operators in the $[\s\s]_0$ family, $a_{[\s\s]_{0, \ell}}$ and, $b_{[\s\s]_{0, \ell}}/\sqrt{c_{[\s\s]_{0, \ell}}}$ respectively. Both coefficients are shown in terms of the unknown thermal coefficients $a_\e$ and $a_T$ and include self-corrections from operators in the $[\s\s]_0$ family and are shown in the normalization in which $f_{\s\s[\s\s]_{0, \ell}}$ is positive.  While in this paper we have not determined $a_\e$ and $a_T$ from the thermal bootstrap, we hope to determine these thermal coefficients in future work by using the KMS constraint.
As explained in section~\ref{sec:freenergydensity}, the Monte Carlo results in \cite{PhysRevE.79.041142, casimir2, casimir3} lead to $b_T = -0.459$, or $a_T = 2.105$. This value is consistent with the estimate obtained in appendix~\ref{app:trytogetbT}.}
\end{table}

\section{Conclusions and future work}
\label{sec:conclusion}

Modern advances in the conformal bootstrap have focused almost entirely on constraining OPE data using CFT correlation functions in flat space. Is there potential for more? A broader perspective on the bootstrap suggests future extensions toward probing dynamical questions in CFT, which are not obviously determined by OPE data in a tractable way. 

As a step toward this end, we have developed an approach to bounding CFT observables at finite temperature.
Treating the thermal two-point function on $S^1_\beta \x \R^{d-1}$ in analogy with the flat-space four-point function, and the KMS condition as the analog of the crossing equations, one extracts constraints on the thermal one-point functions of local operators. 
A key intermediate tool (of independent interest) in realizing this approach was to derive a Lorentzian inversion formula \eqref{eq:inversionformulahigherd} which, given a thermal two-point function, extracts thermal one-point coefficients and operator scaling dimensions. 
We applied this technique to the $d=3$ critical $O(N)$ model, which yielded thermal one-point functions of higher-spin currents \eqref{bell} and some scalar operators \eqref{sigmam}. More generally, we developed a large-spin perturbation theory, applicable to any CFT, in which thermal one-point functions are determined via an analytic expansion in inverse operator spin $J$. This included the universal contributions to thermal coefficients of double-twist operators, $a_{[\f\f]_{0, J}}$, from the presence of the unit operator and the stress tensor in the $\f\x\f$ OPE \eqref{eq:universalcontribution}. 
By summing over entire families of operators and plugging back into the large spin expansion, one can solve for CFT data to increasingly high accuracy. Together with the KMS crossing condition, this suggests an iterative algorithm, discussed further below, with which to ``solve'' the thermal sector of an abstract CFT.

There are many future directions to explore:
\begin{itemize}
\item In this work, we mostly consider a single thermal two-point function. However, the same one-point coefficients appear in the OPE decomposition of {\it every\/} two-point function in a theory (except when forbidden by symmetry). Thus, it might be very constraining to study larger systems of two-point functions simultaneously. 

\item A more straightforward generalization of our work would be to study thermal two-point functions of spinning operators. This is likely easier than studying spinning four-point functions on $\mathbb{R}^d$, due to the simplicity of the spinning thermal conformal blocks \cite{Kravchuk:2017dzd}.

\item Our Lorentzian inversion formula makes it straightforward to compute the perturbative expansion of thermal data to all orders in $1/J$, using the $t$-channel OPE for $z<2$. However, there are also nonperturbative corrections that decay exponentially in $J$, coming from the region $z>2$ (outside the regime of validity of the $t$-channel OPE). How can we compute these corrections? Answering this question may require understanding the full analytic structure of thermal two-point functions better.

\item It would be interesting to study more general compactifications. For example, one could study two-point functions on $T^n \x \R^{d-n}$ for $n\geq 2$. On the other hand, there can also be multiple one-point structures on $T^n$ for $n\geq 2$, so there is more data to compute. 
For recent work on CFTs on spatial tori, see \cite{Belin:2016yll, Belin:2018jtf}.

\item We derived thermal one-point functions of all single-trace operators in the critical $O(N)$ model in $d=3$. A clear target for the future is to generalize these results to other slightly broken higher-spin CFTs, such as the Chern-Simons-fundamental matter theories that are continuously connected to the $O(N)$ model \cite{Giombi:2011kc,Aharony:2011jz}. 
The thermal mass and some current-current correlation functions at nonzero temperature have been computed in the large $N$ limit of these theories, for arbitrary 't Hooft coupling $\lambda$ \cite{Giombi:2011kc,Jain:2012qi,Aharony:2012ns,Gur-Ari:2016xff}. It would be satisfying if the thermal one-point functions $b_{J_\ell}$ in these Chern-Simons-matter theories take the same form as in \eqr{bell}, with the appropriate thermal mass $m_{\rm th}(\lambda)$. More generally, we would like to understand the constraints of slightly broken higher-spin symmetry on thermal correlations, in the spirit of \cite{Maldacena:2011jn,Maldacena:2012sf,Alday:2015ota,turi}. 

\item Through the study of holographic CFTs one can get a better intuition for the applicability of the inversion formula down to small values of the spin. Such a direction would entail studying the holographic thermal two-point function in the regimes discussed in section~\ref{s35}, in which $|w|\rightarrow \infty$. Besides offering better intuition for the applicability of the inversion formula, as discussed in section~\ref{sec:holographic-CFTs}, the study of such a regime would also be illuminating for understanding the thermal properties of the stress-energy tensor as implied by black hole physics. It should also be possible to define geodesic Witten diagrams \cite{Hijano:2015qja,Hijano:2015zsa} for black hole backgrounds in AdS$_{d\geq 4}$, which should define an effective two-point ``thermal conformal block" for $d\geq 3$ CFTs with large higher-spin gap.

\item In section~\ref{sec:large-spin-perturbation-theory}, it proved useful to use diagrams to organize terms in large-spin perturbation theory for thermal correlators. It would be nice to place these diagrams on firmer footing by giving a complete specification of the rules they  satisfy and what terms they correspond to. This problem is already interesting in the context of large-spin perturbation theory for four-point functions \cite{Fitzpatrick:2015qma,Simmons-Duffin:2016wlq}, where the diagrams have an interpretation in terms of physical processes in a special conformal frame \cite{Alday:2007mf}.

\item We have made predictions for thermal one-point functions in the 3d Ising CFT in terms of some unknowns, of which we expect $b_T$ (computed via Monte Carlo in \cite{PhysRevE.79.041142, casimir2, casimir3}) and $b_\e$ are the most important. It would be nice to compute $b_\e$. To our knowledge, it is not present in the literature, but should be straightforward to compute using e.g.\ using Monte Carlo simulation \cite{Hasenbuschprivate}.\footnote{See \cite{Katz:2014rla} where similar quantities were computed for the $O(2)$ model.}

\item While in this paper we have made contact with the 3d Ising model by finding the large-spin expansion for the thermal one-point functions $b_{[\s\s]_{0, J}}$, one can imagine a more involved iterative strategy to solve the thermal bootstrap in the double-lightcone limit. This strategy can be summarized in the following diagram:
\begin{figure}[h!]
\begin{center}
\includegraphics[width=0.5\textwidth]{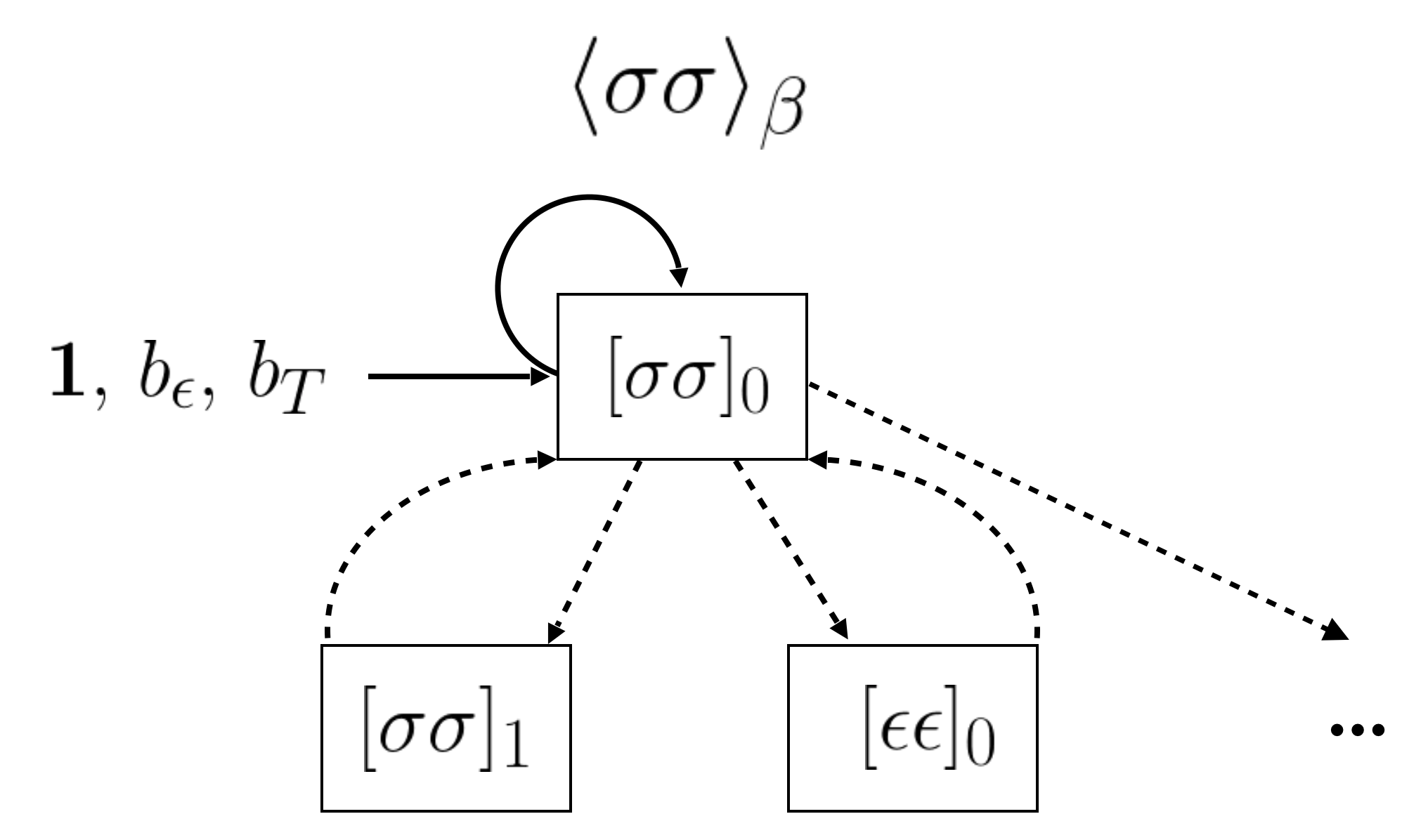}
\vspace{-1.0cm}
\end{center}
\end{figure} \newpage

Following our study in section~\ref{sec:3d-Ising-large-spin-pert-theory}, we start by considering the OPE presentation of the thermal two-point function $\< \s\s\>_\beta$, with the thermal one-point functions of a few low-twist operators as unknowns (in section~\ref{sec:3d-Ising-large-spin-pert-theory}, we consider $b_\e$ and $b_T$ as unknowns). Then, we use the inversion formula on $\<\s\s\>_\beta$ in the double lightcone limit to determine the thermal coefficients of all remaining operators in the $[\s \s]_0$ family as functions of the unknowns. Next, using the technology we developed in section \ref{sec:self-corrections}, we sum over the $[\s\s]_0$ family to determine the self-corrections to the thermal coefficients of the $[\s \s]_0$ family, and also determine the thermal coefficients for the $[\s\s]_1$ and $[\e \e]_0$ families.

In principle, this process can be iterated further by summing over more and more families, and obtaining higher terms in the large-spin expansion. Also, by studying the thermal two-point functions of other operators, we get alternative handles on the thermal coefficients of families of operators. For instance, studying $\langle \e\e\rangle_\beta$ yields more direct information about the $[\e\e]_0$ family. Once the thermal coefficients of families of interest are determined to desired order, we have expressions for a large part of the low-twist spectrum, which still depend only on the unknown thermal one-point functions of the chosen low-twist operators ($b_\e$ and $b_T$). Finally, perhaps these unknowns can be determined by moving away from the double lightcone limit and applying the KMS condition, thus determining the low-twist thermal one-point functions of the 3d Ising CFT.

\item 
The eigenstate thermalization hypothesis (ETH) suggests that we can study thermal correlators as a limit of expectation values in a single eigenstate $|\cO\>$ with sufficiently large dimension. See \cite{Lashkari:2016vgj} for a recent discussion of ETH in the context of CFTs. Assuming ETH, a thermal two-point function $\<\f(x_1)\f(x_2)\>_\b$ is a limit of a family of four-point functions $\<\cO(0)\f(x_1)\f(x_2)\cO(\oo)\>$, where we take $\De_{\cO}\to\oo$, $x_{12}\to 0$ with the product $\De_{\cO}|x_{12}|$ held fixed.\footnote{A similar thermodynamic limit was studied for large-charge correlators in \cite{Jafferis:2017zna}.} It would be interesting to understand whether the ability to view thermal correlators as limits of pure correlators can bring new constraints to the thermal bootstrap. Note that certain properties of vacuum four-point functions may not survive the thermodynamic limit. For example, the analyticity structure changes, with the development of new ``forbidden singularities" reflecting periodicity of the thermal circle \cite{Fitzpatrick:2016ive}.

\item 

One big arena of physics at nonzero temperature that we have not even touched upon in this paper is transport.
Quantities like the diffusivity, viscosity, electrical conductivity, and thermal conductivity are basic experimentally measurable quantities that provide a wealth of information about the low-energy excitations of a system.
These transport coefficients have well-known expressions in terms of two-point functions of components of conserved currents or the stress-energy tensor \cite{kubo1, kubo2, kubo3}.
The most interesting limit of the thermal two-point functions for transport phenomena is the low frequency limit, which translates to large separations in position space. 

Apart from weak coupling expansions, transport has been exhaustively studied from a holographic perspective: For a recent review, see \cite{Hartnoll:2016apf}. 

While the OPE of the thermal two-point function strongly constrains the short distance dynamics in the CFT, it does not directly constrain the long-distance behavior due to the absence of any OPE channel for $|x|>\beta$. 
It is easy to derive functional forms for correlators in the diffusive regime via hydrodynamics, which is the correct low-energy description \cite{kadanoffmartin}.
Can bootstrap techniques allow us to derive this specific form of the diffusive correlator, and the value of the energy diffusion constant for the 3D Ising model?
It would be very interesting to connect the OPE regime to the hydrodynamic regimes in a CFT.

\end{itemize}

\section*{Acknowledgements}

We thank Chris Beem, Simon Caron-Huot, Sergei Gukov, Martin Hasenbusch, Jared Kaplan, Zohar Komargodski, Petr Kravchuk, Juan Maldacena, Alex Maloney, Shiraz Minwalla, Silviu Pufu, Slava Rychkov, Subir Sachdev, Nati Seiberg, Douglas Stanford, and Sasha Zhiboedov for discussions. We especially thank Martin Hasenbusch for sharing unpublished results, and for valuable discussions. DSD, EP, and MK are supported by Simons Foundation grant 488657, and by the Walter Burke Institute for Theoretical Physics. RM is supported by US Department of Energy grant No.\ DE-SC0016244. LVI is supported by Simons Foundation grant 488653.
\appendix

\section{Estimating $b_T$ from $Z_{S^1_\b\x S^{d-1}}$}
\label{app:trytogetbT}

As discussed in section~\ref{sec:spheretoplane}, estimating thermal one-point functions by taking a limit of correlation functions on $S_\b^1 \x S^{d-1}$ is challenging. In general, one needs to know the spectrum $\De_{\cO'}$ and OPE coefficients $f_{\cO\cO'\cO'}$ for arbitrarily high dimension operators $\cO'$ (not to mention the one-point blocks for all tensor structures appearing in $\<\cO\cO'\cO'\>$). In the next appendix, we give slightly more detail in $d=2$. 

However, in any $d$, the observable $b_T$ is special in that it  depends only on the spectrum of the theory.\footnote{We thank Chris Beem, Scott Collier, Liam Fitzpatrick, and Slava Rychkov for discussions that inspired the calculations in this appendix.} This is because the expectation value of the stress-tensor on $S_\beta^1\x S^{d-1}$ is proportional to a derivative of the partition function,
\be
\<T^{00}\>_{S^1_\b \x S^{d-1}} &= \frac{1}{S_d} \frac{\ptl}{\ptl\b} \log Z_{S^1_\b \x S^{d-1}},
\ee
where $S_d=\vol(S^{d-1})=\frac{2\pi^{d/2}}{\G(d/2)}$. Thus, we can compute $b_T$ via the limit
\be
b_T &= \lim_{\b\to 0} \frac{\b^d}{1-1/d} \<T^{00}\>_{S^1_\b \x S^{d-1}} = \frac{1}{S_d(1-1/d)}\lim_{\b\to 0} \b^d \frac{\ptl}{\ptl\b} \log Z_{S^1_\b \x S^{d-1}}.
\ee

The partition function can be expanded in characters
\be
Z_{S^1_\b \x S^{d-1}} &= \sum_{\cO}\chi_{\De,\rho}(e^{-\b}),
\ee
where $\De,\rho$ are the dimension and $\SO(d)$ representation, respectively, of $\cO$ and we sum over primaries only. In practice, even if we don't know the full spectrum of a theory, we can try to estimate $b_T$ by truncating the sum over characters at some $\De_\mathrm{max}$. More precisely, let us define
\be
g_{\De_\mathrm{max}}(\b) &= \frac{1}{S_d(1-1/d)} \b^d \frac{\ptl}{\ptl\b} \log \sum_{\De\leq \De_\mathrm{max}} \chi_{\De,\rho}(e^{-\b}).
\ee
We can then try to extrapolate $g_{\De_\mathrm{max}}(\b)$ towards $\b=0$. The actual value of $g_{\De_\mathrm{max}}(0)$ will always be 0, because $\b^d$ will dominate over the contribution of a finite number of characters at sufficiently small $\b$. However, perhaps we can estimate $b_T$ by evaluating $g_{\De_\mathrm{max}}(\b)$ at a small, nonzero value of $\b$.

As a check on this idea, let us study the free boson, where we know the spectrum exactly. For concreteness, we work in $d=3$. The partition function is given by
\be
Z_\mathrm{free}(q) &= \prod_{j=0}^\oo \frac{1}{(1-q^{j+1/2})^{2j+1}},
\ee
where $q=e^{-\b}$.\footnote{This expression comes from counting states that can be built from arbitrary products of the basic words $\ptl_{\mu_1}\cdots\ptl_{\mu_j}\f$. The dimension of a word is $j+1/2$. Because $\ptl^2 \f=0$, the words transform as traceless symmetric tensors, so there are $2j+1$ of them for a given $j$. This leads to the above product representation of the partition function.}${}^,$\footnote{There is no Casimir energy contribution to the partition function on $S^1\x S^{d-1}$ in odd dimensions. One way to understand this is to start on $S^d$ and perform a Weyl transformation to a long capped cylinder with length $L$. Because there is no Weyl anomaly in odd dimensions, the partition function does not develop any interesting dependence on $L$ at large $L$, and hence the Casimir energy is zero. In 4 and higher even dimensions, the Casimir energy on $S^{d-1}$ is scheme-dependent, since it can be shifted by a counterterm proportional to powers of the scalar curvature. Thus, there is only a universal scheme-independent Casimir energy in $d=2$. (The story is different in supersymmetric theories \cite{Assel:2015nca}.)}
It can be decomposed into conformal characters as
\be
Z_\mathrm{free}(q) &= 1 + \chi^\mathrm{free}(q) + \sum_{\ell=2,4,\dots}\chi^\mathrm{short}_{\ell}(q) + Z_\textrm{long}(q),\nn\\
\chi^\mathrm{free}(q) &= \chi_{1/2,0}(q)-\chi_{1/2+2,0}(q), \nn\\
\chi^\mathrm{short}_\ell(q) &= \chi_{\ell+1,\ell}(q)-\chi_{\ell+2,\ell-1}(q),\nn\\
\chi_{\De,\ell}(q) &= \frac{q^{\De}(2\ell+1)}{(1-q)^3}.
\ee
Here, $\chi_{\De,\ell}(q)$ is the character of a long multiplet, and the first three terms in $Z_\mathrm{free}(q)$ correspond to the unit operator, the boson $\f$ itself, and a tower of higher-spin currents. The long multiplet content is
\be
Z_\mathrm{long}(q) &= \chi_{1,0}(q) + \chi_{3/2,0}(q) + \chi_{2,0}(q) + \dots.
\ee
To determine the quantum numbers and multiplicities of long multiplets, we can include a fugacity for angular momentum and decompose the full partition function with this fugacity into conformal characters. This is a standard exercise and we do not include the details here.

\begin{figure}
    \centering
    \includegraphics[width=0.8\textwidth]{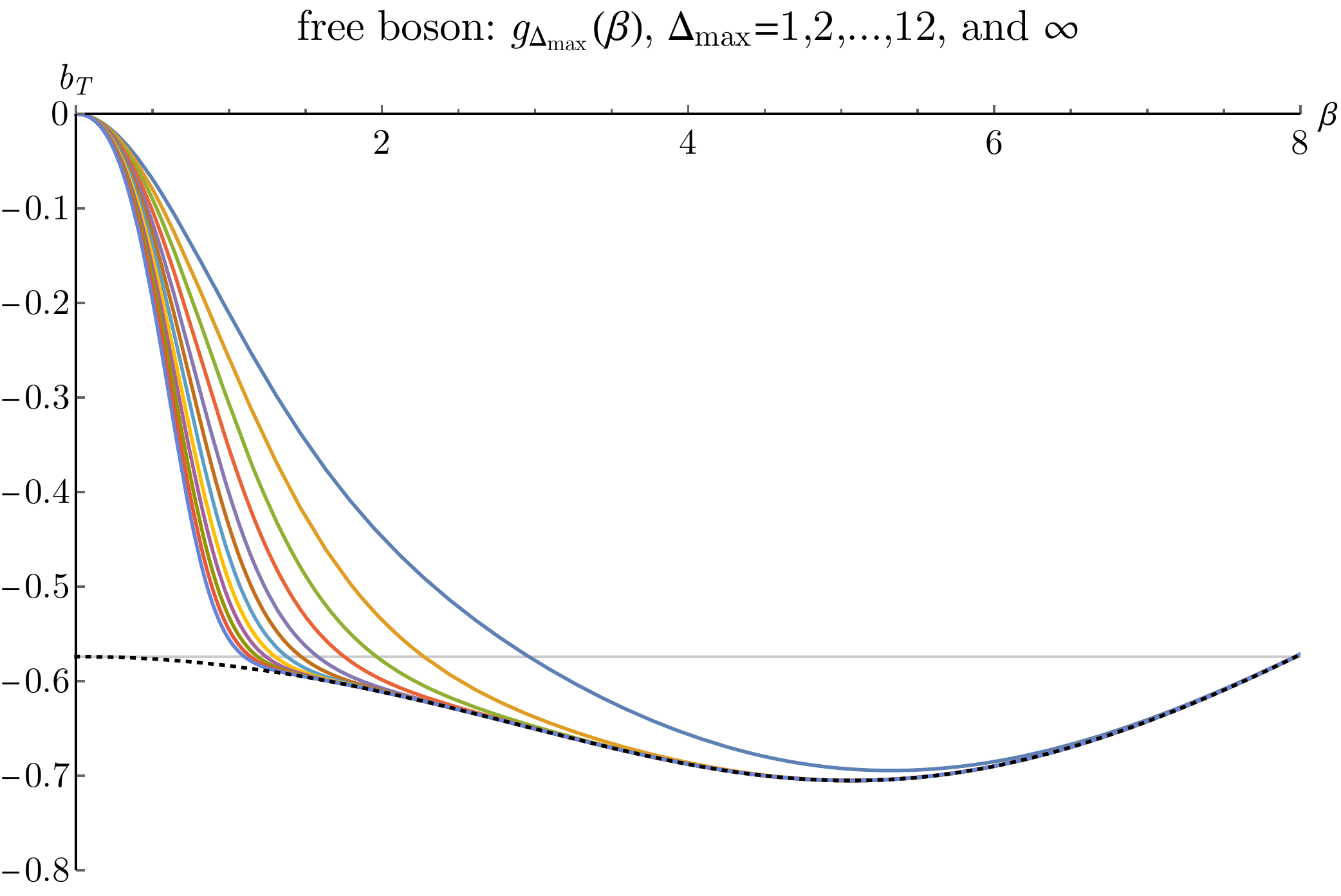}
    \caption{The function $g_{\De_\mathrm{max}}(\b)$ in the free boson theory in 3d, plotted for the values $\De_\mathrm{max}=1,2,\dots,12$ (colors), and $\De_\mathrm{max}=\oo$ (black dotted line). The value of $b_T$ in the free theory (gray horizontal line) is $-\frac{3\z(3)}{2\pi}\approx -0.574$.}
    \label{fig:freetheorebtest}
\end{figure}

Using our knowledge of the spectrum, we can plot $g_{\De_\mathrm{max}}(\b)$ for various values of $\De_\mathrm{max}$ in the free boson theory (figure~\ref{fig:freetheorebtest}). The function with $\De_\mathrm{max}=\oo$ (black dotted line) decays as $e^{-\b/2}$ for large $\b$ (coming from the contribution of the lowest-dimension operator $\f$). It reaches a minimum near $\b=5$, and then smoothly approaches the value $b_T=-\frac{3\z(3)}{2\pi}\approx -0.574$ as $\b\to 0$. The curves with finite $\De_\mathrm{max}$ move closer to the $\De_\mathrm{max}=\oo$ curve, with longer and longer plateaus near $b_T$ before eventually going to $0$ at $\b=0$.

\begin{figure}
    \centering
    \includegraphics[width=0.8\textwidth]{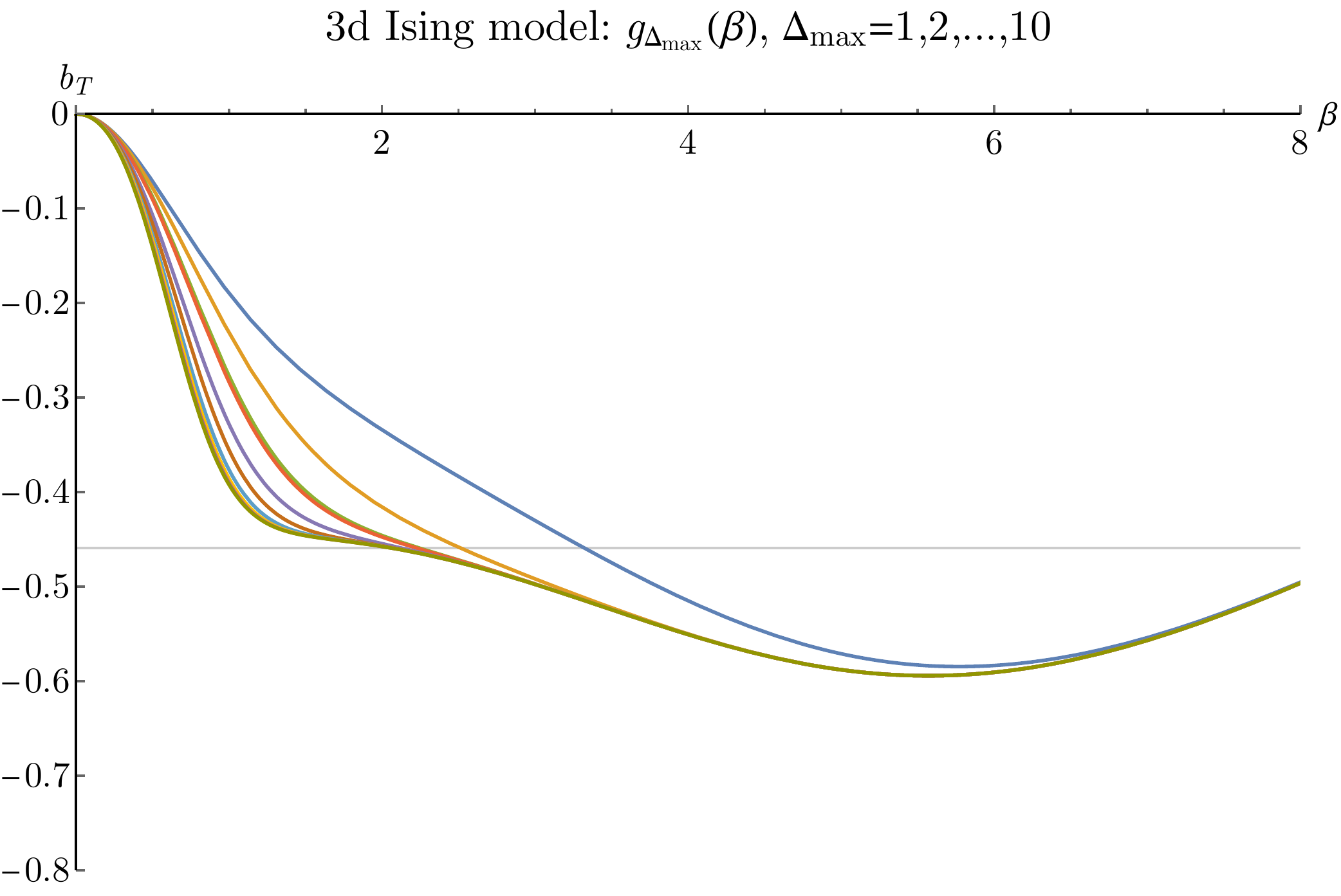}
    \caption{The function $g_{\De_\mathrm{max}}(\b)$ in the 3d Ising model, estimated using known operators only, and plotted for the values $\De_\mathrm{max}=1,2,\dots,10$. The value of $b_T$ determined from Monte Carlo simulations is $-0.459$ (gray horizontal line, \cite{PhysRevE.79.041142, casimir2, casimir3}).}
    \label{fig:isingbtest}
\end{figure}

The 3d Ising model is a nonperturbative theory where we don't know the full spectrum, but we do know a large part of it to reasonable precision from numerical bootstrap computations \cite{ElShowk:2012ht,El-Showk:2014dwa,Kos:2014bka,Simmons-Duffin:2015qma,Komargodski:2016auf,Kos:2016ysd,Simmons-Duffin:2016wlq}. In particular, the spectrum of operators appearing in the $\s\x\s, \s\x\e,$ and $\e\x\e$ OPEs are known up to dimension $8$ \cite{Simmons-Duffin:2016wlq}. Some additional low-twist families are known up to very high dimension, but these are a small portion of the high-dimension spectrum. The lowest-dimension operator not appearing in the above OPEs is expected to be a $\Z_2$-even vector with dimension approximately $6$, though its dimension is not known to high precision \cite{Meneses:2018xpu}. Thus, our knowledge of the spectrum begins to fade when $\De_\mathrm{max}\approx 6$. Nevertheless, in figure~\ref{fig:isingbtest}, we estimate $g_{\De_\mathrm{max}}(\b)$ by including the known operators with dimension $\De\leq \De_\mathrm{max}$. 

Despite our ignorance of the high-dimension spectrum, the plot in figure~\ref{fig:isingbtest} already shows similar structure to the free scalar case, with a plateau beginning to form near $b_T\approx -0.45$, close to the value $b_T\approx -0.459$ determined via Monte Carlo simulations \cite{PhysRevE.79.041142, casimir2, casimir3}. It would be interesting to understand whether figure~\ref{fig:isingbtest} can be turned into a rigorous estimate, perhaps by understanding better the analytic structure of $g_\oo(\b)$. It would also be interesting to understand whether the existence of a minimum, seen as the ``dip" in these plots, is a feature shared by all CFTs.

\section{One-point functions on $S^1_\b\times \R^{d-1}$ from one-point functions on $S^1_\b \x S^{d-1}$}
\label{sec:detailsonthermalsdminus1}

Here we make some basic comments about the challenges in determining $b_\cO$ by passage from $S^1_\b\x S_L^{d-1}$, for generic operators $\cO$. The strategy is to expand the one-point function in $S^1_\b\x S_L^{d-1}$ conformal blocks, and take the large $L$ limit. In \eqr{1ptsphere}, we gave the conformal block expansion for $\la \cO\ra_{S^1_\b\times S_L^{d-1}}$. We focus here on $d=2$ for simplicity. In this case, the thermal blocks factorize,
\eq{global2d}{F(h_{\cO},\hb_{\cO};h_{\cO'},\hb_{\cO'}|\b)= |\mathfrak{g}(h_{\cO},h_{\cO'}|\b)|^2}
where $|f(h)|^2 \equiv f(h)f(\hb)$, and (e.g.\ \cite{gb1, gb2})
\eq{}{\mathfrak{g}(h_{\cO},h_{\cO'}|\b) = {q^{ h_{\cO'}}\o (1-q)^{h_{\cO}}}\,{}_2F_1(2h_{\cO'}-h_{\cO},1-h_{\cO};2h_{\cO'};q)}
where $q\equiv e^{-\b}$. \eqr{global2d} generalizes in the obvious way to unequal left- and right-moving temperatures. Using the connection formulae for hypergeometric functions, we may rewrite the left-moving block (for generic $(h_\cO,h_{\cO'})$) as
\es{}{\mathfrak{g}(h_{\cO},h_{\cO'}|\b) = {q^ {h_{\cO'}}\o (1-q)^{h_{\cO}}}\Bigg[&{\G(2h_{\cO'})\G(2h_{\cO}-1)\o \G(h_{\cO})\G(2h_{\cO'}+h_{\cO}-1)}\,{}_2F_1(2h_{\cO'}-h_{\cO},1-h_{\cO};2-2h_{\cO};1-q)\\+(1-q)^{2h_{\cO}-1}&{\G(2h_{\cO'})\G(1-2h_{\cO})\o \G(2h_{\cO'}-h_{\cO})\G(1-h_{\cO})}\,{}_2F_1(h_{\cO},2h_{\cO'}+h_{\cO}-1;2h_{\cO};1-q)\Bigg]}
As $\b\rar0$, there are two branches:
\eq{gbranch}{\mathfrak{g}(h_{\cO},h_{\cO'}|\b) \sim \b^{-h_{\cO}}{\G(2h_{\cO'})\G(2h_{\cO}-1)\o \G(h_{\cO})\G(2h_{\cO'}+h_{\cO}-1)}+\b^{h_{\cO}-1}{\G(2h_{\cO'})\G(1-2h_{\cO})\o \G(2h_{\cO'}-h_{\cO})\G(1-h_{\cO})}}
Combining the left- and right-moving blocks yields the full scaling behavior of the torus one-point blocks at high temperature.\footnote{The $d>2$ one-point blocks have similar behavior. In fact, there is a third branch in that case. We thank Alex Maloney for sharing the results of \cite{maloney} with us.} 

It is remarkable that, for $h_\cO \geq 1/2$, the leading term in \eqr{gbranch} exhibits the same scaling of the full one-point function on $S^1_\b\times \R$, for all intermediate operators $\cO'$. For instance, for scalar $\cO$ with $\De_\cO=2h_\cO>1$,
\eq{}{F(h_{\cO},\hb_{\cO};h_{\cO'},\hb_{\cO'}|\b) \sim \b^{-2h_\cO}\left|{\G(2h_{\cO'})\G(2h_{\cO}-1)\o \G(h_{\cO})\G(2h_{\cO'}+h_{\cO}-1)}\right|^2}
This leads to a formal expression for the $S^1_\b\times \R^{d-1}$ one-point function $b_\cO$ as a sum over states, in the limit of high temperature:
\eq{cylsum}{b_\cO = {\G^2(2h_\cO-1)\o\G^2(h_{\cO})} \sum_{\text{Primary }\cO'}f_{\cO\cO'\cO'}\left|{\G(2h_{\cO'})\o \G(2h_{\cO'}+h_{\cO}-1)}\right|^2 \qquad \left(h_\cO > {1\o 2}\right)}
Note that the summand is not, in general, sign-definite, and receives contributions from all spins and arbitrarily high energies.\footnote{One can instead use Virasoro conformal blocks, although these are not known in full generality. At large central charge $c$, the torus one-point Virasoro block is simply the the global block \eqr{global2d} times the Virasoro vacuum character, $\chi_{\rm vac}(\b)=q^{(1-c)/24}(1-q)\eta^{-1}(q)$, where $\eta(q)$ is the Dedekind eta function. At high temperature, $\chi_{\rm vac}(\b) \sim \b^{3/2}$. This implies that at $c\rar\infty$, the sum \eqr{cylsum} must, in general, diverge.} On the other hand, for low-dimension operators $\cO$, with $h_\cO <1/2$, the second term of \eqr{gbranch} dominates, and even the recovery of the requisite $\b^{-2h_\cO}$ scaling at high temperature is sensitive to the details of the full sum. For any value of $h_\cO$, one can approximate $b_\cO$ using the asymptotics of $f_{\cO\cO'\cO'}$ for $h_{\cO'}\gg1$, which are often determined by eigenstate thermalization. 

\section{Thermal mass in the $O(N)$ model at large $N$}
\label{app:O(N)-thermal-mass}
Our starting point is the Lagrangian \eqr{eq:O(N)-lagrange}. By integrating out the fields $\phi_i$, the partition function is given by, 
\be
\label{eq:O(N)-path-integral}
Z=  \int D\sigma e^{-\frac{N}2\, \text{Tr} \log (- \nabla^2 + \sigma)} \,.
\ee
As $N\rightarrow \infty$ the partition function is dominated by the saddle-point solution for $\sigma$. On $\mathbb R^3$, due to Poincare symmetry the saddle-point solution can be argued to be $\sigma = 0$. However, on $\mathbb R^2 \times S^1_\beta$, the saddle-point solution is nonzero.
The saddle point equation is
\be
{\partial \o\partial \sigma} \text{Tr} \log (- \nabla^2 + \sigma) = \sum_{n = -\infty}^{\infty} \int \frac{d^2 p}{(2\pi)^2} \frac{1}{\omega_n^2 + p^2 + \sigma} = 0\,,
\ee 
where $\omega_n = 2\pi n/\beta$ are the Matsubara frequencies. 
Doing the sum over $n$, we reduce this equation to
\begin{align}
    \int_0^\Lambda \frac{pdp}{2\pi} \frac{1}{2\sqrt{p^2+\sigma}} \coth \frac{\sqrt{p^2 + \sigma}}{2T}  = 0.
\end{align}
Now make the change of variables $x=\sqrt{p^2 + \sigma}/2T$.
The integral of $\coth x$ is $\log \sinh x$.
The upper limit gives $\log \sinh [\Lambda/2T] \approx \Lambda/2T - \log 2$. 
The linear UV divergence is subtracted out by hand, and the $\log 2$ is left over.
Alternatively, we could replace the integrand $\coth x$ by $\coth x -1$, which is equivalent to a Pauli-Villars regulator.
Thus, overall we get the equation
\begin{align}
    \log \left[ 2\sinh \frac{\sqrt{\sigma}}{2T} \right] = 0,
\end{align}
whose solution is \eqr{eq:O(N)-thermal-mass}. 

\section{Subtleties in dimensional reduction of CFTs}
\label{sec:subtletieswithdimred}

The dimensional reduction of a $d$-dimensional CFT on $S^1$ does not always give a well-defined theory in $d-1$-dimensions. For example, a problem occurs if we try to compactify the free boson CFT in 3d down to 2d (with periodic boundary conditions around the $S^1$).\footnote{Conformal invariance requires that the free boson CFT have a noncompact target space in 3d.} Naively, we should get the 2d free boson with noncompact target space, but this theory is pathological because correlations grow logarithmically with distance. Another way to see the problem is that if we try to compute the propagator using the method of images, the sum over images diverges.\footnote{It is interesting to ask what happens if we have a physical system that has an EFT description in terms of a 3d free boson, and we place it at finite temperature. In this case, the thermal physics can depend on UV details that are not directly captured by the 3d effective CFT. For example, if the 3d boson is the Goldstone boson of a broken $U(1)$ symmetry with symmetry breaking scale $\Lambda$, its dimensional reduction is better described as a 2d boson with compact target space, where the radius is $R\propto \sqrt{\b \Lambda}$.}

We expect this issue to arise whenever we compactify a 3d CFT with a nontrivial moduli space of vacua down to 2d, as long as the boundary conditions do not destroy the moduli space. For example, supersymmetric compactifications of 3d SCFTs with Higgs or Coloumb branches should be treated with care.  One way to study such theories is to introduce twisted boundary conditions that remove the zero mode from the path integral.\footnote{We thank Nati Seiberg for discussions on this point.} Correlation functions in the twisted setting then share many similar properties to those we discuss in this work (for example an OPE and crossing equation). It would be interesting to adapt the techniques in this work to deal with general twisted boundary conditions. In our case, thermal compactification does the job because it breaks supersymmetry and generically lifts the moduli space.

\section{Fixed point of self-corrections of double-twist families}
\label{sec:self-corr-fixed-pt}

Let's consider the self-corrections of the $[\f\f]_0$ family in $\langle \f\f\rangle_\beta$. Given an expression $a_{[\f\f]_0}$ for the one-point functions of the $[\f\f]_0$ family, calculating their self-corrections amount to summing over the family in the $t$-channel starting at the lowest spin operator with $\bar h=\bar h_0$, and collecting the contribution to $a_{[\f\f]_0}$ in the inversion formula.
Therefore, to leading order in large $\bar h$, self-correction is the linear map\footnote{If desired, we could also start the sum at a higher spin, and treat the contributions of the low-spin operators exactly in their anomalous dimensions. The corresponding generalization of our algorithm is straightforward. If the anomalous dimensions are small even for low-spin operators, we can safely start the sum at low spins.}
\begin{align}
S: \quad a_{[\phi\phi]_0}(J) \mapsto a_{[\phi\phi]_0}(J) + (1+(-1)^J) K_J \sum_{n=0}^\infty \alpha_0^{\text{even}}\left[ \frac{\delta^n}{n!} a_{[\phi\phi]_0},\delta,h_{e} \right]\!(\bar h_0) \;  S_{h_f-h_e,h_{e}}^{(n)}(\bar h). \label{eq:def:self-correction-map}
\end{align} As with the main text, $h_f=2h_\f$ stands for the asymptotic $h$ of the family, and $h_e=2h_\f$ stands for the total $h$ of the two external operators.
Self-corrections of $a_{[\phi\phi]_0}$ take the general form
\begin{align}
a_{[\phi\phi]_0}(J)  = a_{[\phi\phi]_0}^0(J) + (1+(-1)^J) K_J \sum_{m=0}^{\infty} f_m S^{(m)}_{h_f-h_{e},h_{e}}(\bar h) \label{eq:self-correction-basis}
\end{align} with some initial $a_{[\phi\phi]_0}^0(J)$, and some coefficients $f_m$. Inserting this form into the self-correction map, we get
\begin{align}
S[a_{[\phi\phi]_0}(J)] =a_{[\phi\phi]_0}(J) +  (1+(-1)^J) K_J \sum_{n=0}^\infty \left( \lambda_n +\sum_{m=0}^\infty f_m T^m_{~~n} \right) S_{h_f-h_{e},h_{e}}^{(n)}(\bar h), \label{eq:self-correction-of-basis}
\end{align} 
where we have defined the vector 
\begin{align}
\lambda_n = \alpha_0^{\text{even}} \left[\frac{\delta^n}{n!} a_{[\phi\phi]_0}^0, \delta, h_{e}\right]\!(\bar h_0) 
\end{align}
and matrix
\begin{align}
T^m_{~~n} = \alpha_0^{\text{even}} \left[\frac{\delta^n}{n!} S^{(m)}_{h_f-h_{e},h_{e}},\delta, h_{e}\right]\!(\bar h_0),
\end{align} of coefficients.  
To evaluate these coefficients, we need to evaluate sums of $S^{(m)}_{c,\Delta} (\bar h)$, which are easily obtained by generalizing our treatment of the sums of $S_{c,\Delta}(\bar h)$ by taking derivatives with respect to $c$. 
The fixed point of the map $S$ satisfies $S[a]=a$, which from \eqref{eq:self-correction-of-basis} is the solution to the  linear equation
\begin{align}
\lambda_n + f_m T^m_{~~n} = 0.
\end{align} Inverting this equation, the fixed point is determined by 
\begin{equation}
f_m = \lambda_n (T^{-1})^n_{~m}. \label{eq:self-correction-fixed-point}
\end{equation} In practice, one can work order-by-order in the small anomalous dimensions of the family, effectively truncating the $m$ and $n$ to some finite order, thus avoiding having to compute an infinite number of coefficients and to invert an infinite matrix. Finally, let's note that it is possible to generalize this method to the $[\phi\phi]_n$ families, and even potentially to considering collections of families at once.

\bibliographystyle{utphys}
\bibliography{Biblio}

\providecommand{\href}[2]{#2}\begingroup\raggedright\begin{thebibliography}{100}

\bibitem{PhysRevB.44.6883}
M.-C. Cha, M.~P.~A. Fisher, S.~M. Girvin, M.~Wallin, and A.~P. Young,
  ``Universal conductivity of two-dimensional films at the
  superconductor-insulator transition,''
  \href{http://dx.doi.org/10.1103/PhysRevB.44.6883}{{\em Phys. Rev. B}
  {\bfseries 44} (Oct, 1991) 6883--6902}.

\bibitem{PhysRevLett.95.180603}
J.~Smakov and E.~Sorensen, ``Universal scaling of the conductivity at the
  superfluid-insulator phase transition,''
  \href{http://dx.doi.org/10.1103/PhysRevLett.95.180603}{{\em Phys. Rev. Lett.}
  {\bfseries 95} (Oct, 2005) 180603}.

\bibitem{Katz:2014rla}
E.~Katz, S.~Sachdev, E.~S. Sorensen, and W.~Witczak-Krempa, ``{Conformal field
  theories at nonzero temperature: Operator product expansions, Monte Carlo,
  and holography},'' \href{http://dx.doi.org/10.1103/PhysRevB.90.245109}{{\em
  Phys. Rev.} {\bfseries B90} no.~24, (2014) 245109},
\href{http://arxiv.org/abs/1409.3841}{{\ttfamily arXiv:1409.3841
  [cond-mat.str-el]}}.

\bibitem{Kos:2015mba}
F.~Kos, D.~Poland, D.~Simmons-Duffin, and A.~Vichi, ``{Bootstrapping the O(N)
  Archipelago},'' \href{http://dx.doi.org/10.1007/JHEP11(2015)106}{{\em JHEP}
  {\bfseries 11} (2015) 106},
\href{http://arxiv.org/abs/1504.07997}{{\ttfamily arXiv:1504.07997 [hep-th]}}.

\bibitem{Nakayama:2016cim}
Y.~Nakayama, ``{Bootstrapping critical Ising model on three-dimensional real
  projective space},''
  \href{http://dx.doi.org/10.1103/PhysRevLett.116.141602}{{\em Phys. Rev.
  Lett.} {\bfseries 116} no.~14, (2016) 141602},
\href{http://arxiv.org/abs/1601.06851}{{\ttfamily arXiv:1601.06851 [hep-th]}}.

\bibitem{Hasegawa:2016piv}
C.~Hasegawa and {\relax Yu}.~Nakayama, ``{$\epsilon$-Expansion in Critical
  $\phi^3$-Theory on Real Projective Space from Conformal Field Theory},''
  \href{http://dx.doi.org/10.1142/S0217732317500456}{{\em Mod. Phys. Lett.}
  {\bfseries A32} no.~07, (2017) 1750045},
\href{http://arxiv.org/abs/1611.06373}{{\ttfamily arXiv:1611.06373 [hep-th]}}.

\bibitem{Hasegawa:2018yqg}
C.~Hasegawa and Y.~Nakayama, ``{Three ways to solve critical $\phi^4$ theory on
  $4-\epsilon$ dimensional real projective space: perturbation, bootstrap, and
  Schwinger-Dyson equation},''
\href{http://arxiv.org/abs/1801.09107}{{\ttfamily arXiv:1801.09107 [hep-th]}}.

\bibitem{Maldacena:1997re}
J.~M. Maldacena, ``{The Large N limit of superconformal field theories and
  supergravity},'' \href{http://dx.doi.org/10.1023/A:1026654312961,
  10.4310/ATMP.1998.v2.n2.a1}{{\em Int. J. Theor. Phys.} {\bfseries 38} (1999)
  1113--1133}, \href{http://arxiv.org/abs/hep-th/9711200}{{\ttfamily
  arXiv:hep-th/9711200 [hep-th]}}.
[Adv. Theor. Math. Phys.2,231(1998)].

\bibitem{Gubser:1998bc}
S.~S. Gubser, I.~R. Klebanov, and A.~M. Polyakov, ``{Gauge theory correlators
  from noncritical string theory},''
  \href{http://dx.doi.org/10.1016/S0370-2693(98)00377-3}{{\em Phys. Lett.}
  {\bfseries B428} (1998) 105--114},
\href{http://arxiv.org/abs/hep-th/9802109}{{\ttfamily arXiv:hep-th/9802109
  [hep-th]}}.

\bibitem{Witten:1998qj}
E.~Witten, ``{Anti-de Sitter space and holography},''
  \href{http://dx.doi.org/10.4310/ATMP.1998.v2.n2.a2}{{\em Adv. Theor. Math.
  Phys.} {\bfseries 2} (1998) 253--291},
\href{http://arxiv.org/abs/hep-th/9802150}{{\ttfamily arXiv:hep-th/9802150
  [hep-th]}}.

\bibitem{ElShowk:2011ag}
S.~El-Showk and K.~Papadodimas, ``{Emergent Spacetime and Holographic CFTs},''
  \href{http://dx.doi.org/10.1007/JHEP10(2012)106}{{\em JHEP} {\bfseries 10}
  (2012) 106},
\href{http://arxiv.org/abs/1101.4163}{{\ttfamily arXiv:1101.4163 [hep-th]}}.

\bibitem{Rattazzi:2008pe}
R.~Rattazzi, V.~S. Rychkov, E.~Tonni, and A.~Vichi, ``{Bounding scalar operator
  dimensions in 4D CFT},''
  \href{http://dx.doi.org/10.1088/1126-6708/2008/12/031}{{\em JHEP} {\bfseries
  12} (2008) 031},
\href{http://arxiv.org/abs/0807.0004}{{\ttfamily arXiv:0807.0004 [hep-th]}}.

\bibitem{Poland:2011ey}
D.~Poland, D.~Simmons-Duffin, and A.~Vichi, ``{Carving Out the Space of 4D
  CFTs},'' \href{http://dx.doi.org/10.1007/JHEP05(2012)110}{{\em JHEP}
  {\bfseries 1205} (2012) 110},
\href{http://arxiv.org/abs/1109.5176}{{\ttfamily arXiv:1109.5176 [hep-th]}}.

\bibitem{El-Showk:2014dwa}
S.~El-Showk, M.~F. Paulos, D.~Poland, S.~Rychkov, D.~Simmons-Duffin, and
  A.~Vichi, ``{Solving the 3d Ising Model with the Conformal Bootstrap II.
  $c$-Minimization and Precise Critical Exponents},''
  \href{http://dx.doi.org/10.1007/s10955-014-1042-7}{{\em J.Stat.Phys.}
  {\bfseries 157} (June, 2014) 869},
\href{http://arxiv.org/abs/1403.4545}{{\ttfamily arXiv:1403.4545 [hep-th]}}.

\bibitem{Paulos:2014vya}
M.~F. Paulos, ``{JuliBootS: a hands-on guide to the conformal bootstrap},''
\href{http://arxiv.org/abs/1412.4127}{{\ttfamily arXiv:1412.4127 [hep-th]}}.

\bibitem{Simmons-Duffin:2015qma}
D.~Simmons-Duffin, ``{A Semidefinite Program Solver for the Conformal
  Bootstrap},'' \href{http://dx.doi.org/10.1007/JHEP06(2015)174}{{\em JHEP}
  {\bfseries 06} (2015) 174},
\href{http://arxiv.org/abs/1502.02033}{{\ttfamily arXiv:1502.02033 [hep-th]}}.

\bibitem{Liendo:2012hy}
P.~Liendo, L.~Rastelli, and B.~C. van Rees, ``{The Bootstrap Program for
  Boundary CFT${}_d$},'' \href{http://dx.doi.org/10.1007/JHEP07(2013)113}{{\em
  JHEP} {\bfseries 1307} (2013) 113},
\href{http://arxiv.org/abs/1210.4258}{{\ttfamily arXiv:1210.4258 [hep-th]}}.

\bibitem{Gliozzi:2015qsa}
F.~Gliozzi, P.~Liendo, M.~Meineri, and A.~Rago, ``{Boundary and Interface CFTs
  from the Conformal Bootstrap},''
  \href{http://dx.doi.org/10.1007/JHEP05(2015)036}{{\em JHEP} {\bfseries 05}
  (2015) 036},
\href{http://arxiv.org/abs/1502.07217}{{\ttfamily arXiv:1502.07217 [hep-th]}}.

\bibitem{Rastelli:2017ecj}
L.~Rastelli and X.~Zhou, ``{The Mellin Formalism for Boundary CFT$_d$},''
  \href{http://dx.doi.org/10.1007/JHEP10(2017)146}{{\em JHEP} {\bfseries 10}
  (2017) 146},
\href{http://arxiv.org/abs/1705.05362}{{\ttfamily arXiv:1705.05362 [hep-th]}}.

\bibitem{Billo:2016cpy}
M.~Billò, V.~Gonçalves, E.~Lauria, and M.~Meineri, ``{Defects in conformal
  field theory},'' \href{http://dx.doi.org/10.1007/JHEP04(2016)091}{{\em JHEP}
  {\bfseries 04} (2016) 091},
\href{http://arxiv.org/abs/1601.02883}{{\ttfamily arXiv:1601.02883 [hep-th]}}.

\bibitem{Gadde:2016fbj}
A.~Gadde, ``{Conformal constraints on defects},''
\href{http://arxiv.org/abs/1602.06354}{{\ttfamily arXiv:1602.06354 [hep-th]}}.

\bibitem{Liendo:2016ymz}
P.~Liendo and C.~Meneghelli, ``{Bootstrap equations for $ \mathcal{N} $ = 4 SYM
  with defects},'' \href{http://dx.doi.org/10.1007/JHEP01(2017)122}{{\em JHEP}
  {\bfseries 01} (2017) 122},
\href{http://arxiv.org/abs/1608.05126}{{\ttfamily arXiv:1608.05126 [hep-th]}}.

\bibitem{Lauria:2017wav}
E.~Lauria, M.~Meineri, and E.~Trevisani, ``{Radial coordinates for defect
  CFTs},''
\href{http://arxiv.org/abs/1712.07668}{{\ttfamily arXiv:1712.07668 [hep-th]}}.

\bibitem{Lemos:2017vnx}
M.~Lemos, P.~Liendo, M.~Meineri, and S.~Sarkar, ``{Universality at large
  transverse spin in defect CFT},''
\href{http://arxiv.org/abs/1712.08185}{{\ttfamily arXiv:1712.08185 [hep-th]}}.

\bibitem{Gliozzi:2014jsa}
F.~Gliozzi and A.~Rago, ``{Critical exponents of the 3d Ising and related
  models from Conformal Bootstrap},''
\href{http://arxiv.org/abs/1403.6003}{{\ttfamily arXiv:1403.6003 [hep-th]}}.

\bibitem{Hikami:2017hwv}
S.~Hikami, ``{Conformal Bootstrap Analysis for Yang-Lee Edge Singularity},''
\href{http://arxiv.org/abs/1707.04813}{{\ttfamily arXiv:1707.04813 [hep-th]}}.

\bibitem{ElShowk:2012ht}
S.~El-Showk, M.~F. Paulos, D.~Poland, S.~Rychkov, D.~Simmons-Duffin, and
  A.~Vichi, ``{Solving the 3D Ising Model with the Conformal Bootstrap},''
  \href{http://dx.doi.org/10.1103/PhysRevD.86.025022}{{\em Phys.Rev.}
  {\bfseries D86} (2012) 025022},
\href{http://arxiv.org/abs/1203.6064}{{\ttfamily arXiv:1203.6064 [hep-th]}}.

\bibitem{Kos:2014bka}
F.~Kos, D.~Poland, and D.~Simmons-Duffin, ``{Bootstrapping Mixed Correlators in
  the 3D Ising Model},'' \href{http://dx.doi.org/10.1007/JHEP11(2014)109}{{\em
  JHEP} {\bfseries 1411} (2014) 109},
\href{http://arxiv.org/abs/1406.4858}{{\ttfamily arXiv:1406.4858 [hep-th]}}.

\bibitem{Kos:2016ysd}
F.~Kos, D.~Poland, D.~Simmons-Duffin, and A.~Vichi, ``{Precision islands in the
  Ising and O(N ) models},''
  \href{http://dx.doi.org/10.1007/JHEP08(2016)036}{{\em JHEP} {\bfseries 08}
  (2016) 036},
\href{http://arxiv.org/abs/1603.04436}{{\ttfamily arXiv:1603.04436 [hep-th]}}.

\bibitem{Simmons-Duffin:2016wlq}
D.~Simmons-Duffin, ``{The Lightcone Bootstrap and the Spectrum of the 3d Ising
  CFT},'' \href{http://dx.doi.org/10.1007/JHEP03(2017)086}{{\em JHEP}
  {\bfseries 03} (2017) 086},
\href{http://arxiv.org/abs/1612.08471}{{\ttfamily arXiv:1612.08471 [hep-th]}}.

\bibitem{Fitzpatrick:2012yx}
A.~L. Fitzpatrick, J.~Kaplan, D.~Poland, and D.~Simmons-Duffin, ``{The Analytic
  Bootstrap and AdS Superhorizon Locality},''
  \href{http://dx.doi.org/10.1007/JHEP12(2013)004}{{\em JHEP} {\bfseries 1312}
  (2013) 004},
\href{http://arxiv.org/abs/1212.3616}{{\ttfamily arXiv:1212.3616 [hep-th]}}.

\bibitem{Komargodski:2012ek}
Z.~Komargodski and A.~Zhiboedov, ``{Convexity and Liberation at Large Spin},''
  \href{http://dx.doi.org/10.1007/JHEP11(2013)140}{{\em JHEP} {\bfseries 1311}
  (2013) 140},
\href{http://arxiv.org/abs/1212.4103}{{\ttfamily arXiv:1212.4103 [hep-th]}}.

\bibitem{Alday:2015eya}
L.~F. Alday, A.~Bissi, and T.~Lukowski, ``{Large spin systematics in CFT},''
  \href{http://dx.doi.org/10.1007/JHEP11(2015)101}{{\em JHEP} {\bfseries 11}
  (2015) 101},
\href{http://arxiv.org/abs/1502.07707}{{\ttfamily arXiv:1502.07707 [hep-th]}}.

\bibitem{Alday:2015ota}
L.~F. Alday and A.~Zhiboedov, ``{Conformal Bootstrap With Slightly Broken
  Higher Spin Symmetry},''
  \href{http://dx.doi.org/10.1007/JHEP06(2016)091}{{\em JHEP} {\bfseries 06}
  (2016) 091},
\href{http://arxiv.org/abs/1506.04659}{{\ttfamily arXiv:1506.04659 [hep-th]}}.

\bibitem{Alday:2015ewa}
L.~F. Alday and A.~Zhiboedov, ``{An Algebraic Approach to the Analytic
  Bootstrap},'' \href{http://dx.doi.org/10.1007/JHEP04(2017)157}{{\em JHEP}
  {\bfseries 04} (2017) 157},
\href{http://arxiv.org/abs/1510.08091}{{\ttfamily arXiv:1510.08091 [hep-th]}}.

\bibitem{Alday:2016njk}
L.~F. Alday, ``{Large Spin Perturbation Theory for Conformal Field Theories},''
  \href{http://dx.doi.org/10.1103/PhysRevLett.119.111601}{{\em Phys. Rev.
  Lett.} {\bfseries 119} no.~11, (2017) 111601},
\href{http://arxiv.org/abs/1611.01500}{{\ttfamily arXiv:1611.01500 [hep-th]}}.

\bibitem{Caron-Huot:2017vep}
S.~Caron-Huot, ``{Analyticity in Spin in Conformal Theories},''
  \href{http://dx.doi.org/10.1007/JHEP09(2017)078}{{\em JHEP} {\bfseries 09}
  (2017) 078},
\href{http://arxiv.org/abs/1703.00278}{{\ttfamily arXiv:1703.00278 [hep-th]}}.

\bibitem{Gribov:1961fr}
V.~N. Gribov, ``{Partial waves with complex orbital angular momenta and the
  asymptotic behavior of the scattering amplitude},'' {\em Sov. Phys. JETP}
  {\bfseries 14} (1962) 1395.
[Zh. Eksp. Teor. Fiz.41,1962(1961)].

\bibitem{Froissart:1961ux}
M.~Froissart, ``{Asymptotic behavior and subtractions in the Mandelstam
  representation},''
\href{http://dx.doi.org/10.1103/PhysRev.123.1053}{{\em Phys. Rev.} {\bfseries
  123} (1961) 1053--1057}.

\bibitem{lspt1}
L.~F. Alday, A.~Bissi, and E.~Perlmutter, ``{Holographic Reconstruction of AdS
  Exchanges from Crossing Symmetry},''
  \href{http://dx.doi.org/10.1007/JHEP08(2017)147}{{\em JHEP} {\bfseries 08}
  (2017) 147},
\href{http://arxiv.org/abs/1705.02318}{{\ttfamily arXiv:1705.02318 [hep-th]}}.

\bibitem{lspt2}
P.~Dey, K.~Ghosh, and A.~Sinha, ``{Simplifying large spin bootstrap in Mellin
  space},'' \href{http://dx.doi.org/10.1007/JHEP01(2018)152}{{\em JHEP}
  {\bfseries 01} (2018) 152},
\href{http://arxiv.org/abs/1709.06110}{{\ttfamily arXiv:1709.06110 [hep-th]}}.

\bibitem{lspt3}
J.~Henriksson and T.~Lukowski, ``{Perturbative Four-Point Functions from the
  Analytic Conformal Bootstrap},''
\href{http://arxiv.org/abs/1710.06242}{{\ttfamily arXiv:1710.06242 [hep-th]}}.

\bibitem{Alday:2017zzv}
L.~F. Alday, J.~Henriksson, and M.~van Loon, ``{Taming the $\epsilon$-expansion
  with Large Spin Perturbation Theory},''
\href{http://arxiv.org/abs/1712.02314}{{\ttfamily arXiv:1712.02314 [hep-th]}}.

\bibitem{Alday:2017vkk}
L.~F. Alday and S.~Caron-Huot, ``{Gravitational S-matrix from CFT dispersion
  relations},''
\href{http://arxiv.org/abs/1711.02031}{{\ttfamily arXiv:1711.02031 [hep-th]}}.

\bibitem{lspt4}
M.~van Loon, ``{The Analytic Bootstrap in Fermionic CFTs},''
  \href{http://dx.doi.org/10.1007/JHEP01(2018)104}{{\em JHEP} {\bfseries 01}
  (2018) 104},
\href{http://arxiv.org/abs/1711.02099}{{\ttfamily arXiv:1711.02099 [hep-th]}}.

\bibitem{lspt5}
J.~Henriksson and M.~van Loon, ``{Critical O(N) model to order $\epsilon^4$
  from analytic bootstrap},''
\href{http://arxiv.org/abs/1801.03512}{{\ttfamily arXiv:1801.03512 [hep-th]}}.

\bibitem{turi}
G.~J. Turiaci and A.~Zhiboedov, ``{Veneziano Amplitude of Vasiliev Theory},''
\href{http://arxiv.org/abs/1802.04390}{{\ttfamily arXiv:1802.04390 [hep-th]}}.

\bibitem{Sachdev:1993pr}
S.~Sachdev, ``{Polylogarithm identities in a conformal field theory in
  three-dimensions},''
  \href{http://dx.doi.org/10.1016/0370-2693(93)90935-B}{{\em Phys. Lett.}
  {\bfseries B309} (1993) 285--288},
\href{http://arxiv.org/abs/hep-th/9305131}{{\ttfamily arXiv:hep-th/9305131
  [hep-th]}}.

\bibitem{Vasiliev:1999ba}
M.~A. Vasiliev, ``{Higher spin gauge theories: Star product and AdS space},''
\href{http://arxiv.org/abs/hep-th/9910096}{{\ttfamily arXiv:hep-th/9910096
  [hep-th]}}.

\bibitem{Klebanov:2002ja}
I.~R. Klebanov and A.~M. Polyakov, ``{AdS dual of the critical O(N) vector
  model},'' \href{http://dx.doi.org/10.1016/S0370-2693(02)02980-5}{{\em Phys.
  Lett.} {\bfseries B550} (2002) 213--219},
\href{http://arxiv.org/abs/hep-th/0210114}{{\ttfamily arXiv:hep-th/0210114
  [hep-th]}}.

\bibitem{Giombi:2012ms}
S.~Giombi and X.~Yin, ``{The Higher Spin/Vector Model Duality},''
  \href{http://dx.doi.org/10.1088/1751-8113/46/21/214003}{{\em J. Phys.}
  {\bfseries A46} (2013) 214003},
\href{http://arxiv.org/abs/1208.4036}{{\ttfamily arXiv:1208.4036 [hep-th]}}.

\bibitem{Alday:2007mf}
L.~F. Alday and J.~M. Maldacena, ``{Comments on operators with large spin},''
  \href{http://dx.doi.org/10.1088/1126-6708/2007/11/019}{{\em JHEP} {\bfseries
  11} (2007) 019},
\href{http://arxiv.org/abs/0708.0672}{{\ttfamily arXiv:0708.0672 [hep-th]}}.

\bibitem{Witczak-Krempa:2015pia}
W.~Witczak-Krempa, ``{Constraining Quantum Critical Dynamics: (2+1)D Ising
  Model and Beyond},''
  \href{http://dx.doi.org/10.1103/PhysRevLett.114.177201}{{\em Phys. Rev.
  Lett.} {\bfseries 114} (2015) 177201},
\href{http://arxiv.org/abs/1501.03495}{{\ttfamily arXiv:1501.03495
  [cond-mat.str-el]}}.

\bibitem{Kravchuk:2017dzd}
P.~Kravchuk, ``{Casimir recursion relations for general conformal blocks},''
  \href{http://dx.doi.org/10.1007/JHEP02(2018)011}{{\em JHEP} {\bfseries 02}
  (2018) 011},
\href{http://arxiv.org/abs/1709.05347}{{\ttfamily arXiv:1709.05347 [hep-th]}}.

\bibitem{Simmons-Duffin:2016gjk}
D.~Simmons-Duffin, \href{http://dx.doi.org/10.1142/9789813149441_0001}{``{The
  Conformal Bootstrap},''} in {\em {Proceedings, Theoretical Advanced Study
  Institute in Elementary Particle Physics: New Frontiers in Fields and Strings
  (TASI 2015): Boulder, CO, USA, June 1-26, 2015}}, pp.~1--74.
\newblock 2017.
\newblock
\href{http://arxiv.org/abs/1602.07982}{{\ttfamily arXiv:1602.07982 [hep-th]}}.
\newblock

\bibitem{Osborn:1993cr}
H.~Osborn and A.~C. Petkou, ``{Implications of conformal invariance in field
  theories for general dimensions},''
  \href{http://dx.doi.org/10.1006/aphy.1994.1045}{{\em Annals Phys.} {\bfseries
  231} (1994) 311--362},
\href{http://arxiv.org/abs/hep-th/9307010}{{\ttfamily arXiv:hep-th/9307010
  [hep-th]}}.

\bibitem{csy}
A.~V. Chubukov, S.~Sachdev, and J.~Ye, ``{Theory of two-dimensional quantum
  Heisenberg antiferromagnets with a nearly critical ground state},''
\href{http://dx.doi.org/10.1103/PhysRevB.49.11919}{{\em Phys. Rev.} {\bfseries
  B49} (1994) 11919--11961}.

\bibitem{PhysRevE.79.041142}
O.~Vasilyev, A.~Gambassi, A.~Macioek, and S.~Dietrich, ``Universal scaling
  functions of critical casimir forces obtained by monte carlo simulations,''
  \href{http://dx.doi.org/10.1103/PhysRevE.79.041142}{{\em Phys. Rev. E}
  {\bfseries 79} (Apr, 2009) 041142}.

\bibitem{casimir2}
M.~Krech and D.~P. Landau, ``Casimir effect in critical systems: A monte carlo
  simulation,'' \href{http://dx.doi.org/10.1103/PhysRevE.53.4414}{{\em Phys.
  Rev. E} {\bfseries 53} (May, 1996) 4414--4423}.

\bibitem{casimir3}
M.~Krech, ``Casimir forces in binary liquid mixtures,''
  \href{http://dx.doi.org/10.1103/PhysRevE.56.1642}{{\em Phys. Rev. E}
  {\bfseries 56} (Aug, 1997) 1642--1659}.

\bibitem{difr}
P.~Di~Francesco, P.~Mathieu, and D.~Senechal,
  \href{http://dx.doi.org/10.1007/978-1-4612-2256-9}{{\em {Conformal Field
  Theory}}}.
\newblock Graduate Texts in Contemporary Physics. Springer-Verlag, New York,
1997.
\newblock

\bibitem{Gaberdiel:1994fs}
M.~Gaberdiel, ``{A General transformation formula for conformal fields},''
  \href{http://dx.doi.org/10.1016/0370-2693(94)90026-4}{{\em Phys. Lett.}
  {\bfseries B325} (1994) 366--370},
\href{http://arxiv.org/abs/hep-th/9401166}{{\ttfamily arXiv:hep-th/9401166
  [hep-th]}}.

\bibitem{Chen:2013dxa}
B.~Chen, J.~Long, and J.-j. Zhang, ``{Holographic Renyi entropy for CFT with W
  symmetry},'' \href{http://dx.doi.org/10.1007/JHEP04(2014)041}{{\em JHEP}
  {\bfseries 04} (2014) 041},
\href{http://arxiv.org/abs/1312.5510}{{\ttfamily arXiv:1312.5510 [hep-th]}}.

\bibitem{maloney}
Y.~Gobeil, A.~Maloney, G.~S. Ng, and J.-q. Wu, ``Thermal conformal blocks,''
  {\em to appear} .

\bibitem{ranga}
F.~M. Haehl, R.~Loganayagam, P.~Narayan, A.~A. Nizami, and M.~Rangamani,
  ``{Thermal out-of-time-order correlators, KMS relations, and spectral
  functions},'' \href{http://dx.doi.org/10.1007/JHEP12(2017)154}{{\em JHEP}
  {\bfseries 12} (2017) 154},
\href{http://arxiv.org/abs/1706.08956}{{\ttfamily arXiv:1706.08956 [hep-th]}}.

\bibitem{Dobrev:1977qv}
V.~K. Dobrev, G.~Mack, V.~B. Petkova, S.~G. Petrova, and I.~T. Todorov,
  ``{Harmonic Analysis on the n-Dimensional Lorentz Group and Its Application
  to Conformal Quantum Field Theory},''
\href{http://dx.doi.org/10.1007/BFb0009678}{{\em Lect. Notes Phys.} {\bfseries
  63} (1977) 1--280}.

\bibitem{Simmons-Duffin:2017nub}
D.~Simmons-Duffin, D.~Stanford, and E.~Witten, ``{A spacetime derivation of the
  Lorentzian OPE inversion formula},''
\href{http://arxiv.org/abs/1711.03816}{{\ttfamily arXiv:1711.03816 [hep-th]}}.

\bibitem{Paulos:2017fhb}
M.~F. Paulos, J.~Penedones, J.~Toledo, B.~C. van Rees, and P.~Vieira, ``{The
  S-matrix Bootstrap III: Higher Dimensional Amplitudes},''
\href{http://arxiv.org/abs/1708.06765}{{\ttfamily arXiv:1708.06765 [hep-th]}}.

\bibitem{Maldacena:2015waa}
J.~Maldacena, S.~H. Shenker, and D.~Stanford, ``{A bound on chaos},''
  \href{http://dx.doi.org/10.1007/JHEP08(2016)106}{{\em JHEP} {\bfseries 08}
  (2016) 106},
\href{http://arxiv.org/abs/1503.01409}{{\ttfamily arXiv:1503.01409 [hep-th]}}.

\bibitem{Aharony:2016dwx}
O.~Aharony, L.~F. Alday, A.~Bissi, and E.~Perlmutter, ``{Loops in AdS from
  Conformal Field Theory},''
  \href{http://dx.doi.org/10.1007/JHEP07(2017)036}{{\em JHEP} {\bfseries 07}
  (2017) 036},
\href{http://arxiv.org/abs/1612.03891}{{\ttfamily arXiv:1612.03891 [hep-th]}}.

\bibitem{Stanford:2015owe}
D.~Stanford, ``{Many-body chaos at weak coupling},''
  \href{http://dx.doi.org/10.1007/JHEP10(2016)009}{{\em JHEP} {\bfseries 10}
  (2016) 009},
\href{http://arxiv.org/abs/1512.07687}{{\ttfamily arXiv:1512.07687 [hep-th]}}.

\bibitem{Fitzpatrick:2011dm}
A.~L. Fitzpatrick and J.~Kaplan, ``{Unitarity and the Holographic
  $S$-Matrix},'' \href{http://dx.doi.org/10.1007/JHEP10(2012)032}{{\em JHEP}
  {\bfseries 1210} (2012) 032},
\href{http://arxiv.org/abs/1112.4845}{{\ttfamily arXiv:1112.4845 [hep-th]}}.

\bibitem{chubukov1994theory}
A.~V. Chubukov, S.~Sachdev, and J.~Ye, ``Theory of two-dimensional quantum
  heisenberg antiferromagnets with a nearly critical ground state,'' {\em
  Physical Review B} {\bfseries 49} no.~17, (1994) 11919.

\bibitem{ZinnJustin:2002ru}
J.~Zinn-Justin, {\em {Quantum Field Theory and Critical Phenomena}}.
\newblock Oxford, England: Clarendon Pr., 2002.
\newblock 1054 p.

\bibitem{Sachdev:1992py}
S.~Sachdev and J.~Ye, ``{Universal quantum critical dynamics of two-dimensional
  antiferromagnets},''
  \href{http://dx.doi.org/10.1103/PhysRevLett.69.2411}{{\em Phys. Rev. Lett.}
  {\bfseries 69} (1992) 2411},
\href{http://arxiv.org/abs/cond-mat/9204001}{{\ttfamily arXiv:cond-mat/9204001
  [cond-mat]}}.

\bibitem{Petkou:1998fb}
A.~C. Petkou and N.~D. Vlachos, ``{Finite size effects and operator product
  expansions in a CFT for d > 2},''
  \href{http://dx.doi.org/10.1016/S0370-2693(98)01530-5}{{\em Phys. Lett.}
  {\bfseries B446} (1999) 306--313},
\href{http://arxiv.org/abs/hep-th/9803149}{{\ttfamily arXiv:hep-th/9803149
  [hep-th]}}.

\bibitem{Petkou:1998fc}
A.~C. Petkou and N.~D. Vlachos, ``{Finite size and finite temperature effects
  in the conformally invariant O(N) vector model for 2 less than d less than
  4},'' in {\em {5th International Workshop on Thermal Field Theories and Their
  Applications Regensburg, Germany, August 10-14, 1998}}.
\newblock 1998.
\newblock
\href{http://arxiv.org/abs/hep-th/9809096}{{\ttfamily arXiv:hep-th/9809096
  [hep-th]}}.
\newblock

\bibitem{skv}
E.~D. Skvortsov, \href{http://dx.doi.org/10.1142/9789813144101_0008}{``{On
  (Un)Broken Higher-Spin Symmetry in Vector Models},''} in {\em {Proceedings,
  International Workshop on Higher Spin Gauge Theories: Singapore, Singapore,
  November 4-6, 2015}}, pp.~103--137.
\newblock 2017.
\newblock
\href{http://arxiv.org/abs/1512.05994}{{\ttfamily arXiv:1512.05994 [hep-th]}}.
\newblock

\bibitem{hs1}
V.~E. Didenko and M.~A. Vasiliev, ``{Static BPS black hole in 4d higher-spin
  gauge theory},'' \href{http://dx.doi.org/10.1016/j.physletb.2013.04.021,
  10.1016/j.physletb.2009.11.023}{{\em Phys. Lett.} {\bfseries B682} (2009)
  305--315}, \href{http://arxiv.org/abs/0906.3898}{{\ttfamily arXiv:0906.3898
  [hep-th]}}.
[Erratum: Phys. Lett.B722,389(2013)].

\bibitem{hs2}
V.~E. Didenko and E.~D. Skvortsov, ``{Exact higher-spin symmetry in CFT: all
  correlators in unbroken Vasiliev theory},''
  \href{http://dx.doi.org/10.1007/JHEP04(2013)158}{{\em JHEP} {\bfseries 04}
  (2013) 158},
\href{http://arxiv.org/abs/1210.7963}{{\ttfamily arXiv:1210.7963 [hep-th]}}.

\bibitem{hs3}
M.~A. Vasiliev, ``{Invariant Functionals in Higher-Spin Theory},''
  \href{http://dx.doi.org/10.1016/j.nuclphysb.2017.01.001}{{\em Nucl. Phys.}
  {\bfseries B916} (2017) 219--253},
\href{http://arxiv.org/abs/1504.07289}{{\ttfamily arXiv:1504.07289 [hep-th]}}.

\bibitem{hs4}
V.~E. Didenko, N.~G. Misuna, and M.~A. Vasiliev, ``{Charges in nonlinear
  higher-spin theory},'' \href{http://dx.doi.org/10.1007/JHEP03(2017)164}{{\em
  JHEP} {\bfseries 03} (2017) 164},
\href{http://arxiv.org/abs/1512.07626}{{\ttfamily arXiv:1512.07626 [hep-th]}}.

\bibitem{hs5}
C.~Iazeolla, E.~Sezgin, and P.~Sundell, ``{On Exact Solutions and Perturbative
  Schemes in Higher Spin Theory},''
  \href{http://dx.doi.org/10.3390/universe4010005}{{\em Universe} {\bfseries 4}
  no.~1, (2018) 5},
\href{http://arxiv.org/abs/1711.03550}{{\ttfamily arXiv:1711.03550 [hep-th]}}.

\bibitem{Camanho:2014apa}
X.~O. Camanho, J.~D. Edelstein, J.~Maldacena, and A.~Zhiboedov, ``{Causality
  Constraints on Corrections to the Graviton Three-Point Coupling},''
  \href{http://dx.doi.org/10.1007/JHEP02(2016)020}{{\em JHEP} {\bfseries 02}
  (2016) 020},
\href{http://arxiv.org/abs/1407.5597}{{\ttfamily arXiv:1407.5597 [hep-th]}}.

\bibitem{Afkhami-Jeddi:2016ntf}
N.~Afkhami-Jeddi, T.~Hartman, S.~Kundu, and A.~Tajdini, ``{Einstein gravity
  3-point functions from conformal field theory},''
  \href{http://dx.doi.org/10.1007/JHEP12(2017)049}{{\em JHEP} {\bfseries 12}
  (2017) 049},
\href{http://arxiv.org/abs/1610.09378}{{\ttfamily arXiv:1610.09378 [hep-th]}}.

\bibitem{Costa:2017twz}
M.~S. Costa, T.~Hansen, and J.~Penedones, ``{Bounds for OPE coefficients on the
  Regge trajectory},'' \href{http://dx.doi.org/10.1007/JHEP10(2017)197}{{\em
  JHEP} {\bfseries 10} (2017) 197},
\href{http://arxiv.org/abs/1707.07689}{{\ttfamily arXiv:1707.07689 [hep-th]}}.

\bibitem{Meltzer:2017rtf}
D.~Meltzer and E.~Perlmutter, ``{Beyond $a=c$: Gravitational Couplings to
  Matter and the Stress Tensor OPE},''
\href{http://arxiv.org/abs/1712.04861}{{\ttfamily arXiv:1712.04861 [hep-th]}}.

\bibitem{Myers:2016wsu}
R.~C. Myers, T.~Sierens, and W.~Witczak-Krempa, ``{A Holographic Model for
  Quantum Critical Responses},''
  \href{http://dx.doi.org/10.1007/JHEP09(2016)066,
  10.1007/JHEP05(2016)073}{{\em JHEP} {\bfseries 05} (2016) 073},
  \href{http://arxiv.org/abs/1602.05599}{{\ttfamily arXiv:1602.05599
  [hep-th]}}.
[Addendum: JHEP09,066(2016)].

\bibitem{Cordova:2017zej}
C.~Cordova, J.~Maldacena, and G.~J. Turiaci, ``{Bounds on OPE Coefficients from
  Interference Effects in the Conformal Collider},''
  \href{http://dx.doi.org/10.1007/JHEP11(2017)032}{{\em JHEP} {\bfseries 11}
  (2017) 032},
\href{http://arxiv.org/abs/1710.03199}{{\ttfamily arXiv:1710.03199 [hep-th]}}.

\bibitem{Intriligator:1998ig}
K.~A. Intriligator, ``{Bonus symmetries of N=4 superYang-Mills correlation
  functions via AdS duality},''
  \href{http://dx.doi.org/10.1016/S0550-3213(99)00242-4}{{\em Nucl. Phys.}
  {\bfseries B551} (1999) 575--600},
\href{http://arxiv.org/abs/hep-th/9811047}{{\ttfamily arXiv:hep-th/9811047
  [hep-th]}}.

\bibitem{gkp2}
S.~S. Gubser, I.~R. Klebanov, and A.~W. Peet, ``{Entropy and temperature of
  black 3-branes},'' \href{http://dx.doi.org/10.1103/PhysRevD.54.3915}{{\em
  Phys. Rev.} {\bfseries D54} (1996) 3915--3919},
\href{http://arxiv.org/abs/hep-th/9602135}{{\ttfamily arXiv:hep-th/9602135
  [hep-th]}}.

\bibitem{Goncalves:2014ffa}
V.~Gonçalves, ``{Four point function of $\mathcal{N}=4$ stress-tensor
  multiplet at strong coupling},''
  \href{http://dx.doi.org/10.1007/JHEP04(2015)150}{{\em JHEP} {\bfseries 04}
  (2015) 150},
\href{http://arxiv.org/abs/1411.1675}{{\ttfamily arXiv:1411.1675 [hep-th]}}.

\bibitem{Iliesiu:2018}
L.~Iliesiu, M.~Kologlu, R.~Mahajan, and D.~Simmons-Duffin, ``{The thermal
  bootstrap of the 3d Ising CFT},'' {\em to appear} .

\bibitem{Belin:2016yll}
A.~Belin, J.~de~Boer, J.~Kruthoff, B.~Michel, E.~Shaghoulian, and M.~Shyani,
  ``{Universality of sparse $d > 2$ conformal field theory at large $N$},''
  \href{http://dx.doi.org/10.1007/JHEP03(2017)067}{{\em JHEP} {\bfseries 03}
  (2017) 067},
\href{http://arxiv.org/abs/1610.06186}{{\ttfamily arXiv:1610.06186 [hep-th]}}.

\bibitem{Belin:2018jtf}
A.~Belin, J.~de~Boer, and J.~Kruthoff, ``{Comments on a state-operator
  correspondence for the torus},''
\href{http://arxiv.org/abs/1802.00006}{{\ttfamily arXiv:1802.00006 [hep-th]}}.

\bibitem{Giombi:2011kc}
S.~Giombi, S.~Minwalla, S.~Prakash, S.~P. Trivedi, S.~R. Wadia, and X.~Yin,
  ``{Chern-Simons Theory with Vector Fermion Matter},''
  \href{http://dx.doi.org/10.1140/epjc/s10052-012-2112-0}{{\em Eur. Phys. J.}
  {\bfseries C72} (2012) 2112},
\href{http://arxiv.org/abs/1110.4386}{{\ttfamily arXiv:1110.4386 [hep-th]}}.

\bibitem{Aharony:2011jz}
O.~Aharony, G.~Gur-Ari, and R.~Yacoby, ``{d=3 Bosonic Vector Models Coupled to
  Chern-Simons Gauge Theories},''
  \href{http://dx.doi.org/10.1007/JHEP03(2012)037}{{\em JHEP} {\bfseries 03}
  (2012) 037},
\href{http://arxiv.org/abs/1110.4382}{{\ttfamily arXiv:1110.4382 [hep-th]}}.

\bibitem{Jain:2012qi}
S.~Jain, S.~P. Trivedi, S.~R. Wadia, and S.~Yokoyama, ``{Supersymmetric
  Chern-Simons Theories with Vector Matter},''
  \href{http://dx.doi.org/10.1007/JHEP10(2012)194}{{\em JHEP} {\bfseries 10}
  (2012) 194},
\href{http://arxiv.org/abs/1207.4750}{{\ttfamily arXiv:1207.4750 [hep-th]}}.

\bibitem{Aharony:2012ns}
O.~Aharony, S.~Giombi, G.~Gur-Ari, J.~Maldacena, and R.~Yacoby, ``{The Thermal
  Free Energy in Large N Chern-Simons-Matter Theories},''
  \href{http://dx.doi.org/10.1007/JHEP03(2013)121}{{\em JHEP} {\bfseries 03}
  (2013) 121},
\href{http://arxiv.org/abs/1211.4843}{{\ttfamily arXiv:1211.4843 [hep-th]}}.

\bibitem{Gur-Ari:2016xff}
G.~Gur-Ari, S.~A. Hartnoll, and R.~Mahajan, ``{Transport in Chern-Simons-Matter
  Theories},'' \href{http://dx.doi.org/10.1007/JHEP07(2016)090}{{\em JHEP}
  {\bfseries 07} (2016) 090},
\href{http://arxiv.org/abs/1605.01122}{{\ttfamily arXiv:1605.01122 [hep-th]}}.

\bibitem{Maldacena:2011jn}
J.~Maldacena and A.~Zhiboedov, ``{Constraining Conformal Field Theories with A
  Higher Spin Symmetry},''
\href{http://arxiv.org/abs/1112.1016}{{\ttfamily arXiv:1112.1016 [hep-th]}}.

\bibitem{Maldacena:2012sf}
J.~Maldacena and A.~Zhiboedov, ``{Constraining conformal field theories with a
  slightly broken higher spin symmetry},''
  \href{http://dx.doi.org/10.1088/0264-9381/30/10/104003}{{\em
  Class.Quant.Grav.} {\bfseries 30} (2013) 104003},
\href{http://arxiv.org/abs/1204.3882}{{\ttfamily arXiv:1204.3882 [hep-th]}}.

\bibitem{Hijano:2015qja}
E.~Hijano, P.~Kraus, E.~Perlmutter, and R.~Snively, ``{Semiclassical Virasoro
  blocks from AdS$_{3}$ gravity},''
  \href{http://dx.doi.org/10.1007/JHEP12(2015)077}{{\em JHEP} {\bfseries 12}
  (2015) 077},
\href{http://arxiv.org/abs/1508.04987}{{\ttfamily arXiv:1508.04987 [hep-th]}}.

\bibitem{Hijano:2015zsa}
E.~Hijano, P.~Kraus, E.~Perlmutter, and R.~Snively, ``{Witten Diagrams
  Revisited: The AdS Geometry of Conformal Blocks},''
  \href{http://dx.doi.org/10.1007/JHEP01(2016)146}{{\em JHEP} {\bfseries 01}
  (2016) 146},
\href{http://arxiv.org/abs/1508.00501}{{\ttfamily arXiv:1508.00501 [hep-th]}}.

\bibitem{Fitzpatrick:2015qma}
A.~L. Fitzpatrick, J.~Kaplan, M.~T. Walters, and J.~Wang, ``{Eikonalization of
  Conformal Blocks},'' \href{http://dx.doi.org/10.1007/JHEP09(2015)019}{{\em
  JHEP} {\bfseries 09} (2015) 019},
\href{http://arxiv.org/abs/1504.01737}{{\ttfamily arXiv:1504.01737 [hep-th]}}.

\bibitem{Hasenbuschprivate}
M.~Hasenbusch, ``private correspondence,''.

\bibitem{Lashkari:2016vgj}
N.~Lashkari, A.~Dymarsky, and H.~Liu, ``{Eigenstate Thermalization Hypothesis
  in Conformal Field Theory},''
\href{http://arxiv.org/abs/1610.00302}{{\ttfamily arXiv:1610.00302 [hep-th]}}.

\bibitem{Jafferis:2017zna}
D.~Jafferis, B.~Mukhametzhanov, and A.~Zhiboedov, ``{Conformal Bootstrap At
  Large Charge},''
\href{http://arxiv.org/abs/1710.11161}{{\ttfamily arXiv:1710.11161 [hep-th]}}.

\bibitem{Fitzpatrick:2016ive}
A.~L. Fitzpatrick, J.~Kaplan, D.~Li, and J.~Wang, ``{On information loss in
  AdS$_{3}$/CFT$_{2}$},'' \href{http://dx.doi.org/10.1007/JHEP05(2016)109}{{\em
  JHEP} {\bfseries 05} (2016) 109},
\href{http://arxiv.org/abs/1603.08925}{{\ttfamily arXiv:1603.08925 [hep-th]}}.

\bibitem{kubo1}
R.~{Kubo}, ``{Statistical-Mechanical Theory of Irreversible Processes. I},''
  \href{http://dx.doi.org/10.1143/JPSJ.12.570}{{\em Journal of the Physical
  Society of Japan} {\bfseries 12} (June, 1957) 570--586}.

\bibitem{kubo2}
R.~{Kubo}, M.~{Yokota}, and S.~{Nakajima}, ``{Statistical-Mechanical Theory of
  Irreversible Processes. II. Response to Thermal Disturbance},''
  \href{http://dx.doi.org/10.1143/JPSJ.12.1203}{{\em Journal of the Physical
  Society of Japan} {\bfseries 12} (Nov., 1957) 1203--1211}.

\bibitem{kubo3}
R.~Kubo, ``The fluctuation-dissipation theorem,'' {\em Reports on Progress in
  Physics} {\bfseries 29} no.~1, (1966) 255.
  \url{http://stacks.iop.org/0034-4885/29/i=1/a=306}.

\bibitem{Hartnoll:2016apf}
S.~A. Hartnoll, A.~Lucas, and S.~Sachdev, ``{Holographic quantum matter},''
\href{http://arxiv.org/abs/1612.07324}{{\ttfamily arXiv:1612.07324 [hep-th]}}.

\bibitem{kadanoffmartin}
L.~P. {Kadanoff} and P.~C. {Martin}, ``{Hydrodynamic equations and correlation
  functions},'' \href{http://dx.doi.org/10.1016/0003-4916(63)90078-2}{{\em
  Annals of Physics} {\bfseries 24} (Oct., 1963) 419--469}.

\bibitem{Assel:2015nca}
B.~Assel, D.~Cassani, L.~Di~Pietro, Z.~Komargodski, J.~Lorenzen, and
  D.~Martelli, ``{The Casimir Energy in Curved Space and its Supersymmetric
  Counterpart},'' \href{http://dx.doi.org/10.1007/JHEP07(2015)043}{{\em JHEP}
  {\bfseries 07} (2015) 043},
\href{http://arxiv.org/abs/1503.05537}{{\ttfamily arXiv:1503.05537 [hep-th]}}.

\bibitem{Komargodski:2016auf}
Z.~Komargodski and D.~Simmons-Duffin, ``{The Random-Bond Ising Model in 2.01
  and 3 Dimensions},'' \href{http://dx.doi.org/10.1088/1751-8121/aa6087}{{\em
  J. Phys.} {\bfseries A50} no.~15, (2017) 154001},
\href{http://arxiv.org/abs/1603.04444}{{\ttfamily arXiv:1603.04444 [hep-th]}}.

\bibitem{Meneses:2018xpu}
S.~Meneses, S.~Rychkov, J.~M. V.~P. Lopes, and P.~Yvernay, ``{A structural test
  for the conformal invariance of the critical 3d Ising model},''
\href{http://arxiv.org/abs/1802.02319}{{\ttfamily arXiv:1802.02319 [hep-th]}}.

\bibitem{gb1}
L.~Hadasz, Z.~Jaskolski, and P.~Suchanek, ``{Recursive representation of the
  torus 1-point conformal block},''
  \href{http://dx.doi.org/10.1007/JHEP01(2010)063}{{\em JHEP} {\bfseries 01}
  (2010) 063},
\href{http://arxiv.org/abs/0911.2353}{{\ttfamily arXiv:0911.2353 [hep-th]}}.

\bibitem{gb2}
P.~Kraus, A.~Maloney, H.~Maxfield, G.~S. Ng, and J.-q. Wu, ``{Witten Diagrams
  for Torus Conformal Blocks},''
  \href{http://dx.doi.org/10.1007/JHEP09(2017)149}{{\em JHEP} {\bfseries 09}
  (2017) 149},
\href{http://arxiv.org/abs/1706.00047}{{\ttfamily arXiv:1706.00047 [hep-th]}}.

\end{thebibliography}\endgroup

\end{document}